\documentclass[aps,prb,twocolumn,amsmath,amsfonts,superscriptaddress]{revtex4-2}
\usepackage{graphicx}% Include figure files
\usepackage{dcolumn}% Align table columns on decimal point
\usepackage{bm}% bold math
\usepackage{boxedminipage}
\usepackage{amsmath}
\usepackage{graphicx}
\usepackage{amsfonts}
\usepackage{feynmf}
\usepackage{mathrsfs}
\usepackage{listings}
\usepackage{algorithm}
\usepackage{algorithmic}
\usepackage{color}
\usepackage{braket}
\usepackage{ulem}
\usepackage{hyperref,cleveref}
\usepackage[maxfloats=256]{morefloats}
\usepackage{txfonts}

\newcommand{\beq}{\begin{eqnarray}}
\newcommand{\eeq}{\end{eqnarray}}

\begin{document}

%\preprint{}

\title{Anomalous acoustoelectric effect induced by clapping modes in chiral superconductors}


\author{Taiki Matsushita}
\affiliation{Department of Materials Engineering Science, Osaka University, Toyonaka, Osaka 560-8531, Japan}
\author{Takeshi Mizushima}
\affiliation{Department of Materials Engineering Science, Osaka University, Toyonaka, Osaka 560-8531, Japan}
\author{Ilya Vekhter}
\affiliation{Department of Physics and Astronomy, Louisiana State University, Baton Rouge, LA 70803-4001}
\author{Satoshi Fujimoto}
\affiliation{Department of Materials Engineering Science, Osaka University, Toyonaka, Osaka 560-8531, Japan}
\affiliation{Center for Quantum Information and Quantum Biology, Osaka University, Toyonaka, Osaka 560-8531, Japan}

\date{\today}
\begin{abstract}
Clapping modes, which are relative amplitude and phase modes between two chiral components of Cooper pairs, are bosonic collective modes inherent to chiral superconductors.
These modes behave as long-lived bosons with masses smaller than the threshold energy, $2\Delta$, for decay into unbound fermion pairs.
Here, we clarify that the real/imaginary clapping modes in chiral superconductors directly couple to acoustic wave propagation when the weak particle-hole asymmetry of the normal state quasiparticle dispersion is taken into account.
The clapping modes driven by an acoustic wave generate an alternating electric current, that is, the acoustoelectric effect in superconductors.
Significantly, the clapping modes give rise to a transverse electric current.
When the sound velocity is comparable to the Fermi velocity, as in heavy fermion compounds, the transverse current is resonantly enhanced at energy below the threshold for continuum excitations.
This resonance provides a smoking-gun evidence of chiral superconductivity.
\end{abstract}

%\keywords{}

\maketitle
\section{Introduction}
Spontaneous breaking of the time-reversal symmetry (TRS) is an important concept in modern condensed matter physics.
The chiral superconducting state is an example of such an ordered state with spontaneously broken TRS, where electrons in the ground state form Cooper pairs with a fixed orbital angular momentum, $\Delta(\bm k)\propto (k_x+ik_y)^\nu\; (\nu \in {\mathbb Z})$~\cite{read2000paired,goswami2013topological,goswami2015topological}.
The integer, $\nu$, in the chiral superconducting order is the chirality of Cooper pairs associated with the orbital angular momentum.
This additional symmetry breaking enriches topological properties and transport phenomena in chiral superconductors~\cite{kallin2016chiral}.
For instance, such systems have been recognized as TRS-broken topological (Weyl) superconductors, where the chirality of the Cooper pairs is a source of non-trivial topology~\cite{volovik}.
In turn, TRS-broken topological superconductors can give rise to chiral Majorana fermion modes, which are essential to the field of fault-tolerant topological quantum computation~\cite{alicea2012new,sato2016majorana,sato2017topological,fu2008superconducting,sanno2021ab}.
Consequently, unequivocal identification of chiral superconducting order in candidate materials remains a high priority.

Over the last decade, chiral superconductivity has been considered in many heavy fermion compounds, such as URu$_2$Si$_2$, UPt$_3$, U$_{1-x}$Th$_x$Be$_{13}$, UCoGe, URhGe and UGe$_2$~\cite{kasahara2009superconducting,kittaka2016evidence,schemm2015evidence,schemm2014observation,tsutsumi2013upt3,yanase2016nonsymmorphic,maclaughlin1984nuclear,PhysRevLett.65.2816,jin1994low,golding1985observation,PhysRevLett.55.1319,machida2018spin,PhysRevB.87.180503,PhysRevB.89.020509}.
In URu$_2$Si$_2$, chiral $d$-wave pairing was supported by the colossal fluctuation-induced Nernst effect above $T_{\rm c}$, which stems from with scatterings of normal electrons through preformed chiral Copper pairs~\cite{sumiyoshi2014giant,yamashita2015colossal}.
Moreover, UPt$_3$ and U$_{1-x}$Th$_x$Be$_{13}$ are spin-triplet superconductors with multiple superconducting phases~\cite{stewart1984possibility,adenwalla1990phase,hasselbach1989critical,ott1983u,smith1984impurities,ott1985phase,kim1991investigation,rauchschwalbe1987phase}.
The TRS-broken superconducting state appears at low temperatures and low magnetic fields for UPt$_3$, and at low temperatures and in the range $0.019\lesssim x\lesssim 0.045$ for U$_{1-x}$Th$_x$Be$_{13}$~\cite{sauls1994order,schemm2014observation,tsutsumi2012spin,izawa2014pairing,avers,shimizu17,machidaJPSJ18,mizushima2018topology}.
The ferromagnetic materials, UCoGe, URhGe, and UGe$_2$, are also candidates for chiral superconductors~\cite{aoki2019review}.
These materials have strong magnetic Ising anisotropy and non-unitary spin-triplet Cooper pairing compatible with ferromagnetism~\cite{PhysRevLett.99.067006,mineev2017phase,aoki2001coexistence,saxena2000superconductivity,PhysRevB.66.134504}.
The angle-resolved NMR measurements in UCoGe suggest non-unitary chiral order with the $d$-vector represented by ${\bm d}(\bm k)\sim (a_1k_a+ia_2k_b,a_3k_b+ia_4k_a,0)$, where $a_i\; (i=1,2,3,4)$ are real coefficients~\cite{PhysRevLett.108.066403,tada2013spin}.
Recently, non-unitary chiral superconductivity was also proposed in UTe$_2$,
where the normal state is paramagnetic, but superconductivity survives even at extremely high magnetic fields over 40 T~\cite{Aoki_UTe2,ran2019nearly,ishihara2021chiral,hayes2020weyl}.
The superconducting state from the paramagnetic normal state shares many common features with ferromagnetic superconductors, including strong magnetic Ising anisotropy and the reentrant superconducting transition~\cite{knebel_Ute2,miyake2019metamagnetic}.

The chirality of Cooper pairs, $\nu$, is reflected in the anomalous transverse transport coefficients.
It is responsible for the anomalous thermal Hall effect and the fluctuation-driven Nernst effect~\cite{read2000paired,sumiyoshi2014giant,yamashita2015colossal}.
The mechanisms of the anomalous thermal Hall effect are classified into (i) intrinsic, which arises from the Berry curvature, and (ii) extrinsic, via asymmetric impurity scattering~\cite{sumiyoshi_ATHE,nomura2012cross,Arfi1988,ngampruetikorn2020impurity,yip2016low,yilmaz2020spontaneous}.
In chiral superconductors, the fluctuation-driven Nernst effect stems from skew scattering via preformed chiral Cooper pairs, qualitatively different from the conventional fluctuation-induced Nernst effect in superconductors~\cite{ussishkin2002gaussian}.
Here, we propose transport phenomena mediated by long-lived massive bosonic collective modes of the superconducting order parameter to identify chiral superconductors.
We show that the coupling of the acoustic waves traveling through a chiral superconductor to these modes generates a transverse alternating (ac) current.
This is reminiscent of the acoustoelectric effect (AEE), generation of an ac electric current by propagating acoustic waves in metals that was extensively studied since the 1950s~\cite{parmenter1953acousto,weinreich1957observation,weinreich1959acoustoelectric,kalameitsev2019valley,sukhachov2020acoustogalvanic}.
Since the transverse current we find is due solely to the chirality of the superconducting order, with no applied magnetic field, we refer to this effect as ``{\it anomalous acoustoelectric effect}'' (AAEE).
This effect is enhanced at resonant frequencies in heavy fermion materials, where the sound velocity is comparable to the Fermi velocity.

Bosonic collective modes directly reflect the symmetry of the order parameter, providing another probe of chiral superconductivity~\cite{vollhardt1990superfluid}.
In chiral superconductors, the characteristic modes are relative amplitude and phase oscillations between the two chiral components.
In analogy with superfluid $^3$He-A, these are referred to as real and imaginary clapping modes, respectively~\cite{vollhardt1990superfluid}.
In the weak coupling limit, the clapping modes always exist in any chiral superconductors with orbital angular momentum $|\nu| \ge 1$.
Coupling of collective modes to external fields depends on the symmetry of the order parameter.
In chiral superconductors, electromagnetic waves directly couple to the clapping mode, providing high resolution spectroscopy of bosonic excitation spectra in chiral ground states~\cite{PJH:1992,tewordt,higashitani,balatsky00,balatskyPRL00,kee,miura,sauls2015anisotropy}.

Here, we consider the response of clapping modes to an acoustic wave, which is a dynamical crystal deformation, and study the resulting transport phenomena inherent to chiral superconductors.
The advantages of acoustic waves are twofold; (i) the acoustic wave is free from the screening effect by the Meissner current, and hence can be utilized as a bulk probe, and (ii) linear coupling to the clapping modes depends on the effective mass of normal electrons.
Indeed, coupling of the sound waves to clapping modes was studied in Refs.~\onlinecite{Kee2000,higashitani} in the context of Sr$_2$RuO$_4$, with the conclusion that the effects are weak due to the mismatch between the speed of sound, $\bm v_s$,  and the Fermi velocity, $\bm v_{\rm F}$.
Large effective mass in heavy fermion materials makes the two velocities comparable.
We show that in this limit the clapping modes are resonantly excited by the acoustic waves.

\begin{figure}[t]
    \includegraphics[width=10cm]{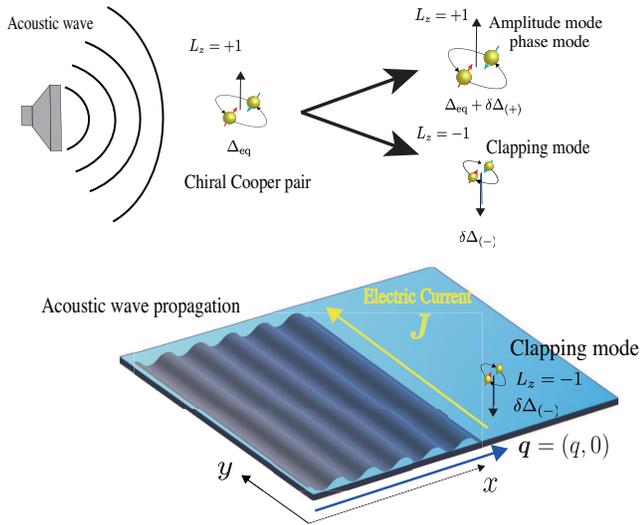}
    \caption{
        Schematic image of the AAEE in chiral superconductors: Clapping excitations of chiral Cooper pairs are driven by propagating acoustic waves, leading to transverse electric current.
    }
    \label{fig:AAEE}
\end{figure}

Using the augmented quasiclassical transport theory incorporating the weak particle-hole asymmetry (PHA) of normal electrons, we demonstrate that the acoustic waves propagating in chiral superconductors linearly couple to clapping modes through the PHA, and the clapping modes generate a transverse electric current characteristic of the AAEE, see Fig.~\ref{fig:AAEE}~\cite{SERENE1983221}.
The AAEE is a direct consequence of the formation of chiral Cooper pairs in the superconducting ground state, and the resonant behavior of the transverse current provides a direct bulk spectroscopy of chiral superconductivity in heavy fermion systems.

%%%%%%%%%%%%FIGURE
%%%%%%%%%%%%FIGURE

The organization of this paper is as follows.
In Sec.~\ref{chiral_fluc}, we introduce a model of chiral superconductors and clapping modes as the low-lying collective excitations.
In Sec.~\ref{quasiclassical limit_theory}, we present the quasiclassical transport theory incorporating the weak PHA, which is a powerful tool for studying transport phenomena in superconductors.
In Sec.~\ref{Response}, the linear response theory for acoustic waves is described on the basis of the Keldysh Green's function, and the acoustoelectric conductivity tensor is given in terms of the contributions from Bogoliubov quasiparticles and collective modes.
In Sec.~\ref{OP_fluc}, we present the numerical results on the dispersions of bosonic collective modes and the acoustelectric conductivity tensor in chiral $p$-wave superconductors.
We demonstrate that the acoustic wave propagation linearly couples to the clapping modes through the PHA of the density of states (DOS), giving rise to anomalous transverse current.
We summarize our results in Sec.~\ref{sec:conclusion}.
We describe the technical details on how to incorporate the weak PHA of the quasiparticle DOS into the quasiclassical Green's function, and how to obtain the collective modes, and physical observables in Appendices~\ref{sec:PHA} and \ref{sec:derivation}.

%%%%%%%%%%%%%%%%%%%%%%%%%%%%%%%%%%%%%%%%%%%%%%%%%%%%%%
%%%%%%%%%%%%%%%%%%%%%%%%%%%%%%%%%%%%%%%%%%%%%%%%%%%%%%

\section{Model and basics of Chirality fluctuations}
\label{chiral_fluc}
Let us introduce a simple model of chiral superconductors and the clapping modes as low-lying bosonic excitations.
In this paper, we consider the two-dimensional spinless chiral $p$-wave state on the cylindrical Fermi surface in equilibrium,
\begin{eqnarray}
    \Delta(\bm k)=\frac{\Delta_{\rm eq} (k_x+ik_y)}{k_{\rm F}},
    \label{eq:deltaeq}
\end{eqnarray}
where without loss of generality, we choose the equilibrium gap amplitude to be real, $\Delta_{\rm eq}\in \mathbb{R}$ by the gauge transformation.
The pairing state in Eq.~\eqref{eq:deltaeq} has a definite chirality $\nu = +1$ associated with the orbital angular momentum.
For the cylindrical Fermi surface, the formation of the chiral Cooper pairs opens an isotropic excitation gap in the fermionic energy spectrum and generates the nontrivial Berry flux in the momentum space~\cite{RevModPhys.82.1959,volovik,sato2016majorana,mizushimaJPSJ16}.
This is a simple model of TRS-broken topological superconductors with a non-trivial Chern number.

The chiral ground state is degenerate with respect to chiralities $\nu = \pm 1$, and there exists another ground state, $\Delta(\bm k)\propto (k_x-ik_y)$ with $\nu = -1$.
We assume here that in equilibrium a uniform chiral state of Eq.~\eqref{eq:deltaeq} without chiral domains is realized.
Then, the linear fluctuations of the order parameter around equilibrium are represented by $\delta \Delta({\bm k},Q) \equiv \Delta({\bm k},Q) - \Delta({\bm k})$,
\begin{eqnarray}
  \label{sc_fluc1}
  \delta \Delta({\bm k},Q)=\frac{\delta \Delta_{(+)}(Q) (k_x+ik_y)}{k_{\rm F}}+\frac{\delta \Delta_{(-)}(Q) (k_x-ik_y)}{k_{\rm F}},
\end{eqnarray}
where $Q\equiv (\omega,{\bm q})$ is the frequency ($\omega$) and the center-of-mass momentum (${\bm q}$) of Cooper pairs.
The subscript of the order parameter fluctuations, $\delta \Delta_{(\pm )}$, represents the chirality of Cooper pairs.
Note that $\delta \Delta_{(+)}$ corresponds to the amplitude and phase fluctuations in the equilibrium order with chirality $\nu =+1$, and $\delta \Delta_{(-)}$ has the opposite chirality to the equilibrium order and can be understood as a Berry phase fluctuation.

The collective modes in homogeneous superconductors are classified in terms of  parity under the particle-hole exchange, $\mathcal{C}=\pm 1$.
This classification gives four modes in the chiral superconductors,
\begin{gather}
    \label{sc_fluc_Dmode}
    \delta\mathcal{D}_\mathcal{C}(Q)\equiv\delta \Delta_{(+)}(Q)+\mathcal{C}\delta \Delta_{(+)}^*(Q),\\
    \label{sc_fluc_Emode}
    \delta\mathcal{E}_\mathcal{C}(Q)\equiv\delta \Delta_{(-)}(Q)+\mathcal{C}\delta \Delta_{(-)}^*(Q).
\end{gather}
The $\delta \mathcal{D}_{\pm}$-modes are collective modes of Cooper pairs with chirality $\nu=+1$, corresponding to the conventional amplitude and the phase modes, respectively.
These modes appear due to the broken $U(1)$ gauge symmetry and exist even in conventional superconductors.
The odd-parity $\delta \mathcal{D}_{\mathcal{-}}$-mode, the phase mode, is the Nambu-Goldstone mode associated with the broken $U(1)$ gauge symmetry,
which is gapped out by the Anderson-Higgs mechanism~\cite{hirashima1987p1,hirashima1987p2}.
The even-parity $\delta \mathcal{D}_{\mathcal{+}}$-mode, the amplitude mode, is known as the Higgs mode with the mass gap, $2|\Delta_{\rm eq}|$.
The mass gap corresponds to the fermionic continuum edge, beyond which Cooper pairs decay into two fermions.
The depairing of Cooper pairs introduces an intrinsic lifetime of the bosonic modes.
The $\delta \mathcal{E}_{\pm}$-modes are collective modes of Cooper pairs with opposite chirality $\nu=-1$, inherent to chiral superconductors.
The $\delta\mathcal{E}_\mathcal{+}$ and $\delta \mathcal{E}_\mathcal{-}$ modes in chiral superconductors are known as real and imaginary clapping modes, respectively, in analogy with the A-phase of superfluid $^3$He.
The mass gaps of two modes are degenerate at $\sqrt{2}|\Delta_{\rm eq}|$, in the cylindrical Fermi surface and the weak coupling limit, and
this degeneracy is lifted by applying the anisotropy of the superconducting gap~\cite{sauls2015anisotropy}.
Note that the mass gaps of two $\delta\mathcal{E}_{\pm}$-modes are always smaller than the depairing energy $2|\Delta_{\rm eq}|$ and thus these modes behave as long-lived bosons.
The long lifetime of the $\delta\mathcal{E}_{\pm}$-modes enables one to capture the signal of the $\delta\mathcal{E}_{\pm}$-modes by spectroscopies or transport measurements.
As the two clapping modes are inherent to chiral superconductors, a direct probe of the $\delta\mathcal{E}_{\pm}$ modes provides a fingerprint of chiral Cooper pairs.

To capture an essence of the interplay of collective modes and electric charge transport, in this paper, we consider the simple model for chiral $p$-wave superconductors.
We would however like to emphasize that the clapping modes, which are the key ingredients responsible for the AAEE, exist in any chiral superconductors with orbital angular momentum $|\nu|\ge 1$, and their mass gaps are always degenerate at $\omega=\sqrt{2}\Delta_{\rm eq}$ in the weak coupling limit~\cite{hsiao}.
Hence, our theory on the AAEE induced by the clapping modes is applicable to a variety of chiral superconductors.

%%%%%%%%%%%%%%%%%%%%%%%%%%%%%%%%%%%%%%%%%%%%%%%%%%%%%%
%%%%%%%%%%%%%%%%%%%%%%%%%%%%%%%%%%%%%%%%%%%%%%%%%%%%%%

\section{Quasiclassical Transport theory}
\label{quasiclassical limit_theory}

In our calculations, we employ the quasiclassical theory, which provides a powerful tool for describing superconducting phenomena~\cite{SERENE1983221}.
Typical length and energy scales in the superconducting state are the coherence length, $\xi_0 \equiv v_{\rm F}/2\pi k_{\rm B}T_{\rm c}$, and the excitation gap, $\Delta_{\rm eq} \sim k_{\rm B}T_{\rm c}$.
At weak coupling, both of these and other relevant parameters, such as temperature $T$, external potentials $V$, and characteristic frequencies $\omega$, are very small relative to the atomic scales, which are given by Fermi temperature $T_{\rm F}$, Fermi energy $\epsilon_{\rm F}$, and Fermi momentum $k_{\rm F}$.
This difference in scales allows one to perform an asymptotic expansion of full many-body propagators in small parameters $T/T_{\rm F}$, $V/\epsilon_{\rm F}$, $\omega/\epsilon_{\rm F}$, integrating out all quantities that vary rapidly on the atomic scales, determining the envelope functions that contain information about the observables. Similarly, for external fields varying at wave vectors
such that $q^{-1} \gtrsim \xi_0 \gg k^{-1}_{\rm F}$, the corresponding quasiclassical theory is local.

Traditionally, quasiclassical methods ignored the PHA near the Fermi surface.
It is, however, essential for our analysis, and below we show that the leading-order correction from the weak PHA results in the linear coupling of the clapping modes to acoustic waves, which drive the transverse electric current.

\subsection{Quasiclassical transport theory}

The central object of the quasiclassical theory is the quasiclassical Green's function, $\check{g}(\epsilon,\bm k_{\rm F},\bm x,t)$.
It can be thought of as the envelope of the full Green's function, which does not account for the rapid oscillations at the Fermi wavelength and time scales of the order of the inverse bandwith, but, instead, gives an effective low-energy description of transport phenomena.
Technically, it is obtained from the full Green's function, $\check{G}(\epsilon,\bm k,\bm x,t)$, where $\bm x$ and $t$ are the center of mass coordinate and time, and $\bm k$ and $\epsilon$ are the relative momentum and frequency, respectively, by integrating $\check{G}$ over a momentum shell $|\xi_k| = v_{\rm F}|k-k_{\rm F}|<\epsilon_{\rm c}\ll \epsilon _{\rm F}$, so that
\begin{eqnarray}
\label{Keldysh_prop0}
    \check{g}(\epsilon,\bm k_{\rm F},\bm x,t)=\int^{+\epsilon_{\rm c}}_{-\epsilon_{\rm c}} d\xi_k \frac{N(\xi_k+\epsilon_{\rm F})}{N(\epsilon_{\rm F})} \check{\tau}_z \check{G}(\epsilon,\bm k,\bm x,t)\,.
\end{eqnarray}
Since the Green's function is strongly peaked near the Fermi surface, its quasiclassical counterpart depends only weakly on the cutoff energy, $\epsilon_{\rm c}$, and the high energy contribution simply renormalizes the coupling constants (such as the effective mass or the superconducting pairing) that enter the low energy description.

For superconductors, Eq.~\eqref{Keldysh_prop0} is a matrix in the spin, particle-hole (Nambu), and Keldysh (retarded/advanced) space.
Hereafter, we assume that the spin structure of the superconducting order, described by the $\bm d$-vector, is fixed by the spin-orbit interaction, and therefore do not explicitly consider the spin degrees of freedom.
We denote $4\times 4\; (2\times 2)$ matrices in Keldysh (Nambu) space as $\check{a}\;(\underline{a})$.
If a matrix $\underline{a}$ is only defined in the Nambu space, the corresponding matrix in the Keldysh space is assumed to be $\check{a} = \underline{a}\otimes \openone$.
In Eq.~\eqref{Keldysh_prop0}, $\check{\tau}_z$ is the $z$-component of the Pauli matrix in the Nambu (particle-hole) space.
In the same equation, $N(\epsilon)$ is the DOS in the normal state, and $\xi_k$ is the kinetic energy of electrons measured with respect to the Fermi energy.

To take account of the leading-order correction to the quasiclassical limit, we expand the Green's function in the small quantity $(k_{\rm F}\xi_0)^{-1}\ll 1$,
\begin{eqnarray}
\label{Keldysh_prop}
    \check{g}(\epsilon,\bm k_{\rm F},\bm x,t)=\sum_{n=0}^\infty \check{g}_{(n)}(\epsilon,\bm k_{\rm F},\bm x,t),
\end{eqnarray}
At each order, we denote the Green's function as,
\begin{align}
    \check{g}_{(n)}=
    \begin{pmatrix}
        \underline{g}^{\rm R}_{(n)}&&\underline{g}^{\rm K}_{(n)}\\
        0&&\underline{g}^{\rm A}_{(n)}
    \end{pmatrix},\quad
    \underline{g}_{(n)}^{\rm X}=
    \begin{pmatrix}
        g_{(n)}^{\rm X}&&f_{(n)}^{\rm X}\\
        -\overline{f}_{(n)}^{\rm X}&&\overline{g}_{(n)}^{\rm X}
    \end{pmatrix},
\end{align}
where the superscript $\rm X=R, A, K$ represents the retarded, advanced, and Keldysh functions, respectively.

The $n=0$ component in Eq.~\eqref{Keldysh_prop} corresponds to the ``standard'' quasiclassical propagator,
\begin{eqnarray}
    \check{g}_{(0)}(\epsilon,\bm k_{\rm F},\bm x,t)&\equiv&\int d\xi_k  \check{\tau}_z \check{G}(\epsilon,\bm k,\bm x,t),
    \label{qcl_exp}
\end{eqnarray}
and describes the dynamics of quasiparticles and condensates in the quasiclassical limit, $(k_{\rm F}\xi_0)^{-1}=0$.
This quasiclassical limit propagator obeys the Eilenberger equation,
\begin{eqnarray}
    \label{Eilenberger1}
    \left[\epsilon \check{\tau}_z-\check{\Delta}-\check{\sigma}_{\rm imp}-\check{v}_{\rm ex},\check{g}_{(0)}\right]_\circ+i {\bm v}_{\rm F}\cdot {\bm \nabla} \check{g}_{(0)}=0\,.
\end{eqnarray}
In Eq.~\eqref{Eilenberger1}, we introduced the short-hand notation,  $\left[A,B\right]_\circ \equiv A\circ B-B\circ A$, defined with the $\circ$-product~\cite{eschrig2000distribution},
\begin{eqnarray}
	A \circ B \equiv \exp\left[ \frac{i}{2}\left(\partial_\epsilon \partial_{t'}-\partial_{\epsilon'} \partial_{t}\right) \right]A(\epsilon,t)B(\epsilon',t')\big|_{\epsilon=\epsilon',t=t'}.
\end{eqnarray}
In this paper, we set $\hbar=k_{\rm B}=e=1$.
The Green's function in the quasiclassical limit is supplemented by the normalization condition, $\check{g}^2_{(0)}=-\pi^2$ since $\check{g}^2_{(0)}$ also satisfies Eq.~(\ref{Eilenberger1}).
The superconducting order parameter matrix, $\check{\Delta}$, is defined as
\begin{eqnarray}
    \label{SCgap_mat}
    \underline{\Delta}(\bm k_{\rm F},\bm x,t)&=&
    \begin{pmatrix}
        0&&\Delta(\bm k_{\rm F},\bm x,t)\\
        -\Delta^\dag(\bm k_{\rm F},\bm x,t)&&0
    \end{pmatrix}.
\end{eqnarray}
In the following, we assume a clean limit and set the impurity self-energy, $\check{\sigma}_{\rm imp}=0$ since we focus on the (long wavelength) collective modes.

The external potential, $\check{v}_{\rm ex}$  in Eq.~\eqref{Eilenberger1}, results from the dynamical crystal deformation induced by an acoustic wave.
Crystal deformation changes the interatomic length and modifies the hopping integral of normal electrons and the electron energy~\cite{parmenter1953acousto,Shapourianstrain}.
In the long-wavelength limit (typical wavelength of the sound wave in crystals is $1.0 \times 10^{-2}\;{\rm cm}$, much longer than the lattice constants $\mathcal{O}( 1 \rm \AA)$),  the effects on the electrons can be described by the effective one-particle {\it deformation} potential,
$v_{\rm ex}(\bm x,t)$,  proportional to the symmetrized strain tensor, $u_{ij}=\frac{1}{2}(\partial_iu_j+\partial_ju_i)$, where $\bm u$ is the displacement vector~\cite{sukhachov2020acoustogalvanic}.
The deformation potential induced by the acoustic wave is given by,
\begin{eqnarray}
    \label{def_pot}
    \underline{v}_{\rm ex}(\bm x,t)=v_{\rm ex}(\bm x,t)\underline{\tau}_0
    =v_{\rm ex0}\exp \left[i(\bm q \cdot \bm x-\omega t)\right]\underline{\tau}_0,
\end{eqnarray}
where $\underline{\tau}_0$ is the $2\times 2$ identity matrix in the Nambu space,
$\omega=v_s |\bm q|$ is the frequency of the acoustic wave with the wave vector $\bm q$,
and $v_{\rm s}$ is the sound velocity.

\subsection{Particle-hole asymmetry in the density of state}
We described above the quasiclassical transport theory of superconductors in the limit $(k_{\rm F}\xi_0)^{-1}=0$.
The quasiclassical limit, $(k_{\rm F}\xi_0)^{-1}=0$, postulates that the Fermi surface of normal electrons is sufficiently large and thus the DOS in Eq.~\eqref{Keldysh_prop} is replaced by $N(\epsilon_{\rm F})$, and the superconducting order parameter, and all the potentials have been pinned to the Fermi surface values.
In this work, we focus on the leading-order correction from the PHA in the DOS of normal electrons to the transport coefficients due to the collective modes.
This corresponds to accounting for the slope of the DOS at the Fermi energy, $\frac{\partial N(\epsilon)}{\partial \epsilon}\big|_{\epsilon=\epsilon_{\rm F}}$,  in evaluating the propagator, Eq.~\eqref{Keldysh_prop}, which, in turn, induces the weak PHA in superconducting states~\cite{yip1992circular,ueki2018charging,masaki2019vortex}.
Below, we demonstrate that the PHA drastically changes the linear coupling between external fields and collective modes.

While the PHA is very small in conventional superconductors, the large DOS peak in heavy fermion superconductors means that the PHA becomes appreciable.
In addition, we emphasize that even small PHA may lead to appreciable observable consequences. Superfluid $^3$He is a typical example.
The bulk normal $^3$He has a large Fermi surface, and the PHA contribution can be roughly estimated as $\Delta/\epsilon_{\rm F}\sim 10^{-3}$.
In the B-phase of superfluid $^3$He, however, it has been predicted that the PHA correction alters the coupling of the stress tensor to the order parameter fluctuations, and the {\it real squashing mode} significantly contributes to the attenuation of the longitudinal zero sound even though the linear coupling of such mode to zero sound is suppressed by the approximate particle-hole symmetry~\cite{koch,sauls}.
This mode has been detected as a sharp resonant peak in the absorption spectrum of longitudinal sound~\cite{mast,giannetta,Avenel,movshovich}.
Hence, the PHA correction to the collective dynamics of Cooper pairs makes a significant contribution in clean superconductors and superfluids even when the factor $\Delta/\epsilon_{\rm F} \sim 1/(k_{\rm F}\xi_0)$ is small.

The leading-order correction due to the PHA appears in the term, $\check{g}_{(1)}$, in Eq.~\eqref{Keldysh_prop}, and hence we keep this term but ignore the higher-order corrections, $\check{g}_{(n \ge 2)}$.
A seeming problem with expressing the quasiclassical propagator as,
\begin{eqnarray}
    \label{ex_QC_gr}
     \check{g} \simeq  \check{g}_{(0)}+\check{g}_{(1)}\,,
\end{eqnarray}
is that, generally, $\check{g}_{(n\ge 1)}$ breaks the normalization condition, and thus the correction to the Eilenberger equation for $\check{g}_{(n\ge 1)}$, which is a homogeneous equation, cannot have a unique solution.
As shown in Appendix~\ref{sec:PHA}, however, the leading-order correction is obtained from the Green's function in the quasiclassical limit, $\check{g}_{(0)}$, as
\begin{eqnarray}
    \label{PHA_Keldysh}
    \check{g}_{(1)}=\frac{a}{2\epsilon_{\rm F}}\left[ \epsilon \check{\tau}_z-\check{\Delta}-\check{v}_{\rm ex},\check{g}_{(0)} \right]_{\circ +},
\end{eqnarray}
where we have introduced the dimensionless material parameter, $a\equiv \frac{\epsilon_{\rm F}}{N(\epsilon_{\rm F})}\frac{\partial N(\epsilon_{\rm F})}{\partial \epsilon}\big|_{\epsilon=\epsilon_{\rm F}}\sim \mathcal{O}(1)$, and
the anti-commutator $[\check{A},\check{B}]_{\circ +}\equiv \check{A}\circ \check{B}+\check{B}\circ\check{A}$.

With the quasiclassical Green's function, the electric current is expressed as,
\begin{eqnarray}
          \label{electric_current1}
	{\bm J}&\simeq&-N(\epsilon_{\rm F})\int \frac{d\epsilon}{4\pi i}\left\langle \frac{1}{2}{\bm v}_{\rm F}{\rm Tr}\left[\underline{\tau}_z\left( \underline{g}^{\rm K}_{(0)}+ \underline{g}^{\rm K}_{\rm (1)}\right)\right]\right\rangle_{{\rm FS},{\bm k}_{\rm F}}.
\end{eqnarray}
where $\rm Tr \left[ \cdots \right]$ represents the trace in the Nambu space, the bracket, $\braket{\cdots}_{{\rm FS},\bm k_{\rm F}}$, denotes the Fermi surface average~\cite{ueki2018charging,com2}. %, and $\kappa=1+a_v/a$, with $a_v=\frac{\epsilon_{\rm F}}{v_{\rm F}}\frac{\partial v_(\epsilon)}{\partial \epsilon}\big|_{\epsilon=\epsilon_{\rm F}}$ appearing due to the expansion of the velocity in the quasiparticle energy near the Fermi surface.
For the details on the derivation, see Appendix~\ref{sec:PHA}.
The effect of the PHA is now included (to leading order) into the second term of Eq.~(\ref{electric_current1}).
We now calculate the quasiclassical Keldysh Green's function as the linear response to the deformation potential, and obtain the electric current from Eq.~\eqref{electric_current1}.
Then, the acoustoelectric conductivity, $\chi_{ij}$, is defined as,
\begin{eqnarray}
	J_i=\chi_{ij}\left(-\frac{\partial v_{\rm ex}}{\partial x_j} \right).
\end{eqnarray}

We note that, in addition to the correction considered above, there also exists a quantum correction to the quasiclassical transport, Eq.~\eqref{Eilenberger1}.
It is obtained from the higher order contribution of the gradient expansion, which brings about quasiparticle transport phenomena mediated by nontrivial geometric structures in real and momentum spaces~\cite{kobayashi2018negative}.
The quantum correction is responsible, for example, for the intrinsic anomalous thermal Hall effect and negative thermal magnetoresistivity~\cite{kobayashi2018negative}.
However, the geometric phase is not the primary for the collective dynamics of condensates.
In this paper, therefore, we focus on the effects arising from the weak PHA.

%%%%%%%%%%%%%%%%%%%%%%%%%%%%%%%%%%%%%%%%%
%%%%%%%%%%%%%%%%%%%%%%%%%%%%%%%%%%%%%%%%%

{\section{Electric current and collective excitations induced by the acoustic wave}}
\subsection{Transverse electric current induced by the acoustic wave}
\label{AAEE_Higgs}

{We are now in the position to compute
%In this section, we describe
the linear response of the electric current to the deformation potential, and %present the expression of
the anomalous electric conductivity tensor. We outline the calculation below, while giving the technical details in Appendix B.}

{Our starting point is the equilibrium version of the Eilenberger equation, Eq.~\eqref{Eilenberger1}, which reads
\begin{eqnarray}
    \label{Eilenberger_eq_main}
    \left[\epsilon \check{\tau}_z-\check{\Delta}_{\rm eq},\check{g}_{\rm (0)eq}\right]=0,
\end{eqnarray}
subject to $\check{g}_{\rm (0)eq}^2=-\pi^2$. The solution of this equation is well-known, and is given in App.~\ref{GF_eq}. This allows us to compute the correction to the Green's function due to the normal state particle-hole asymmetry from the equilibrium version of Eq.~\eqref{PHA_Keldysh}, with the result given in Eq.~\eqref{PHA_Keldysh_eq}. }

{%For the simple calculations, we assume the homogeneous equilibrium gap function without the domain structures of the order parameter, and compute
We now derive the acoustoelectric conductivity as the linear response to the deformation potential.
Once again we separate the contribution without particle-hole anisotropy by solving first the non-equilibrium equation
\begin{align}
    \label{Eilenberger_noneq_main}
    \left[\epsilon \check{\tau}_z-\check{\Delta}_{\rm eq},\delta \check{g}_{(0)}\right]_\circ-\left[\delta \check{\Delta}+\check{v}_{\rm ex},\check{g}_{\rm (0) eq}\right]_\circ %\nonumber\\
   +i {\bm v}_{\rm F}\cdot {\bm \nabla} \delta \check{g}_{(0)}=0\,,
\end{align}
which includes the dynamical fluctuations of the order parameter, $\delta \check{\Delta}$, which have to be determined self-consistently by solving the gap equation. We then use this solution to obtain leading order corrections due to PHA. The details are given in Appendix~\ref{app:linear}.}

{Using this solution to evaluate Eq.~(\ref{electric_current1}), we obtain the electric current as the sum of two contributions, due to Bogoliubov quasiparticles (QP) and collective modes (CM) espectively.
\begin{eqnarray}
    \label{electric_current_main}
    {\bm J}&=&{\bm J}_{\rm QP}+{\bm J}_{\rm CM}\,,
\end{eqnarray}
where the first term is proportional to $v_{\rm ex0}$ explicitly, while the second depends on the deformation potential via the order parameter fluctuations, $\delta\Delta_\pm=\delta \Delta\pm \delta \Delta^{\ast}$, see Eqs.~\eqref{electric_currentQP}-\eqref{electric_currentCM}.
In our model, without loss of generality, we consider the acoustic wave propagating along the $x$-direction, $\bm q=(q,0)$.
Expressing the order parameter fluctuations via the collective modes using Eqs.~\eqref{sc_fluc1}-\eqref{sc_fluc_Emode}, 
\begin{gather}
    \label{sc_D+_main}
    k_F\delta\Delta_+=\left[\delta\mathcal{D}_++\delta\mathcal{E}_+\right]k_x+i\left[\delta\mathcal{D}_- - \delta\mathcal{E}_-\right]k_y\,,
    \\
    \label{sc_D-_main}
    k_F\delta\Delta_-=\left[\delta\mathcal{D}_- + \delta\mathcal{E}_-\right]k_x+i\left[\delta\mathcal{D}_+ - \delta\mathcal{E}_+\right]k_y\,,
\end{gather}
we connect the current to the collective mode propagators.}

% using the quasiclassical Keldysh theory with the PHA corrections.
{Similarly to the current,
the acoustoelectric conductivity tensor is decomposed into the contributions from the Bogoliubov quasiparticles ($\chi_{ij}^{\rm QP}$) and collective modes ($\chi_{ij}^{\rm CM}$) as}
\begin{eqnarray}
	\label{AE_conductivity}
	\chi_{ij}=\chi_{ij}^{\rm QP}+\chi_{ij}^{\rm CM},
\end{eqnarray}
where
\begin{eqnarray}
	\label{chixx_QP}
   	\chi_{xx}^{\rm QP}&=&\frac{iN(\epsilon_{\rm F})v_{\rm F}^2(X_0+X_1)}{4\omega},\\
    	\chi_{yx}^{\rm QP}&=&0,\\
	\label{chixx_CM}
    	\chi_{xx}^{\rm CM}&=&-\frac{iN(\epsilon_{\rm F})v_{\rm F}^2\Delta_{\rm eq}}{8}\left[\left(\overline{\lambda}_1+\overline{\lambda}_2\right)\frac{\delta \mathcal{E}_-}{v_{\rm ex0}} +\frac{a}{2\epsilon_{\rm F}}\left(\varphi_0+\varphi_1\right)\frac{\delta \mathcal{E}_+}{v_{\rm ex0}}\right]\nonumber\\\\
	\label{chiyx_CM}
         \chi_{yx}^{\rm CM}&=&\frac{N(\epsilon_{\rm F})v_{\rm F}^2\Delta_{\rm eq}}{8}\left[\left(\overline{\lambda}_0-\overline{\lambda}_2\right)\frac{\delta \mathcal{E}_+}{v_{\rm ex0}}-\frac{a}{2\epsilon_{\rm F}} \left(\varphi_0-\varphi_2\right)\frac{\delta \mathcal{E}_-}{v_{\rm ex0}}\right]
         \,.\nonumber\\
\end{eqnarray}
{Here $\lambda=|\Delta_{\rm eq}|^2 \overline{\lambda}$ is the generalized Tsuneto function, Eq.~\eqref{gen:Tsuneto}, the function $\varphi$ is defined in Eq.~\eqref{phi}, and we introduced the moments of those functions, $\overline{\lambda}_n=\braket{(\hat{k}_x^2-\hat{k}_y^2)^n\overline{\lambda}}_{{\rm FS},\bm k_{\rm F}}$, $X_n=\braket{(\hat{k}_x^2-\hat{k}_y^2)^n \frac{2\omega^2}{\omega^2-\eta^2}(\lambda-1)}_{{\rm FS},\bm k_{\rm F}}$, as well as $\varphi_n=\braket{(\hat{k}_x^2-\hat{k}_y^2)^n\varphi}_{{\rm FS},\bm k_{\rm F}}$. As above, angle brackets denote the normalized Fermi surface average, and $\eta=\bm v_F\cdot\bm q$.
The second terms of Eqs.~\eqref{chixx_CM} and \eqref{chiyx_CM} arise from the PHA term of the Keldysh response function.
The PHA effect is also incorporated into the clapping modes ($\delta \mathcal{E}_\pm$).}

Eq.~(\ref{chixx_QP}) shows that a propagating acoustic wave generates quasiparticle-mediated longitudinal current.
This is an extension of the AEE in normal metals to the superconducting state, which always exists regardless of the symmetry of the superconducting order~\cite{parmenter1953acousto}.
However, the Bogoliubov quasiparticles carry no transverse current in the clean limit.
In addition to the quasiparticle current, Eqs.~\eqref{chixx_CM} and \eqref{chiyx_CM} show that the clapping modes ($\delta \mathcal{E}_\pm$) carry the electric current.
In particular, the clapping modes ($\delta \mathcal{E}_\pm$) lead to a transverse electric current, flowing perpendicular to the direction of propagation of the acoustic wave.
Hence, this anomalous transverse current provides a direct probe of the clapping modes, and carries a fingerprint of chiral Cooper pairs.

%%%%%%%%%%%%%%%%%%%%%%%%%%%%%%%%%%%%%%%%%
%%%%%%%%%%%%%%%%%%%%%%%%%%%%%%%%%%%%%%%%%

{\subsection{Excitation of the collective modes by the acoustic waves}
\label{OP_fluc}
In Eqs.~\eqref{chixx_CM} and \eqref{chiyx_CM}, we expressed the acoustoelectric conductivity tensor using the clapping modes induced by the acoustic wave.
These order parameter fluctuations have to be separately determined by solving the superconducting gap equation under the perturbing potential.}

{We now proceed to determine the dispersion of the collective modes in chiral superconductors, and show that the clapping mode linearly couples to the acoustic wave in the presence of the PHA.
Substituting the nonequilibrium Keldysh pair amplitudes into the gap equation, we obtain the matrix form of the equation for the order parameter fluctuations as,
\begin{widetext}
\begin{eqnarray}
    \label{eigeneq}
    &\left(\begin{smallmatrix}
        \frac{\omega^2-4\Delta_{\rm eq}^2}{2}\overline{\lambda}_0-\frac{v_{\rm F}^2q^2 (\overline{\lambda}_0+\overline{\lambda}_1)}{4}&&
        \frac{\omega^2-4\Delta_{\rm eq}^2}{2}\overline{\lambda}_1-\frac{v_{\rm F}^2q^2(\overline{\lambda}_1+\overline{\lambda}_2)}{4}&&
        \frac{a\omega}{2\epsilon_{\rm F}}\left[\frac{\omega^2-4\Delta_{\rm eq}^2}{2}\overline{\lambda}_1-\frac{v_{\rm F}^2q^2(\overline{\lambda}_1+\overline{\lambda}_2)}{4}\right]\\
        \frac{\omega^2-4\Delta_{\rm eq}^2}{2}\overline{\lambda}_1-\frac{v_{\rm F}^2q^2(\overline{\lambda}_1+\overline{\lambda}_2)}{4}&&
        \frac{\omega^2\overline{\lambda}_0-4\Delta_{\rm eq}^2\overline{\lambda}_2}{2}-\frac{v_{\rm F}^2q^2 (\overline{\lambda}_0+\overline{\lambda}_1)}{4}&&
        \frac{a\omega}{2\epsilon_{\rm F}}\left[\frac{\omega^2-4\Delta_{\rm eq}^2}{2}\overline{\lambda}_0-\frac{v_{\rm F}^2q^2(\overline{\lambda}_0+\overline{\lambda}_1)}{4}+\gamma \right]\\
        \frac{a\omega}{2\epsilon_{\rm F}}\left[\frac{\omega^2-4\Delta_{\rm eq}^2}{2}\overline{\lambda}_1-\frac{v_{\rm F}^2q^2(\overline{\lambda}_1+\overline{\lambda}_2)}{4}\right]&&
        \frac{a\omega}{2\epsilon_{\rm F}}\left[\frac{\omega^2-4\Delta_{\rm eq}^2}{2}\overline{\lambda}_0-\frac{v_{\rm F}^2q^2(\overline{\lambda}_0+\overline{\lambda}_1)}{4}+\gamma\right]&&
        \frac{\omega^2\overline{\lambda}_0-4\Delta_{\rm eq}^2(\overline{\lambda}_0-\overline{\lambda}_2)}{2}-\frac{v_{\rm F}^2q^2 (\overline{\lambda}_0+\overline{\lambda}_1)}{4}
    \end{smallmatrix}\right)
    \begin{pmatrix}
        \delta \mathcal{D}_+\\
        \delta \mathcal{E}_+\\
        \delta \mathcal{E}_-
    \end{pmatrix}
    %\nonumber\\
    =
    \begin{pmatrix}_
       \frac{av_{\rm ex0}}{\epsilon_{\rm F}}\left(\omega\Delta_{\rm eq}\overline{\lambda}_0+ 2\Delta_{\rm eq}(\gamma-X_0)\right)\\
       \frac{av_{\rm ex0}}{\epsilon_{\rm F}}\left(\omega\Delta_{\rm eq}\overline{\lambda}_1- 2\Delta_{\rm eq}X_1\right)\\
        2\omega \Delta_{\rm eq}\overline{\lambda}_1 v_{\rm ex0}
    \end{pmatrix},
\end{eqnarray}
\end{widetext}
where $\gamma \simeq\frac{2}{N(\epsilon_{\rm F})V_{\rm pair}}=2\ln \left(\frac{1.13\epsilon_c}{T_c}\right)$.
The details of the derivation of Eq.~\eqref{eigeneq} are given in Appendix~\ref{sec:derivation}.}
The kernel of the matrix in Eq.~(\ref{eigeneq}) gives the eigenfrequencies of bosonic excitations, $\omega^C_{\Gamma}(\bm q)$ ($\Gamma=\delta \mathcal{D},~\delta \mathcal{E}$), and the damping rates of each mode~\cite{uematsu2019chiral}.
The right-hand side of Eq.~\eqref{eigeneq} represents the driving force from external perturbations, such as acoustic waves.
{In Eq.~\eqref{eigeneq}, the phase mode ($\delta \mathcal{D}_-$) is neglected since the phase mode is gapped out by the Anderson-Higgs mechanism.}

It is instructive to first consider the collective modes of the order parameter in the absence of the driving potential $v_{\rm ex0}=0$.
If we ignore the PHA and set $a=0$, the matrix in Eq.~\eqref{eigeneq} is block-diagonalized to the $C=+$ and $C=-$ subsectors.
In the long wavelength limit, $\bm q\rightarrow 0$, Eq.~\eqref{eigeneq} reduces to
\begin{align}
    \label{gapeq_lwl}
    \left(\begin{matrix}
        \omega^2-4\Delta_{\rm eq}^2&&
        0&&0&\\
        0&&
        \omega^2-2\Delta_{\rm eq}^2
        &&0\\
        0&&0&&
        \omega^2-2\Delta_{\rm eq}^2
    \end{matrix}\right)
    \begin{pmatrix}
        \delta \mathcal{D}_+\\
        \delta \mathcal{E}_+\\
        \delta \mathcal{E}_-
    \end{pmatrix}=
    \begin{pmatrix}
        0\\
        0\\
        0
    \end{pmatrix},%\nonumber\\
\end{align}
where we use the fact that in this limit the moments of the Tsuneto functions become $\overline{\lambda}_0=\overline{\lambda},\;\overline{\lambda}_1=0$ and $\overline{\lambda}_2=\overline{\lambda}/2$~\cite{com4}.
We find therefore that the eigenfrequency of the amplitude Higgs mode is $\omega^+_{\delta \mathcal{D}}=2|\Delta_{\rm eq}|$, while the real/imaginary clapping modes are degenerate with $\omega^{\pm}_{\delta \mathcal{E}}=\sqrt{2}|\Delta_{\rm eq}|$.

For ${\bm q}\neq{\bm 0}$ and $a\neq 0$, the matrix in Eq.~\eqref{eigeneq}
is not diagonal, and therefore the amplitude and the clapping modes hybridize.
We denote the corresponding eigenmodes $\delta \mathcal{D}_1$ and $\delta \mathcal{E}_{1,2}$, and
define them as being smoothly connected to one of the original modes, namely
\begin{gather}
	\lim_{a,\bm q\to 0}\delta \mathcal{D}_{1}=\delta  \mathcal{D}_{+},\\
	\lim_{a,\bm q\to 0}\delta \mathcal{E}_{1}=\delta  \mathcal{E}_{+},\\
	\lim_{a,\bm q\to 0}\delta \mathcal{E}_{2}=\delta  \mathcal{E}_{-}.
\end{gather}
In principle, these modes can be described in the framework of the time-dependent Ginzburg-Landau formalism~\cite{sauls2015anisotropy}.

\begin{figure}[t!]
    \includegraphics[width=6cm]{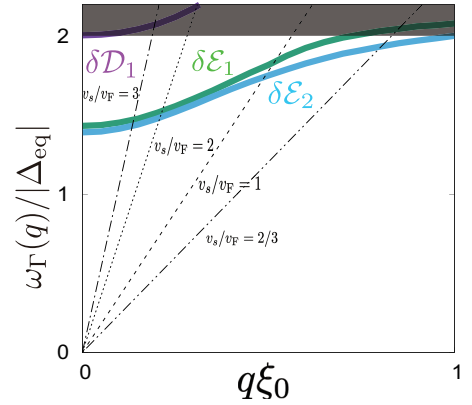}
    \caption{
        The dispersions of the $\delta \mathcal{D}_1$ and $\delta \mathcal{E}_{1,2}$ eigenmodes (thick curves) and the phonon (dashed/dotted curve) in the quasiclassical limit, $a=0$.
        The phonon dispersions are plotted for the sound velocity $v_s/v_{\rm F}=2/3,~1,~2,~3$.
        The shaded area for $\omega\ge 2|\Delta_{\rm eq}|$ corresponds to the conitnuum excitations of Bogoliubov quasiparticles, where the collective modes may acquire a finite damping rate.
    }
    \label{fig2_AAEE}
\end{figure}

Figure~\ref{fig2_AAEE} shows the dispersions of the the $\delta \mathcal{D}_1$ and $\delta\mathcal{E}_{1,2}$ eigenmodes in the quasiclassical limit, $a=0$, where the mixing is solely due to finite $\bm q$.
Note that, as is clear from Eq,~\eqref{eigeneq}, in this limit  $\delta\mathcal{E}_-$ remains an eigenmode, while $\delta\mathcal{E}_{+}$ hybridizes with the amplitude Higgs mode $\delta \mathcal{D}_+$.
The increased splitting between the real and the imaginary clapping modes with increased $q$ is due to this hybridization.

%%%%%%%%%%%%FIGURE
%%%%%%%%%%%%FIGURE

It is worth noting that the quasiclassical approximation is most reliable for $q\xi_0 \lesssim 1$, and therefore we restrict our consideration to this range.
Since Eq.~\eqref{eigeneq} and the acoustoelectric conductivity tensor contain $\gamma$, $\varphi_n$, and the bandwidth, $\epsilon_{\rm F}$, we need to choose the parameters consistent with the hierarchy of the energy scales in superconductors, $T_c< \epsilon_c < \epsilon_{\rm F}$.
For this purpose, we introduce phenomenological material parameters, $b=\frac{\pi T_c}{\epsilon_{\rm F}}(k_{\rm F}\xi_0)$ and $c=\frac{\epsilon_c}{\epsilon_{\rm F}}(k_{\rm F}\xi_0)$. In the simplest estimate, where $\epsilon_{\rm F}=k_{\rm F}v_{\rm F}$, and $\xi_0\simeq v_{\rm F}/2\pi T_c$, we have $b \sim O(1)$ at low temperatures.
At the weak coupling, we have to choose $b< c$, and then the parameter $\gamma\sim\ln c/b$.

%%%%%%%%%%%%%%%%%%%%%%%%%%%%%%%%%%%%%%%%%
%%%%%%%%%%%%%%%%%%%%%%%%%%%%%%%%%%%%%%%%%

\section{Anomalous acoustoelectric effect.}
\label{result}

Let us now return to the analysis of the collective modes under a driving force in the right-hand side of Eq,~\eqref{eigeneq}.
In the absence of the PHA ($a=0$), only the imaginary clapping mode, $\delta \mathcal{E}_-$, can be driven by propagating acoustic waves.
As discussed above, this mode also decouples from the Higgs, and the real clapping modes.
According to Eqs.~\eqref{chixx_CM} and \eqref{chiyx_CM}, the $\delta \mathcal{E}_-$ mode contributes to both the longitudinal and transverse conductivities, but the transverse current carried by the $\delta \mathcal{E}_-$ mode vanishes when $a=0$.
Therefore, the collective modes do not yield the anomalous, transverse, response when the PHA is neglected. %negligibly small.

The situation is different when the PHA is included.
Now the other two modes are also driven by the deformation potential, albeit with the coefficient that depends on the PHA of the normal state, $ \delta \mathcal{D}_+/\delta v_{\rm ex} \sim \mathcal{O}(T_c/\epsilon_{\rm F})$ and $ \delta \mathcal{E}_+/\delta v_{\rm ex} = \mathcal{O}(T_c/\epsilon_{\rm F})$.
These modes also hybridize with the $\delta \mathcal{E}_-$ with coefficients $~\mathcal{O}(T_c/\epsilon_{\rm F})$. Therefore, the last term in the longitudinal conductivity, Eq.~\eqref{chixx_CM}, is the second order in the PHA, and does not contribute significantly.
On the other hand, as Eq.~\eqref{chiyx_CM} shows,  the $\delta \mathcal{E}_-$ mode also carries transverse electric current when the PHA is included.
The electric current carried by the $\delta \mathcal{E}_-$ mode is the same order as that carried by the $\delta \mathcal{E}_+$ mode.
Therefore, we expect the transverse current to be linear in the PHA.

Consequently, in most materials, the resulting effect is very weak.
However, in heavy fermion systems, the Fermi velocity may be comparable to the speed of sound, $v_s$.
In Fig.~\ref{fig2_AAEE}, we plot the acoustic phonon dispersion for  $v_s/v_{\rm F}=2/3,~1,~2,~3$.
As $v_s/v_{\rm F}$ increases, the phonon dispersion and the collective mode dispersion intersect at a finite momentum ${\bm q}_{\rm c}$, which satisfies for each mode $\omega_{\Gamma}(\bm q_c)=v_s|\bm q_c|\;(\Gamma=\delta \mathcal{D}_1,\;\delta \mathcal{E}_1,\;\delta \mathcal{E}_2)$.
Importantly, for $v_s/v_{\rm F} \sim \mathcal{O}(1)$, the intersection with the $\delta \mathcal{E}_{1,2}$ modes occurs at energies below the particle-hole continuum.
At that point, the collective modes can be resonantly excited by propagating acoustic waves, leading to resonant amplification of the acoustoelectric effect both in the longitudinal and in the transverse channels.
The resonance between the eigenmodes and phonons also leads to the characteristic $\omega$-dependence of the response.

%%%%%%%%%%%%FIGURE
\begin{figure}[t!]
    \includegraphics[width=8.6cm]{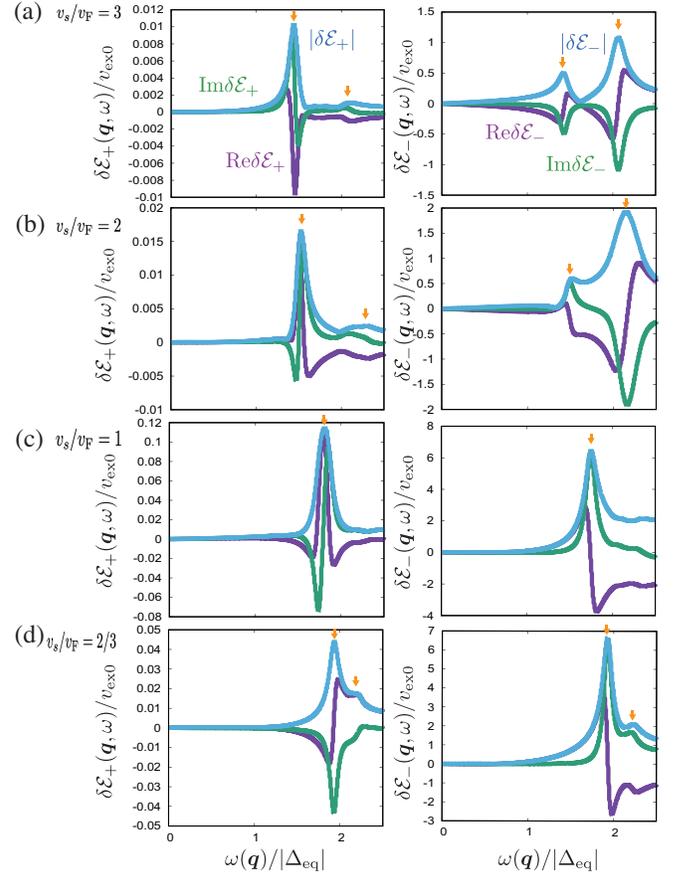}
    \caption{
	The $\omega$-dependence of the real and imaginary clapping modes ($\delta\mathcal{E}_+$ and $\delta\mathcal{E}_-$) induced by the acoustic wave with $v_s/v_{\rm F}=2/3,~1,~2,~3$.
	We set the parameters as $T=0.1T_c,\; a=1,\; b=0.5,\; c=5$, and  $k_{\rm F}\xi_0=100$.
	The orange arrows indicate resonant peaks of the corrective mode current.}
    \label{fig3_AAEE}
\end{figure}
%%%%%%%%%%%%%%%%

Fig.~\ref{fig3_AAEE} shows the linear response of the real and imaginary clapping modes ($\delta\mathcal{E}_+$ and $\delta\mathcal{E}_-$) to the acoustic wave, obtained from Eq.~\eqref{eigeneq}.
The spectra of $|\delta \mathcal{E}_{\pm}(\omega(\bm q))|$ have sharp peaks at the resonant frequencies, $\omega_{\Gamma}(\bm q_c)=v_s|\bm q_c|$.
We note that the amplitude at resonance of the $\delta \mathcal{E}_+$ mode is determined by the PHA correction, which is the order of $1/(k_{\rm F}\xi_0)$, and hence two order smaller than the resonance of the $\delta \mathcal{E}_-$ mode for our choice of $k_{\rm F}\xi_0 = 100$.
As $v_s/v_{\rm F}$ decreases, the resonance peak shifts to the higher frequency and approaches the edge of the continuum of the fermionic excitations.
After crossing that threshold, the finite lifetime of the collective modes leads to the broadening and amplitude reduction of the resonance peaks.
Also note that the modes $\delta \mathcal{E}_{1,2}$ are nearly degenerate for $q\xi_0 \in [0,1]$, and therefore we do not resolve the difference in the resonance energies in our calculations.

%the two resonant peaks at the eigenenergies of $\delta \mathcal{E}_{1,2}$ are indistinguishable.

%%%%%%%%%%%%FIGURE
\begin{figure}[t]
   \includegraphics[width=8.5cm]{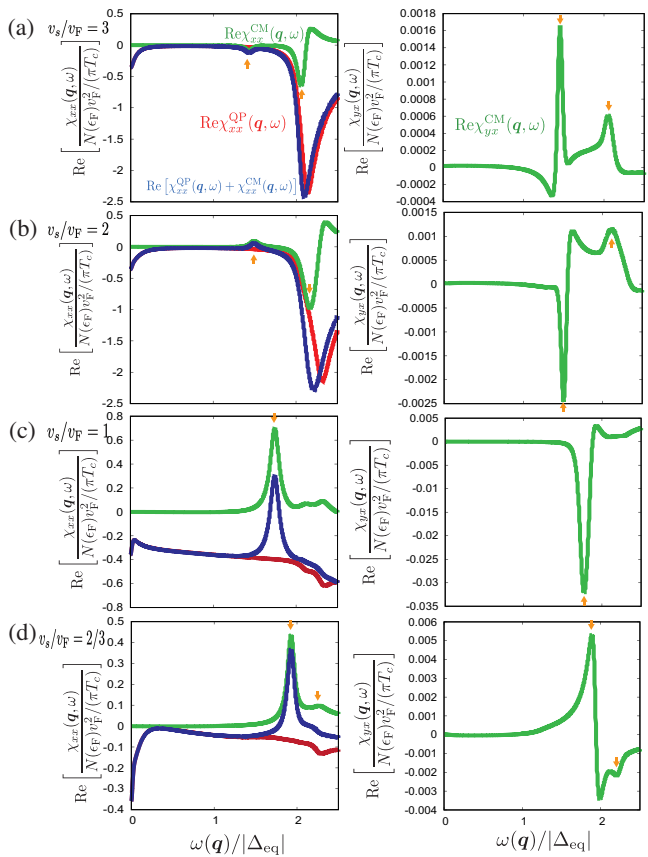}
    \caption{
	The $\omega$-dependence of the acoustoelectric conductivity.
	The left (right) panels correspond to the longitudinal (transverse) acoustoelectric conductivity, $\chi_{xx}$ ($\chi_{yx}$).
	We take the same parameters as those in Fig.~\ref{fig3_AAEE}.
	The orange arrows indicate resonant peaks of the collective mode current.
	}
    \label{fig4_AAEE}
\end{figure}
%%%%%%%%%%%%FIGURE

We also find that, as the resonance shifts to shorter wavelength, the resonance amplitude of the real and imaginary clapping modes increases. This occurs because the driving forces of $\delta \mathcal{E}_{\pm}$ are proportional to $q^3$ through $X_1$ or $\overline{\lambda}_1$ in Eq.~(\ref{eigeneq}), and therefore vanish in the long wavelength limit when the superconducting gap in equilibrium is isotropic. Therefore as the ratio $v_s/v_F$ decreases, the crossing point in Fig.~\ref{fig3_AAEE} moves to higher $q$, and the resonance amplitude of $|\delta\mathcal{E}_{\pm}|$ grows. This growth, of course, is cut off by merging of the resonance frequency with the continuum at $2|\Delta_{\rm eq}|$ for small values of the $v_s/v_{\rm F}$ ratio, and hence the values $v_s/v_{\rm F}\simeq 1$ provide the optimal range for driving of the collective modes by the acoustic waves.

%It is also found that the shift of the intersection point to the shorter wavelength enhances the contributions of real and imaginary clapping modes.
%This is because the driving forces of $\delta \mathcal{E}_{\pm}$ are proportional to $q^3$ through $X_1$ or $\overline{\lambda}_1$, and vanish in the long wavelength limit when the superconducting gap in equilibrium is isotropic [see Eq.~(\ref{eigeneq})].
%The enhancement of resonance of clapping modes is confirmed in Fig.~\ref{fig3_AAEE}, where the intensity of $|\delta\mathcal{E}_{\pm}|$ increases with decreasing the sound velocity $v_s$.

Finally, in  Fig.~\ref{fig4_AAEE}, we plot the frequency-dependent longitudinal and transverse acoustoelectric conductivities.
The left panels demonstrate that both the quasiparticle and the collective mode ($\delta \mathcal{E}_{\pm}$) generate the longitudinal ac electric current, while only the collective modes contribute to the transverse conductivity.
Reflecting the resonances between the collective modes and the acoustic wave, all components of $\chi_{ij}^{\rm CM}$ also show sharp peaks at the resonant frequencies.
When the sound velocity is comparable to the Fermi velocity, e.g.  due to the large effective mass of electrons, the chiral superconducting fluctuations resonate with the acoustic wave below the fermionic continuum edge, which results in the pronounced peak in the $\omega$-dependence of the electric current. This resonance peak dominates the response in both longitudinal and transverse channels, and therefore can be detected experimentally.

%%%%%%%%%%%%%%%%%%%%%%%%%%%%%%%%%%%%%%%%%
%%%%%%%%%%%%%%%%%%%%%%%%%%%%%%%%%%%%%%%%%

\section{Conclusion}
\label{sec:conclusion}
In this paper, we theoretically investigated the acoustoelectric effect in chiral superconductors, focusing especially on the interplay between the collective modes and acoustic waves.
Using the quasiclassical transport theory and incorporating the weak particle-hole asymmetry of the low-energy excitations of normal electrons, we found that the real/imaginary clapping modes can be driven by propagating acoustic waves, and are coupled by the particle-hole asymmetry factor of order of $\Delta_{\rm eq}/\epsilon_{\rm F}\sim 1/(k_{\rm F}\xi_0)$. These modes, in turn, drive both the longitudinal and the transverse electric currents. In the longitudinal current, the collective mode contribution is additive to that of the quasiparticles. However, in chiral superconductors, the collective modes also drive the transverse, anomalous, acoustoelectric effect. This effect is inherent to chiral superconductors, and reflects spontaneous breaking of the time-reversal symmetry, and the chirality degrees of freedom of the Cooper pairs.

For systems, where the sound velocity is comparable to the Fermi velocity, the contribution of the collective modes to the acoustoelectric effect is resonantly enhanced when the phonon and the collective mode energies coincide.
This generates the resonant contributions to longitudinal and transverse electric currents.
The transverse electric current carried by the  clapping modes is reduced by the particle-hole asymmetry factor, compared to the longitudinal current mediated by Bogoliubov quasiparticles and the imaginary clapping mode, but the resonance nature allows its experimental determination.

%{In our analysis, we do not considered the Meissner diamagnetic current induced by the electric current generated by the acoustic wave; however, the Meissner diamagnetic current is not serious for the acoustoelectric effect in chiral superconductors since the electric current induced by the acoustic wave is the alternating electric current, which can not be screened by the supercurrent.}

We stress again that the clapping modes always exist in any chiral superconductors with orbital angular momentum $|\nu| \ge 1$, at least in the weak coupling limit.
While above we considered the chiral $p$-wave superconducting state as a simple model, our main result is independent of the spin and orbital stats of Cooper pairs, and thus applicable to other (non $p$-wave) chiral superconductors.  {For example, in two-dimensional chiral superconductors, the mass gaps of the clapping modes are universal and take $\sqrt{2}|\Delta_{\rm eq}|$ regardless of the chirality, $\nu$, of the chiral order parameters, $\Delta_{\rm eq}(\bm k) \propto (k_x+ik_y)^\nu$~\cite{hsiao}.}

{There have been extensive investigations of the dc anomalous transport phenomena in chiral superconductors  ~\cite{Arfi1988,ngampruetikorn2020impurity,yip2016low,yilmaz2020spontaneous,PhysRevLett.128.097001}. These transport coefficients are affected by the impurity scattering and the particle-hole anisotropy induced in the impurity band at energies much below the gap, and may be made more complex by the presence of nodal quasiparticles. It is therefore important to emphasize that the resonance acoustoelectric response occurs at finite frequencies $\omega\lesssim 2\Delta$. In this range the impurities and nodal excitations broaden and slightly shift the resonance, but do not qualitatively change our analysis above in clean systems.}

{For the same reason, that we consider the ac signal, Meissner currents do not screen the field generated by the acoustic wave, and therefore diamagnetic screening by the condensate gives small corrections to our results.}

%{In this paper, we focused on the full gap chiral superconductors; however, three-dimensional chiral superconductors in real materials accompany nodal structures.
%Generally, nodal quasiparticle excitations affect the collective excitations in superconductors.
%These nodal excitations modify the width and positions of the resonant peaks in the anomalous acoustoelectric effect in chiral superconductors.
%But, these low-energy quasiparticle excitations do not cancel the anomalous acoustoelectric effect due to the clapping excitations.
%Nodal quasiparticle excitations introduce the intrinsic damping rate in the collective modes with the lower energy than $2|\Delta_{\rm eq}|$, broadening the resonant peaks of the transverse current.
%The positions of the resonant peaks are also affected by the nodal structures; however, they are expected not to be sensitive to symmetries of the order parameters in chiral superconductors.
%For two-dimensional chiral superconductors, the mass gaps of the clapping modes are universal and take $\sqrt{2}|\Delta_{\rm eq}|$ regardless of the chirality, $\nu$, of the chiral order parameters, $\Delta_{\rm eq}(\bm k) \propto (k_x+ik_y)^\nu$~\cite{hsiao}.
%It suggests that the positions of the resonant peaks of the transverse current weakly depend on symmetries of the chiral order parameter.}

{The overall magnitude of the effect depends on the details of
%It is quite difficult to estimate the specific size of
 deformation potential induced by the acoustic wave. It is difficult to estimate it reliably in heavy-fermion superconductors since the electron-electron correlations renormalize the electron-phonon coupling significantly. %interaction between electrons and the acoustic wave.
%We, therefore, do not present the size of the transverse electric current in real materials.
We note that the ultrasonic attenuation, which relies on the same coupling,
%It is known that the strong electron-electron correlation suppresses the interaction between electrons and the acoustic vibration; however, some experiments
has been measured in a number of such materials, %successfully observed the ultrasonic attenuations in heavy-fermion superconductors,
including UBe$_{13}$ and UPt$_3$~\cite{PhysRevB.33.5906}.
Therefore, the experimental detection of the anomalous acoustoelectric effect is feasible in heavy-fermion chiral superconductors. Moreover, our calculations predict a resonance behavior of the ac acoustoelectric effect, and, in clean systems with moderate broadening, we expect the resonance signature to be easily observable.}

{Consequently, the anomalous acoustoelectric effect is a generic feature of chiral superconductors, and is expected  in a wide range of superconducting materials.
The anomalous acoustoelectric effect and its resonant behavior provide a smoking gun evidence of chiral superconductivity in heavy fermion materials and superconducting materials with small Fermi surfaces, where the particle-hole asymmetry is appreciable.}

\begin{acknowledgments}
    The authors are grateful to K. Kimura, Y. Nagai and Y. Yanase for fruitful discussions.
    T. Matsushita was supported by a  Japan Society for the Promotion of Science (JSPS) Fellowship for Young Scientists and by JSPS KAKENHI Grant No.~JP19J20144.
    This work was also supported by JST CREST Grant No. JPMJCR19T5, Japan, and the Grant-in-Aid for Scientific Research on Innovative Areas ``Quantum Liquid Crystals (JP20H05163)'' from JSPS of Japan,
    and JSPS KAKENHI (Grant No.~JP20K03860, No.~JP20H01857, and No.~JP21H01039).
    I. V. acknowledges partial support from the NSF Grant No DMR-1410741.
\end{acknowledgments}

%\begin{appendix}
\appendix

\section{Particle-Hole Asymmetry Corrections to Quasiclassical Transport Theory}
\label{sec:PHA}

\subsection{Quasiclassical Transport Equation}

Here we derive the Keldysh transport equation in the quasiclassical limit and the auxiliary equation, which is used to obtain the leading-order correction to the Green's functions and physical quantities due to the particle-hole asymmetry.
We begin with the Green's function in the Wigner representation.
\begin{eqnarray}
\check{G}(\epsilon,\bm k, \bm x,t)&=&
\begin{pmatrix}
\underline{G}^{\rm R}(\epsilon,\bm k, \bm x,t)&&\underline{G}^{\rm K}(\epsilon,\bm k, \bm x,t)\\
0&&\underline{G}^{\rm A}(\epsilon,\bm k, \bm x,t)
\end{pmatrix},\\
\underline{G}^{\rm X}(\epsilon,\bm k, \bm x,t)&=&
\begin{pmatrix}
G^{\rm X}(\epsilon,\bm k, \bm x,t)&&F^{\rm X}(\epsilon,\bm k, \bm x,t)\\
\overline{F}^{\rm X}(\epsilon,\bm k, \bm x,t)&&\overline{G}^{\rm X}(\epsilon,\bm k, \bm x,t)
\end{pmatrix}.
\end{eqnarray}
These obey the left-hand Gor'kov equation,
\begin{align}
\label{l_gorkov}
\left(\epsilon \check{\tau}_z-\check{\Delta}-\check{v}_{\rm ex}\right)\otimes \check{\tau}_z\check{G}+\frac{i}{2}{\bm v}(\bm k)\cdot {\bm \nabla}\check{\tau}_z\check{G}-\xi_k\check{\tau}_z\check{G}=1,%\nonumber\\
\end{align}
and the corresponding right-hand %Gor'kov
equation,
\begin{align}
\label{r_gorkov}
\check{\tau}_z\check{G}\otimes \left(\epsilon \check{\tau}_z-\check{\Delta}-\check{v}_{\rm ex}\right)+\frac{i}{2}{\bm v}(\bm k)\cdot {\bm \nabla}\check{\tau}_z\check{G}-\xi_k\check{\tau}_z\check{G}=1,%\nonumber\\
\end{align}
where $\rm X=R,~A,~K$ and ${\bm v}(\bm k)={\bm \nabla}_{\bm k}\xi_k$. The Groenewold-Moyal product is given by $A\otimes B(X,p)=e^{i(\partial^A_X\partial^B_p-\partial^A_p\partial^B_X)/2}A(X,p)B(X,p)$, where we have introduced abbreviated notation, $\partial^A_X\partial^B_p = -\partial^A_t\partial^B_{\epsilon}+{\bm \nabla}^A_{\bm R}\cdot{\bm \nabla}^B_{\bm p}$, $X\equiv(t,{\bm x})$, and $p\equiv(\epsilon,{\bm p})$.
Subtracting and adding Eq.~(\ref{l_gorkov}) and Eq.~(\ref{r_gorkov}), we obtain
\begin{gather}
\left[ \epsilon \check{\tau}_z-\check{\Delta}-\check{v}_{\rm ex}, \check{\tau}_z\check{G}\right]_\otimes+i{\bm v}(\bm k)\cdot {\bm \nabla}\check{\tau}_z\check{G}=0,\label{eq:G1}\\
\left[ \epsilon \check{\tau}_z-\check{\Delta}-\check{v}_{\rm ex}, \check{\tau}_z\check{G}\right]_{\otimes +}-\xi_k\check{\tau}_z\check{G}-1=0.
\label{eq:G2}
\end{gather}
We now take the quasiclassical limit, $(k_{\rm F}\xi)^{-1}=0$.
The quasiclassial approximation postulates the slow variation of superconducting order parameters in the space-time (compared to $k_{\rm F}$ and $\epsilon_{\rm F}$ scales respectively), and accounts for quasiparticles confined to a low-energy shell near the Fermi surface.
Then, the quasiclassical Green's functions, $\check{g}$, are obtained by integrating $\check{G}$ over a small shell in momentum space near the Fermi surface as in Eq.~\eqref{Keldysh_prop0}, and replacing the DOS of normal electrons to $N(\xi_k+\epsilon_{\rm F})\simeq N(\epsilon_{\rm F})$.
We finally obtain the quasiclassical transport equation from Eq.~\eqref{eq:G1} as
\begin{eqnarray}
&\left[ \epsilon \check{\tau}_z-\check{\Delta}-\check{v}_{\rm ex}, \check{g}_{(0)}\right]_\otimes+i{\bm v}_{\rm F}\cdot {\bm \nabla}\check{g}_{(0)}=0.
\end{eqnarray}
Equation~\eqref{eq:G2} reduces to the auxiliary relation
\begin{eqnarray}
\label{auxiliary}
\int d\xi_k (\xi_k \check{\tau}_z\check{G}+1)=\frac{1}{2}\left[ \epsilon \check{\tau}_z-\check{\Delta}-\check{v}_{\rm ex}, \check{g}_{(0)}\right]_{\otimes+}.
\end{eqnarray}
This is used to derive the leading-order correction due to the particle-hole asymmetry, such as Eq.~(\ref{electric_current1}).

\subsection{Particle-Hole Asymmetry effect on Physical Quantities}
Using the Green's function in the quasiclassical limit and the auxiliary equation (\ref{auxiliary}), we now derive the particle-hole asymmetry-driven corrections to the physical quantities.
We first derive the correction to the quasiclassical Green' function in Eq.~(\ref{PHA_Keldysh}).
The particle-hole asymmetry appears as the leading-order correction to the DOS at the Fermi energy, $N(\xi_k+\epsilon_{\rm F})\simeq N(\epsilon_{\rm F})+N^{\prime}(\epsilon_{\rm F})\xi_k$, where $N^{\prime}(\epsilon_{\rm F})\equiv [\partial N(\epsilon)/\partial \epsilon]_{\epsilon=\epsilon_{\rm F}}$.
Substituting this expansion and utilizing the auxiliary relation~\eqref{auxiliary}, we obtain
\begin{eqnarray}
\label{Keldysh_prop1}
    \check{g}(\epsilon,\bm k_{\rm F},\bm x,t)&\simeq &\int d\xi_k\left(1+\frac{a}{\epsilon_{\rm F}}\xi_k\right)\ \check{\tau}_z \check{G}(\epsilon,\bm k,\bm x,t)\nonumber \\
    &\simeq &\check{g}_{(0)}+\frac{a}{2\epsilon_{\rm F}}\left[ \epsilon \check{\tau}_z-\check{\Delta}-\check{v}_{\rm ex},\check{g}_{(0)} \right]_{\otimes +}.
\end{eqnarray}
The second term in Eq.~(\ref{Keldysh_prop1}) describes the particle-hole asymmetry of the DOS, $a\equiv \epsilon_{\rm F}N^{\prime}(\epsilon_{\rm F})/N(\epsilon_{\rm F})$.

The electric current is defined with the Keldysh component of the Gor'kov Green's function as
\begin{eqnarray}
	\label{electric_current3}
	{\bm J}=-\int \frac{d\epsilon}{4\pi i}\int \frac{d{\bm k}}{(2\pi)^2}{\bm v}(\bm k)G^{\rm K}(\epsilon,\bm k,\bm x,t).
\end{eqnarray}
The Keldysh Green's function obeys the following relation~\cite{SERENE1983221},
\begin{eqnarray}
G^{\rm K}(\epsilon,\bm x,\bm k,t)=-\overline{G}^{\rm K}(-\epsilon,\bm x,-\bm k,t).
\end{eqnarray}
By using this relation, Eq.~(\ref{electric_current3}) is recast into
\begin{eqnarray}
	\label{electric_current4}
	{\bm J}=-\frac{1}{2}\int \frac{d\epsilon}{4\pi i}\int \frac{d{\bm k}}{(2\pi)^2}{\rm Tr}\left[{\bm v}(\bm k)\underline{G}^{\rm K}(\epsilon,\bm k,\bm x,t)\right].
\end{eqnarray}

The standard quasiclassical approximation is effective when the Fermi energy is sufficiently large, and hence assumes particle-hole symmetry in the quasiparticle density of states.
The physical quantities in this limit are thus computed in the approximation $\int \frac{d{\bm k}}{(2\pi)^D}=\int d\epsilon N(\epsilon) \simeq N({\epsilon}_{\rm F})\int d\epsilon $, where $D$ is the dimension of the system.
The contribution of the particle-hole asymmetry is incorporated by including
the leading-order correction to $N({\epsilon}_{\rm F})$.
Substituting the expansion of $N(\epsilon)$, and utilizing the auxiliary equation (\ref{auxiliary}), we obtain the electric current in terms of the quasiclassical Green's function $\check{g}_{(0)}$ as
\begin{align}
	\label{electric_current5}
	{\bm J}\simeq& -\frac{1}{2}\int \frac{d\epsilon}{4\pi i}\int d\xi_k\left(N(\epsilon_{\rm F})+\frac{\partial N(\epsilon)}{\partial \epsilon}\bigg|_{\epsilon=\epsilon_{\rm F}}\xi_k\right)\nonumber\\	
	&\times\left\langle {\rm Tr}\left[{\bm v}_{\rm F}\underline{G}^{\rm K}(\epsilon,\bm k,\bm x,t)\right]\right\rangle_{{\rm FS},{\bm k}_{\rm F}}\nonumber\\
	=&-N(\epsilon_{\rm F})\int \frac{d\epsilon}{4\pi i}\left\langle \frac{1}{2}{\rm Tr}\left[{\bm v}_{\rm F}\underline{\tau}_z\underline{g}^{\rm K}_{(0)}(\epsilon,\bm k_{\rm F},\bm x,t)\right]\right\rangle_{{\rm FS},{\bm k}_{\rm F}}\nonumber\\
	&-N^{\prime}(\epsilon_{\rm F})\int \frac{d\epsilon}{4\pi i}\left\langle \frac{1}{4}{\rm Tr}\left[{\bm v}_{\rm F}\underline{\tau}_z\left[ \epsilon \check{\tau}_z-\check{\Delta}-\check{v}_{\rm ex}, \check{g}_{(0)}\right]_{\otimes +}\right]\right\rangle_{{\rm FS},{\bm k}_{\rm F}}.
\end{align}
In the first line of Eq.~(\ref{electric_current5}), we apply the quasiclassical approximation and expand the DOS.
Then, the auxiliary relation in Eq.~(\ref{auxiliary}) is used to derive the second line of Eq.~(\ref{electric_current5}).
Using Eq.~(\ref{PHA_Keldysh}), we finally obtain Eq.~(\ref{electric_current1}) as
\begin{eqnarray}
	{\bm J}&\simeq&-N(\epsilon_{\rm F})\int \frac{d\epsilon}{4\pi i}\left\langle \frac{1}{2}{\bm v}_{\rm F}{\rm Tr}\left[\underline{\tau_z} \left(\underline{g}^{\rm K}_{(0)}+ \underline{g}^{\rm K}_{\rm (1)}\right)\right]\right\rangle_{{\rm FS},{\bm k}_{\rm F}}\,.
\end{eqnarray}

{\section{Derivation of Keldysh Response function}
\label{Response}
We next derive the nonequilibrium Keldysh Green's function by the linear response to the deformation potential.
We denote the equilibrium quantities as $\check{x}_{\rm eq}\;(x=g_{(n)},\Delta)$ and the linear deviation from equilibrium  as $\delta \check{x}\;(x=g_{(n)},\Delta)$.}

{\subsection{Equilibrium Green's Function}
\label{GF_eq}
In the absence of external perturbations, the system is translationally invariant, and the equilibrium quasiclassical Green's function, $\check{g}_{(0) \rm eq}$, obeys the homogeneous Eilenberger equation,
\begin{eqnarray}
    \label{Eilenberger_eq}
    \left[\epsilon \check{\tau}_z-\check{\Delta}_{\rm eq},\check{g}_{\rm (0)eq}\right]=0,
\end{eqnarray}
subject to $\check{g}_{\rm (0)eq}^2=-\pi^2$.
The solutions are given by
\begin{gather}
    \underline{g}_{\rm (0)eq}^{\rm R,A}=-\pi \frac{\epsilon \underline{\tau}_z-\underline{\Delta}_{\rm eq}}{D^{\rm R,A}(\epsilon)},\\
    \underline{g}^{\rm K}_{\rm (0)eq}=\left(\underline{g}^{\rm R}_{\rm (0)eq}-\underline{g}^{\rm A}_{\rm (0)eq}\right)\tanh \left(\frac{\epsilon}{2T}\right)
    =\alpha(\epsilon)\underline{\tau}_z+\beta(\epsilon)\underline{\Delta}_{\rm eq},
\end{gather}
where $D^{\rm R}(\epsilon)=D^{\rm A \ast}(\epsilon)=\sqrt{\Delta_{\rm eq}^2-(\epsilon+i0^+)^2}\; (0^+>0)$,
and
\begin{eqnarray}
    \alpha(\epsilon)=-\epsilon \beta(\epsilon)=-2\pi i n_s(\epsilon)\tanh \left(\frac{\epsilon}{2T}\right),
\end{eqnarray}
with the spectral function $n_s(\epsilon)$,
\begin{eqnarray}
    n_s(\epsilon)=\frac{|\epsilon|}{\sqrt{\epsilon^2-\Delta_{\rm eq}^2}}\Theta(\epsilon^2-\Delta_{\rm eq}^2)\,.
\end{eqnarray}
Here, $\Theta(x)$ is the Heaviside function.}

{Substituting $\check{g}_{\rm (0)eq}$ into Eq.~(\ref{PHA_Keldysh}), one obtains the leading-order correction from the PHA as
\begin{align}
    \label{PHA_Keldysh_eq}
    \underline{g}_{(1){\rm eq}}^{\rm K}=\frac{a}{2\epsilon_{\rm F}}
  \left[2\epsilon g_{\rm (0) eq}^{\rm K}-\Delta_{\rm eq}(\bm k_{\rm F})\overline{f}_{\rm (0) eq}^{\rm K}-\Delta_{\rm eq}^\ast(\bm k_{\rm F})f_{\rm (0) eq}^{\rm K}\right]\underline{\tau}_0.
\end{align}
Note that the equilibrium pair amplitudes in the PHA correction, $f^{\rm K}_{\rm (1)eq}$ and $\bar{f}^{\rm K}_{\rm (1)eq}$, vanishes since the Keldysh propagator in the quasiclassical limit obeys the relation, $g_{\rm (0) eq}^{\rm K}+\overline{g}_{\rm (0) eq}^{\rm K}=0$.
This implies that the weak PHA in the DOS does not renormalize the equilibrium gap function.
Hence, in evaluating the temperature response below, we assume a BCS-like temperature dependence of the equilibrium gap function, $\Delta_{\rm eq}(T)=1.765 T_{\rm c} \tanh(1.74\sqrt{T_{\rm c}/T-1})$~\cite{tinkham}.}

{\subsection{Nonequilibrium Keldysh Green's function}
\label{app:linear}
We now derive the derive the Keldysh response function to the propagating acoustic wave.
%Since we are interested in linear response, we can expand the response function, $\delta\check{g}$, in powers of $1/(k_F\xi_0)$ as above.
The nonequilibrium Green's function in the leading order, i.e. the quasiclassical limit, $\delta \check{g}_{(0)}$, obeys the Eilenberger equation,
\begin{align}
    \label{Eilenberger_noneq}
    \left[\epsilon \check{\tau}_z-\check{\Delta}_{\rm eq},\delta \check{g}_{(0)}\right]_\circ-\left[\delta \check{\Delta}+\check{v}_{\rm ex},\check{g}_{\rm (0) eq}\right]_\circ %\nonumber\\
   +i {\bm v}_{\rm F}\cdot {\bm \nabla} \delta \check{g}_{(0)}=0.
\end{align}
It is important to note that, since we are looking at finite frequencies, we included the dynamical fluctuations of the order parameter, $\delta \check{\Delta}$, which are determined by self-consistently solving the gap equation.}

{Equation~(\ref{Eilenberger_noneq}) includes the $\circ$-product of equilibrium and nonequilibrium quantities, such as $\check{A}_{\rm eq}\circ \delta \check{B}(t)$ and $\delta \check{A}(t) \circ \check{B}_{\rm eq}$.
These $\circ$-products can be cast into a more convenient form by performing the Fourier transformation in $\bm x$ and  $t$~\cite{eschrig2000distribution},
which gives for the Keldysh part of Eq.~(\ref{Eilenberger_noneq})
\begin{align}
    \label{Eilenberger2}
&\epsilon_+\underline{\tau}_z\delta \underline{g}_{(0)}^{\rm K}-\epsilon_-\delta \underline{g}_{(0)}^{\rm K}\underline{\tau}_z-\underline{\Delta}_{\rm eq}({\bm k}_{\rm F})\delta\underline{g}_{(0)}^{\rm K}
    +\delta  \underline{g}_{(0)}^{\rm K}\underline{\Delta}_{\rm eq}({\bm k}_{\rm F}) \nonumber\\
&    -\delta \underline{\Delta}\underline{g}_{\rm (0)eq}^{\rm K}(\epsilon_-)+\underline{g}_{\rm (0)eq}^{\rm K}(\epsilon_+)\delta \underline{\Delta}\nonumber\\
    &+v_{\rm ex 0}\left(\underline{g}_{\rm (0)eq}^{\rm K}(\epsilon_+)-\underline{g}_{\rm (0)eq}^{\rm K}(\epsilon_-)\right)-\eta \delta \underline{g}_{(0)}^{\rm K}=0,
\end{align}
where we introduced shorthand notation, $\delta \underline{g}_{(0)}^{\rm K}\equiv\delta \underline{g}_{(0)}^{\rm K}(\epsilon,\bm k_{\rm F},\bm q,\omega)$, $\delta \underline{\Delta}\equiv \delta \underline{\Delta}({\bm k}_{\rm F},\bm q,\omega)$, and $\eta={\bm v}_{\rm F}\cdot \bm q$.
The frequency shift, $\epsilon \to \epsilon_\pm=\epsilon \pm \frac{\omega+i0^+}{2}$, arises from the finite frequency of the acoustic mode in $\underline{v}_{\rm ex}(\bm x, t)$.
We also introduced a small imaginary part, $\omega \to\omega+i0^+$, to obtain the causal response function.}

{\begin{widetext}
For solving Eq.~(\ref{Eilenberger2}), it is convenient to introduce the following quantities,
\begin{gather}
    \delta g_{(n)\pm}^{\rm K}=\delta g_{(n)}^{\rm K}\pm \delta \overline{g}_{(n)}^{\rm K},\\
    \delta f_{(n)\pm}^{\rm K}=\delta f_{(n)}^{\rm K}\pm \delta \overline{f}_{(n)}^{\rm K},\\
    \delta \Delta_{\pm}=\delta \Delta\pm \delta \Delta^{\ast}.
\end{gather}
Using these quantities, Eq.~(\ref{Eilenberger2}) becomes
\begin{eqnarray}
    \label{Eilenberger3}
    \begin{pmatrix}
        -\eta&&\omega&&2i\Delta_{\rm eq}\hat{k}_y&&-2\Delta_{\rm eq}\hat{k}_x\\
        \omega&&-\eta&&0&&0\\
        2i\Delta_{\rm eq}\hat{k}_y&&0&&-\eta&&2\epsilon\\
        2\Delta_{\rm eq}\hat{k}_x&&0&&2\epsilon&&-\eta
    \end{pmatrix}
    \begin{pmatrix}
        \delta g_{(0)-}^{\rm K}\\
        \delta g_{(0)+}^{\rm K}\\
        \delta f_{(0)+}^{\rm K}\\
        \delta f_{(0)-}^{\rm K}
    \end{pmatrix}
    +
    \begin{pmatrix}
        0&&\alpha_-&&-i\Delta_{\rm eq}\hat{k}_y\beta_+&&\Delta_{\rm eq}\hat{k}_x\beta_+\\
        \alpha_-&&0&&-\Delta_{\rm eq}\hat{k}_x\beta_-&&i\Delta_{\rm eq}\hat{k}_y\beta_-\\
        -i\Delta_{\rm eq}\hat{k}_y\beta_+&&\Delta_{\rm eq}\hat{k}_x\beta_-&&0&&\alpha_+\\
        -\Delta_{\rm eq}\hat{k}_x\beta_+&&i\Delta_{\rm eq}\hat{k}_y\beta_-&&\alpha_+&&0
    \end{pmatrix}
    \begin{pmatrix}
        0\\
        2v_{\rm ex 0}\\
        \delta \Delta_+\\
        \delta \Delta_-
    \end{pmatrix}=0,%\nonumber\\
\end{eqnarray}
with $\hat{\bm {k}}={\bm k}_{\rm F}/k_{\rm F}$.
We now obtain the nonequilibrium Keldysh Green's function in the quasiclassical limit,
\footnotesize
\begin{eqnarray}
    \label{Keldysh_QCL}
    \begin{pmatrix}
        \delta g_{(0)-}^{\rm K}\\
        \delta g_{(0)+}^{\rm K}\\
        \delta f_{(0)+}^{\rm K}\\
        \delta f_{(0)-}^{\rm K}
    \end{pmatrix}
    &=&
    \frac{1}{(4\epsilon^2-\eta^2)(\omega^2-\eta^2)+4\Delta_{\rm eq}^2\eta^2}
    \begin{pmatrix}
    \eta(4\epsilon^2-\eta^2)&&\omega(4\epsilon^2-\eta^2)&&2\eta \Delta_{\rm eq} (2\epsilon \hat{k}_x-i\eta \hat{k}_y) &&-2\eta \Delta_{\rm eq}(2i\epsilon \hat{k}_y-\eta \hat{k}_x) \\
    \omega(4\epsilon^2-\eta^2)&&\eta(4\epsilon^2-\eta^2-4\Delta_{\rm eq}^2)&&2\omega \Delta_{\rm eq} (2\epsilon \hat{k}_x-i\eta \hat{k}_y)  &&-2\omega \Delta_{\rm eq}(2i\epsilon \hat{k}_y-\eta \hat{k}_x) \\
    -2\eta \Delta_{\rm eq} (2\epsilon \hat{k}_x+i\eta \hat{k}_y)&&-2\omega \Delta_{\rm eq} (2\epsilon \hat{k}_x+i\eta \hat{k}_y) &&\eta (4\epsilon^2-\eta^2-4\Delta_{\rm eq}^2 \hat{k}_x^2)&&2\epsilon(\omega^2-\eta^2)+4i\Delta_{\rm eq}^2 \hat{k}_x\hat{k}_y\\
    -2\eta \Delta_{\rm eq}(2i\epsilon \hat{k}_y+\eta \hat{k}_x) && -2\omega \Delta_{\rm eq}(2i\epsilon \hat{k}_y+\eta \hat{k}_x)&&2\epsilon(\omega^2-\eta^2)-4i\Delta_{\rm eq}^2 \hat{k}_x\hat{k}_y&&\eta (4\epsilon^2-\eta^2-4\Delta_{\rm eq}^2 \hat{k}_y^2)
    \end{pmatrix}
    \nonumber\\
    &\times&
    \begin{pmatrix}
        0&&\alpha_-&&-i\Delta_{\rm eq}\hat{k}_y\beta_+&&\Delta_{\rm eq}\hat{k}_x\beta_+\\
        \alpha_-&&0&&-\Delta_{\rm eq}\hat{k}_x\beta_-&&i\Delta_{\rm eq}\hat{k}_y\beta_-\\
        -i\Delta_{\rm eq}\hat{k}_y\beta_+&&\Delta_{\rm eq}\hat{k}_x\beta_-&&0&&\alpha_+\\
        -\Delta_{\rm eq}\hat{k}_x\beta_+&&i\Delta_{\rm eq}\hat{k}_y\beta_-&&\alpha_+&&0
    \end{pmatrix}
    \begin{pmatrix}
        0\\
        2v_{\rm ex 0}\\
        \delta \Delta_+\\
        \delta \Delta_-
    \end{pmatrix}
\end{eqnarray}
\normalsize
where $\alpha_\pm=\alpha(\epsilon_+)\pm \alpha(\epsilon_-)$, $\beta_\pm=\beta(\epsilon_+)\pm \beta(\epsilon_-)$.
We also need the frequency integral of the nonequilibrium Keldysh Green's function to compute the order parameter fluctuation and the electric current.
The integrated Keldysh Green's functions are given by,
\begin{align}
    \label{Eilenberger3}
    \int \frac{d\epsilon}{2\pi i}
    \begin{pmatrix}
        \delta g_{(0)-}^{\rm K}\\
        \delta g_{(0)+}^{\rm K}\\
        \delta f_{(0)+}^{\rm K}\\
        \delta f_{(0)-}^{\rm K}
    \end{pmatrix}
    =
    -\begin{pmatrix}
        2+\left(\frac{\eta^2}{\omega^2-\eta^2}\right)(\lambda-1)&&\left(\frac{2\omega \eta}{\omega^2-\eta^2}\right)(\lambda-1)&&i\eta \Delta_{\rm eq}\hat{k}_y\overline{\lambda}&&-\eta \Delta_{\rm eq}\hat{k}_x\overline{\lambda}\\
        \left(\frac{2\omega \eta}{\omega^2-\eta^2}\right)(\lambda-1)&&\left(\frac{\omega^2}{\omega^2-\eta^2}\right)(\lambda-1)&&i\omega \Delta_{\rm eq}\hat{k}_y\overline{\lambda}&&-\omega \Delta_{\rm eq}\hat{k}_x\overline{\lambda}\\
        i\eta \Delta_{\rm eq}\hat{k}_y\overline{\lambda}&&i\omega \Delta_{\rm eq}\hat{k}_y\overline{\lambda}&&-\gamma-\frac{\omega^2-\eta^2-4\Delta_{\rm eq}^2\hat{k}_x^2}{2}\overline{\lambda}&&-2i\Delta_{\rm eq}^2\hat{k}_x\hat{k}_y\overline{\lambda}\\
        \eta \Delta_{\rm eq}\hat{k}_x\overline{\lambda}&&\omega \Delta_{\rm eq}\hat{k}_x\overline{\lambda}&&2i\Delta_{\rm eq}^2\hat{k}_x\hat{k}_y\overline{\lambda}&&-\gamma-\frac{\omega^2-\eta^2-4\Delta_{\rm eq}^2\hat{k}_y^2}{2}\overline{\lambda}
    \end{pmatrix}
    \begin{pmatrix}
        0\\
        2v_{\rm ex0}\\
        \delta \Delta_+\\
        \delta \Delta_-
    \end{pmatrix}.
\end{align}
\end{widetext}
  The $\gamma$-function in Eq.~(\ref{Eilenberger3}) is defined as
\begin{align}
    \gamma=\int_{-\epsilon_{\rm c}}^{\epsilon_{\rm c}}\frac{d\epsilon}{4\pi i}\beta_+
    =2\int^{\epsilon_{\rm c}}_{|\Delta_{\rm eq}|}\frac{d\epsilon}{\sqrt{\epsilon^2-\Delta_{\rm eq}^2}}\tanh\left(\frac{\epsilon}{2T}\right)+\mathcal{O}\left(\frac{\omega}{\epsilon_{\rm c}}\right)^2.
\end{align}
where $\epsilon_c$ is the frequency cut-off associated with pairing interaction.
For $\omega\ll \epsilon_{\rm c}$, the $\gamma$-function reduces to the equilibrium gap equation, $\frac{\gamma}{2}\simeq\frac{1}{N(\epsilon_{\rm F})V_{\rm pair}}$, where we consider the $p$-wave pairing interaction, $V({\bm k}_{\rm F},{\bm k}_{\rm F}')=V_{\rm pair}\hat{\bm k}\cdot \hat{\bm k}'$.
%\sout{{We use the standard methods to eliminate the dependence on the density of states and the pairing interaction in favor of a measurable quantity, {\it i.e.}, the bulk transition temperature $T_{\rm c}$, via,}}
The gap equation has a logarithmic divergence on $\epsilon_c$~\cite{RevModPhys.63.239}.
To regularize the ultraviolet divergence in the gap equation, we utilize the fact that the cutoff energy, $\epsilon_c$, and the pairing interaction, $V_{\rm pair}$, are related to a measurable quantity, i.e., the bulk transition temperature, $T_c$,
through linearized gap equation,
\begin{eqnarray}
    \label{gamma}
    \frac{\gamma}{2}\simeq\frac{1}{N(\epsilon_{\rm F})V_{\rm pair}}=\ln \left(\frac{1.13\epsilon_c}{T_c}\right).
\end{eqnarray}}

\begin{figure}[t]
    \includegraphics[width=6cm]{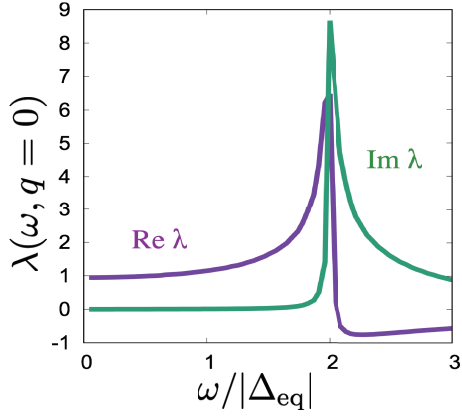}
    \caption{
        {The frequencies dependence of the Tsuneto function in the limit of the long-wavelength and the zero-temperature.}
    }
    \label{fig1_AAGE}
\end{figure}

{The $\lambda$-function in Eq.~(\ref{Eilenberger3}) is the generalized Tsuneto function,
\begin{align}
    \lambda=\Delta_{\rm eq}^2\overline{\lambda}
    =&\int^\infty_{|\Delta_{\rm eq}|}d\epsilon \frac{2\tanh \left(\frac{\epsilon}{2T}\right)}{\sqrt{\epsilon^2-\Delta_{\rm eq}^2}}\bigg[\frac{\eta^2-2\omega \epsilon_+}{(4\epsilon_+^2-\eta^2)(\omega^2-\eta^2)+4\eta^2\Delta_{\rm eq}^2}\nonumber\\
    &+\frac{\eta^2-2\omega \epsilon_-}{(4\epsilon_-^2-\eta^2)(\omega^2-\eta^2)+4\eta^2\Delta_{\rm eq}^2}\bigg]\,,
    \label{gen:Tsuneto}
\end{align}
which characterizes the phase stiffness of the condensate~\cite{moores1993transverse}.
In the long-wavelength limit ($\eta \rightarrow 0$) and the zero-temperature limit, the $\lambda$-function reduces to the Tsuneto function,
\begin{eqnarray}
  \lambda(\omega)=
  \begin{cases}
    \frac{\sin^{-1}(x)}{x\sqrt{1-x^2}}, & \mbox{if } x=\frac{\omega}{2|\Delta_{\rm eq}|}<1\,,
    \vspace*{0.2cm}
    \\
    -\frac{\ln (x+\sqrt{x^2-1})}{2x\sqrt{x^2-1}}+\frac{i\pi}{2x\sqrt{x^2-1}}, & \mbox{if } x=\frac{\omega}{2|\Delta_{\rm eq}|}>1\,.
  \end{cases}
\end{eqnarray}
Note that the Tsuneto function has an imaginary part when $\omega/2|\Delta_{\rm eq}|>1$.
This imaginary part describes intrinsic damping due to breaking of Cooper pairs into two Bogoliubov quasiparticles (see Fig.~\ref{fig1_AAGE}).}

%%%%%%%%%%%%FIGURE
%%%%%%%%%%%%FIGURE

{Substituting Eq.~(\ref{Eilenberger3}) into Eq.~(\ref{PHA_Keldysh}), we straightforwardly obtain the PHA part of the nonequilibrium Keldysh Green's function,
\begin{align}
    \label{PHA_res1}
        \delta g_{\rm (1)+}^{\rm K}=&\frac{a}{2\epsilon_{\rm F}}\big[2\left(\epsilon \delta g_{(0)-}^{\rm K}
       +\Delta_{\rm eq}\hat{k}_x\delta f_{(0)-}^{\rm K}-i\Delta_{\rm eq}\hat{k}_yf_{(0)+}^{\rm K}\right)\nonumber\\
         &+\Delta_{\rm eq}\left(\hat{k}_x\delta \Delta_+ -i\hat{k}_y\delta \Delta_-\right)\beta_+\big],\\
        \delta g_{\rm (1)-}^{\rm K}=&\frac{a}{2\epsilon_{\rm F}}\big[2\epsilon \delta g_{(0)+}^{\rm K}%\nonumber\\
        -2v_{\rm ex0}\alpha_+ + \Delta_{\rm eq}\left(i\hat{k}_y\delta \Delta_{+} -\hat{k}_x\delta \Delta_-\right)\beta_-\big], \label{PHA_res2}
\\
        \delta f_{\rm (1)+}^{\rm K}=&\frac{a}{2\epsilon_{\rm F}}\big[\omega \delta f_{(0)-}^{\rm K}-2\Delta_{\rm eq}\hat{k}_x\delta g_{\rm (0)+}^{\rm K}\label{eq:deltaf1}-\delta \Delta_- \alpha_- -2\Delta_{\rm eq}\hat{k}_x \beta_+ v_{\rm ex 0}\big],\\
        \delta f_{\rm (1)-}^{\rm K}=&\frac{a}{2\epsilon_{\rm F}}\big[\omega \delta f_{(0)+}^{\rm K}-2i\Delta_{\rm eq}\hat{k}_y\delta g_{(0)+}^{\rm K}
        -\delta \Delta_+ \alpha_- -2i\Delta_{\rm eq}\hat{k}_y \beta_+ v_{\rm ex0}\big]\label{eq:deltaf2}.
\end{align}
In contrast with the equilibrium Green's function, the PHA in the DOS affects the pair amplitudes in the nonequilibrium Keldysh Green's function.
As seen later, the PHA correction to the pair amplitude drastically changes the linear coupling between the acoustic wave and collective modes.
In order to evaluate the order parameter fluctuation and the electric current, it is necessary to integrate the PHA correction to the nonequilibrium Keldysh Green's function.
Specifically, $\int \frac{d\epsilon}{2\pi i} \delta g_{\rm (1)-}^{\rm K}$ directly modifies the electric current, and $\int \frac{d\epsilon}{2\pi i} \delta f_{\rm (1)\pm}^{\rm K}$ affects the order parameter fluctuation.
Using Eq.~\eqref{Keldysh_QCL}, we obtain,
\begin{eqnarray}
\label{g1-}
\int \frac{d\epsilon}{2\pi i} \delta g_{\rm (1)-}^{\rm K}&=&\frac{a}{\epsilon_{\rm F}}\bigg[ \varphi \eta \Delta_{\rm eq}(\hat{k}_x\delta \Delta_+ - i\hat{k}_y\delta \Delta_-)-\int \frac{d\epsilon}{2\pi i}v_{\rm ex0}\alpha_+\bigg]\nonumber\\\\
       \int \frac{d\epsilon}{2\pi i} \delta f_{\rm (1)+}^{\rm K}&=&\frac{a}{2\epsilon_{\rm F}}\bigg[\omega \int \frac{d\epsilon}{2\pi i}\delta f_{(0)-}^{\rm K}-2\Delta_{\rm eq}\hat{k}_x\int \frac{d\epsilon}{2\pi i}\delta g_{\rm (0)+}^{\rm K}\label{eq:deltaf1}\nonumber\\
       &&-4\Delta_{\rm eq}\hat{k}_x \gamma v_{\rm ex0}\bigg],\\
        \int \frac{d\epsilon}{2\pi i}\delta f_{\rm (1)-}^{\rm K}&=&\frac{a}{2\epsilon_{\rm F}}\bigg[\omega \int \frac{d\epsilon}{2\pi i}\delta f_{(0)+}^{\rm K}-2i\Delta_{\rm eq}\hat{k}_y\int \frac{d\epsilon}{2\pi i}\delta g_{(0)+}^{\rm K}\nonumber\\
        &-&4i\Delta_{\rm eq}\hat{k}_y \gamma v_{\rm ex0}\bigg]\label{eq:deltaf2}.
\end{eqnarray}
The $\varphi$-function in Eq.~\eqref{g1-} is defined as,
\begin{widetext}
\begin{eqnarray}
\varphi&=&-2\int^{\epsilon_c}_{|\Delta_{\rm eq}|}d\epsilon \frac{\tanh\left(\frac{\epsilon}{2T}\right)}{\sqrt{\epsilon^2-\Delta_{\rm eq}^2}}\bigg[
\left(\frac{2\omega\epsilon_+^2-\epsilon_+(4\epsilon_+^2+\omega^2-\eta^2-4\Delta^2)}{(4\epsilon^2_+-\eta^2)(\omega^2-\eta^2)+4\Delta_{\rm eq}^2\eta^2}+\frac{2\omega\epsilon_-^2+\epsilon_-(4\epsilon_-^2+\omega^2-\eta^2-4\Delta^2)}{(4\epsilon^2_--\eta^2)(\omega^2-\eta^2)+4\Delta_{\rm eq}^2\eta^2}\right)
\bigg].
\label{phi}
\end{eqnarray}
\end{widetext}
%where we introduced the cut-off in Eq.~\eqref{phi} to avoid  the logarithmic divergence.
%Eq.~\eqref{g1-} generates the PHA correction to the electric current.
The $\varphi$-function also exhibits a logarithmic divergence on $\epsilon_c$, and thus the cut-off frequency is necessary to regularize
the integral.
Note that the second term in Eq.~\eqref{g1-} is even in the momentum and does not modify the electric current. Therefore, the contribution to the current from the terms that reflect the PHA is due to the order parameter fluctuations. }

{\subsection{Anomalous acoustoelectric effect induced by clapping modes}
We here derive the expression of the electric current with the use of the obtained Keldysh response function.
Substituting the Keldysh response function from Eqs.~\eqref{Eilenberger3} and \eqref{g1-} into Eq.~(\ref{electric_current1}), we obtain the electric current as the sum of two contributions,
\begin{eqnarray}
    \label{electric_current2}
    {\bm J}&=&{\bm J}_{\rm QP}+{\bm J}_{\rm CM},\\
    \label{electric_currentQP}
    {\bm J}_{\rm QP}&=&\frac{1}{4}N(\epsilon_{\rm F})\left\langle {\bm v}_{\rm F}\left( \frac{4\omega \eta}{\omega^2-\eta^2}\right)\left(\lambda-1\right)v_{\rm ex0}
    \right\rangle_{{\rm FS},{\bm k}_{\rm F}},\\
    \label{electric_currentCM}
    {\bm J}_{\rm CM}&=&\frac{1}{4}N(\epsilon_{\rm F})\big\langle {\bm v}_{\rm F}\eta \Delta_{\rm eq}\overline{\lambda}\big(i\hat{k}_y \delta \Delta_+-\hat{k}_x\delta \Delta_-\big)\big\rangle_{{\rm FS},{\bm k}_{\rm F}}\nonumber\\
    &-&\frac{a}{8\epsilon_{\rm F}}N(\epsilon_{\rm F})\big\langle {\bm v}_{\rm F}\eta \Delta_{\rm eq}\varphi \big(i\hat{k}_x \delta \Delta_+-\hat{k}_y\delta \Delta_-\big)\big\rangle_{{\rm FS},{\bm k}_{\rm F}}
\end{eqnarray}
The first term, ${\bm J}_{\rm QP}$, represents the electric current carried by the Bogoliubov quasiparticles, while the second term, ${\bm J}_{\rm CM}$, is the electric current carried by the collective modes, $\delta\Delta_{\pm}$.}
%The second term of the collective mode current arises from the PHA correction to the nonequilibrium Keldysh Green's function.
%The PHA effect is also incorporated into the order parameter fluctuations.}

{As discussed in the main text, we consider the acoustic wave propagating along the $x$-direction, $\bm q=(q,0)$, and 
express the order parameter fluctuations, $\delta\Delta_\pm$, via the collective modes using Eqs.~\eqref{sc_fluc1}-\eqref{sc_fluc_Emode},
\begin{gather}
    \label{sc_D+}
    k_F\delta\Delta_+=\left[\delta\mathcal{D}_++\delta\mathcal{E}_+\right]k_x+i\left[\delta\mathcal{D}_- - \delta\mathcal{E}_-\right]k_y\,,
    \\
    \label{sc_D-}
    k_F\delta\Delta_-=\left[\delta\mathcal{D}_- + \delta\mathcal{E}_-\right]k_x+i\left[\delta\mathcal{D}_+ - \delta\mathcal{E}_+\right]k_y\,.
\end{gather}
Substituting this into Eq.~\eqref{electric_currentCM}, we obtain the 
 acoustoelectric conductivity tensor, which is also decomposed into the contributions from the Bogoliubov quasiparticles ($\chi_{ij}^{\rm QP}$) and collective modes ($\chi_{ij}^{\rm CM}$) , }
{\begin{eqnarray}
	\label{AE_conductivity}
	\chi_{ij}=\chi_{ij}^{\rm QP}+\chi_{ij}^{\rm CM},
\end{eqnarray}
where
\begin{eqnarray}
	\label{chixx_QP_sup}
   	\chi_{xx}^{\rm QP}&=&\frac{iN(\epsilon_{\rm F})v_{\rm F}^2(X_0+X_1)}{4\omega},\\
    	\chi_{yx}^{\rm QP}&=&0,\\
	\label{chixx_CM_sup}
    	\chi_{xx}^{\rm CM}&=&-\frac{iN(\epsilon_{\rm F})v_{\rm F}^2\Delta_{\rm eq}}{8}\left[\left(\overline{\lambda}_0+\overline{\lambda}_1\right)\frac{\delta \mathcal{D}_-}{v_{\rm ex0}} +\left(\overline{\lambda}_1+\overline{\lambda}_2\right)\frac{\delta \mathcal{E}_-}{v_{\rm ex0}}\right]\nonumber\\
	&-&\frac{iaN(\epsilon_{\rm F})v_{\rm F}^2\Delta_{\rm eq}}{16\epsilon_{\rm F}} \left[\left(\varphi_1+\varphi_2\right)\frac{\delta \mathcal{D}_+}{v_{\rm ex0}}+\left(\varphi_0+\varphi_1\right)\frac{\delta \mathcal{E}_+}{v_{\rm ex0}}\right]\nonumber\\
	&\approx&-\frac{iN(\epsilon_{\rm F})v_{\rm F}^2\Delta_{\rm eq}}{8}\left[\left(\overline{\lambda}_1+\overline{\lambda}_2\right)\frac{\delta \mathcal{E}_-}{v_{\rm ex0}} +\frac{a}{2\epsilon_{\rm F}}\left(\varphi_0+\varphi_1\right)\frac{\delta \mathcal{E}_+}{v_{\rm ex0}}\right]\nonumber\\\\
	\label{chiyx_CM_sup}
         \chi_{yx}^{\rm CM}&=&\frac{N(\epsilon_{\rm F})v_{\rm F}^2\Delta_{\rm eq}}{8}\left[\left(\overline{\lambda}_0-\overline{\lambda}_2\right)\frac{\delta \mathcal{E}_+}{v_{\rm ex0}}-\frac{a}{2\epsilon_{\rm F}} \left(\varphi_0-\varphi_2\right)\frac{\delta \mathcal{E}_-}{v_{\rm ex0}}\right]
         \,.\nonumber\\
\end{eqnarray}
Here we introduced the moments of the Tsuneto function, $\overline{\lambda}_n=\braket{(\hat{k}_x^2-\hat{k}_y^2)^n\overline{\lambda}}_{{\rm FS},\bm k_{\rm F}}$ and $X_n=\braket{(\hat{k}_x^2-\hat{k}_y^2)^n \frac{2\omega^2}{\omega^2-\eta^2}(\lambda-1)}_{{\rm FS},\bm k_{\rm F}}$, as well as the moments of $\varphi$-function, $\varphi_n=\braket{(\hat{k}_x^2-\hat{k}_y^2)^n\varphi}_{{\rm FS},\bm k_{\rm F}}$.
In the third line of Eq.~(\ref{chixx_CM_sup}), we neglected the contribution of the phase mode, ($\delta \mathcal{D}_-$), since the long-range Coulomb interaction shifts the frequency of this mode to the plasmon energy, above the pair-breaking continuum $\omega > 2|\Delta_{\rm eq}|$ (the Anderson-Higgs mechanism).
In the second equality of Eq.~(\ref{chixx_CM_sup}), we also neglected the Higgs mode($\delta \mathcal{D}_+$), since its energy lies above the threshold energy, $2|\Delta_{\rm eq}|$.
Also, as shown in the Appendix C, the induced the Higgs mode is the order of $\delta \mathcal{D}_+/v_{\rm ex0}\sim \mathcal{O}(T_c/\epsilon_{\rm F})$,
and thus its contribution  is the second order in the PHA.}

{This yields the expressions in Eqs.\eqref{chixx_QP}-\eqref{chiyx_CM} in the main text.}
%Hence, these contributions are negligibly small.}

\section{Derivation of the matrix gap equation~(\ref{eigeneq}) for the order parameter fluctuations}
\label{sec:derivation}
In this appendix, we derive the matrix equation~(\ref{eigeneq}) for the order parameter fluctuations.
We begin with the gap equations for the linear response of the Keldysh pair amplitudes,
\begin{eqnarray}
    \label{gapeq3}
    \delta \Delta(\bm k_{\rm F})&=&N(\epsilon_{\rm F})\int \frac{d\epsilon}{4\pi i}\left\langle V({\bm k}_{\rm F},{\bm k}_{\rm F}')\delta  f^{\rm K}(\epsilon,{\bm k}_{\rm F}') \right\rangle_{{\rm FS},\bm k_{\rm F}'},\\
    \label{gapeq4}
    \delta \Delta^*(\bm k_{\rm F})&=&N(\epsilon_{\rm F})\int \frac{d\epsilon}{4\pi i}\left\langle V({\bm k}_{\rm F},{\bm k}_{\rm F}')\delta  \overline{f}^{\rm K}(\epsilon,{\bm k}_{\rm F}') \right\rangle_{{\rm FS},\bm k_{\rm F}'},
\end{eqnarray}
where $V({\bm k}_{\rm F},{\bm k}_{\rm F}')=V_{\rm pair}\hat{\bm k}\cdot \hat{\bm k}'$ is the $p$-wave pairing interaction.
The Keldysh pair amplitudes are decomposed into the Green's function in the quasiclassical limit and the PHA correction as $\delta  f^{\rm K}=\delta  f^{\rm K}_{(0)}+\delta  f^{\rm K}_{(1)}$.
Using Eqs.~\eqref{Eilenberger3}, \eqref{eq:deltaf1}, and \eqref{eq:deltaf2}, we recast Eq.~(\ref{gapeq3}) and (\ref{gapeq4}) into,
\begin{widetext}
\begin{align}
    \label{gapeq5}
    \frac{\delta \Delta(\bm k_{\rm F})}{N(\epsilon_{\rm F})V_{\rm pair}}=&\frac{1}{2}\bigg\langle \hat{\bm k}\cdot \hat{\bm k}'\bigg[\left(1+\frac{\omega a}{2\epsilon_{\rm F}}\right)
    \bigg\{-\omega \Delta_{\rm eq}({\bm k}_{\rm F}')\overline{\lambda}v_{\rm ex0}+\left(\gamma+\frac{\omega^2-\eta^2-2\Delta_{\rm eq}^2}{2}\overline{\lambda}\right)\delta\Delta({\bm k}_{\rm F}')-\Delta_{\rm eq}({\bm k}_{\rm F}')^2\overline{\lambda}\delta\Delta^*({\bm k}_{\rm F}')\bigg\}\nonumber\\
    &-\frac{a\gamma}{\epsilon_{\rm F}}\Delta({\bm k}_{\rm F}')v_{\rm ex0}-\frac{a}{2\epsilon_{\rm F}}\Delta({\bm k}_{\rm F}')\bigg\{-\frac{4\omega^2}{\omega^2-\eta^2}(\lambda-1)v_{\rm ex0}+\omega\overline{\lambda}
    \left(\Delta^*({\bm k}_{\rm F}')\delta \Delta({\bm k}_{\rm F}')-\Delta({\bm k}_{\rm F}')\delta \Delta^*({\bm k}_{\rm F}')
    \right)\bigg\}
    \bigg]\bigg\rangle_{{\rm FS},\bm k_{\rm F}'},\\
    \label{gapeq6}
    \frac{\delta \Delta^*(\bm k_{\rm F})}{N(\epsilon_{\rm F})V_{\rm pair}}=&\frac{1}{2}\bigg\langle \hat{\bm k}\cdot \hat{\bm k}'\bigg[\left(1-\frac{\omega a}{2\epsilon_{\rm F}}\right)
    \bigg\{\omega \Delta^*_{\rm eq}({\bm k}_{\rm F}')\overline{\lambda}v_{\rm ex0}+\left(\gamma+\frac{\omega^2-\eta^2-2\Delta_{\rm eq}^2}{2}\overline{\lambda}\right)\delta\Delta^*({\bm k}_{\rm F}')-\Delta^*_{\rm eq}({\bm k}_{\rm F}')^2\overline{\lambda}\delta\Delta({\bm k}_{\rm F}')\bigg\}\nonumber\\
    &-\frac{a\gamma}{\epsilon_{\rm F}}\Delta^*({\bm k}_{\rm F}')v_{\rm ex0}-\frac{a}{2\epsilon_{\rm F}}\Delta^*({\bm k}_{\rm F}')\bigg\{-\frac{4\omega^2}{\omega^2-\eta^2}(\lambda-1)v_{\rm ex0}+\omega\overline{\lambda}
    \left(\Delta^*({\bm k}_{\rm F}')\delta \Delta({\bm k}_{\rm F}')-\Delta({\bm k}_{\rm F}')\delta \Delta^*({\bm k}_{\rm F}')
    \right)\bigg\}
    \bigg]\bigg\rangle_{{\rm FS},\bm k_{\rm F}'}.
\end{align}
Let us consider the $p$-wave order parameter fluctuation, $\delta \Delta({\bm k}_{\rm F}')=\delta \Delta_{(+)} (\hat{k}_x+i\hat{k}_y)+\delta \Delta_{(-)} (\hat{k}_x-i\hat{k}_y)$.
Multiplying $(\hat{k}_x\pm i\hat{k}_y)$ with Eqs.~(\ref{gapeq6}-\ref{gapeq7}) and performing momentum integral with $\bm k_{\rm F}$, we then obtain
\begin{align}
    \label{gapeq7}
    \frac{\delta \Delta_{(\pm)}}{N(\epsilon_{\rm F})V_{\rm pair}}=&\frac{1}{2}\bigg\langle (\hat{k}_x'\mp i\hat{k}_y')\bigg[\left(1+\frac{\omega a}{2\epsilon_{\rm F}}\right)
    \bigg\{-\omega \Delta_{\rm eq}({\bm k}_{\rm F}')\overline{\lambda}v_{\rm ex0}+\left(\gamma+\frac{\omega^2-\eta^2-2\Delta_{\rm eq}^2}{2}\overline{\lambda}\right)\delta\Delta({\bm k}_{\rm F}')-\Delta_{\rm eq}({\bm k}_{\rm F}')^2\overline{\lambda}\delta\Delta^*({\bm k}_{\rm F}')\bigg\}\nonumber\\
    &-\frac{a\gamma}{\epsilon_{\rm F}}\Delta({\bm k}_{\rm F}')v_{\rm ex0}-\frac{a}{2\epsilon_{\rm F}}\Delta({\bm k}_{\rm F}')\bigg\{-\frac{4\omega^2}{\omega^2-\eta^2}(\lambda-1)v_{\rm ex0}+\omega\overline{\lambda}
    \left(\Delta^*({\bm k}_{\rm F}')\delta \Delta({\bm k}_{\rm F}')-\Delta({\bm k}_{\rm F}')\delta \Delta^*({\bm k}_{\rm F}')
    \right)\bigg\}
    \bigg]\bigg\rangle_{{\rm FS},\bm k_{\rm F}'},\\
    \label{gapeq8}
    \frac{\delta \Delta_{(\pm)}^*}{N(\epsilon_{\rm F})V_{\rm pair}}=&\frac{1}{2}\bigg\langle (\hat{k}_x'\mp i\hat{k}_y')\bigg[\left(1-\frac{\omega a}{2\epsilon_{\rm F}}\right)
    \bigg\{\omega \Delta^*_{\rm eq}({\bm k}_{\rm F}')\overline{\lambda}v_{\rm ex0}+\left(\gamma+\frac{\omega^2-\eta^2-2\Delta_{\rm eq}^2}{2}\overline{\lambda}\right)\delta\Delta^*({\bm k}_{\rm F}')-\Delta^*_{\rm eq}({\bm k}_{\rm F}')^2\overline{\lambda}\delta\Delta({\bm k}_{\rm F}')\bigg\}\nonumber\\
    &-\frac{a\gamma}{\epsilon_{\rm F}}\Delta^*({\bm k}_{\rm F}')v_{\rm ex0}-\frac{a}{2\epsilon_{\rm F}}\Delta^*({\bm k}_{\rm F}')\bigg\{-\frac{4\omega^2}{\omega^2-\eta^2}(\lambda-1)v_{\rm ex0}+\omega\overline{\lambda}
    \left(\Delta^*({\bm k}_{\rm F}')\delta \Delta({\bm k}_{\rm F}')-\Delta({\bm k}_{\rm F}')\delta \Delta^*({\bm k}_{\rm F}'')
    \right)\bigg\}
    \bigg]\bigg\rangle_{{\rm FS},\bm k_{\rm F}'}.
\end{align}
We consider the acoustic wave propagation along the $x$-direction, $\bm q=(q,0)$.
By using $\overline{\lambda}_n=\braket{(\hat{k}_x^2-\hat{k}_y^2)^n\overline{\lambda}}_{{\rm FS},\bm k_{\rm F}}$ and $X_n=\braket{(\hat{k}_x^2-\hat{k}_y^2)^n \frac{2\omega^2}{\omega^2-\eta^2}(\lambda-1)}_{{\rm FS},\bm k_{\rm F}}$, and performing the momentum integral, Eqs.~\eqref{gapeq7} and \eqref{gapeq8} are reduced to the linear equations for $\delta \Delta_{(\pm)}$ and $\delta\Delta^{\ast}_{(\pm)}$,
\begin{align}
    \label{gapeq9}
    \left[\left(1+\frac{a\omega}{2\epsilon_{\rm F}}\right)\omega \Delta_{\rm eq}\overline{\lambda}_0+\frac{a}{\epsilon_{\rm F}}(\gamma-X_0)
    \right]v_{\rm ex0}
    =&\left(1+\frac{a\omega}{2\epsilon_{\rm F}}\right)\bigg[ \mathcal{A}\delta \Delta_{(+)}
    +\mathcal{B}\delta \Delta_{(-)}-\Delta_{\rm eq}^2(\overline{\lambda}_0\delta \Delta^*_{(+)}+\overline{\lambda}_1\delta \Delta^*_{(-)})\bigg]\nonumber\\
    &-\frac{a}{2\epsilon_{\rm F}}\Delta_{\rm eq}^2\omega\left(\overline{\lambda}_0\delta \mathcal{D}_-+\overline{\lambda}_1\delta \mathcal{E}_- \right)+\frac{a\omega}{2\epsilon_{\rm F}}\gamma \delta \Delta_{(+)},\\
    \label{gapeq10}
    \left[-\left(1-\frac{a\omega}{2\epsilon_{\rm F}}\right)\omega \Delta_{\rm eq}\overline{\lambda}_0+\frac{a}{\epsilon_{\rm F}}(\gamma-X_0)
    \right]v_{\rm ex0}
    =&\left(1-\frac{a\omega}{2\epsilon_{\rm F}}\right)\bigg[ \mathcal{A}\delta \Delta^*_{(+)}
    +\mathcal{B}\delta \Delta^*_{(-)}-\Delta_{\rm eq}^2(\overline{\lambda}_0\delta \Delta_{(+)}+\overline{\lambda}_1\delta \Delta_{(-)})\bigg]\nonumber\\
    &-\frac{a}{2\epsilon_{\rm F}}\Delta_{\rm eq}^2\omega\left(\overline{\lambda}_0\delta \mathcal{D}_-+\overline{\lambda}_1\delta \mathcal{E}_- \right)-\frac{a\omega}{2\epsilon_{\rm F}}\gamma \delta \Delta^*_{(+)},\\
     \label{gapeq11}
    \left[\left(1+\frac{a\omega}{2\epsilon_{\rm F}}\right)\omega \Delta_{\rm eq}\overline{\lambda}_1-\frac{a}{\epsilon_{\rm F}}X_1
    \right]v_{\rm ex0}
    =&\left(1+\frac{a\omega}{2\epsilon_{\rm F}}\right)\bigg[ \mathcal{B}\delta \Delta_{(+)}
    +\mathcal{A}\delta \Delta_{(-)}-\Delta_{\rm eq}^2(\overline{\lambda}_1\delta \Delta^*_{(+)}+(2\overline{\lambda}_2-\overline{\lambda}_0)\delta \Delta^*_{(-)})\bigg]\nonumber\\
    &-\frac{a}{2\epsilon_{\rm F}}\Delta_{\rm eq}^2\omega\left(\overline{\lambda}_1\delta \mathcal{D}_-+\overline{\lambda}_1\delta \mathcal{E}_+-2\overline{\lambda}_2\delta \Delta_{(-)}^* \right)+\frac{a\omega}{2\epsilon_{\rm F}}\gamma \delta \Delta_{(-)},\\
    \label{gapeq12}
    \left[-\left(1-\frac{a\omega}{2\epsilon_{\rm F}}\right)\omega \Delta_{\rm eq}\overline{\lambda}_1-\frac{a}{\epsilon_{\rm F}}X_1
    \right]v_{\rm ex0}
    =&\left(1-\frac{a\omega}{2\epsilon_{\rm F}}\right)\bigg[ \mathcal{B}\delta \Delta^*_{(+)}
    +\mathcal{A}\delta \Delta^*_{(-)}-\Delta_{\rm eq}^2(\overline{\lambda}_1\delta \Delta_{(+)}+(2\overline{\lambda}_2-\overline{\lambda}_0)\delta \Delta_{(-)})\bigg]\nonumber\\
    &-\frac{a}{2\epsilon_{\rm F}}\Delta_{\rm eq}^2\omega\left(\overline{\lambda}_1\delta \mathcal{D}_-+\overline{\lambda}_1\delta \mathcal{E}_++2\overline{\lambda}_2\delta \Delta_{(-)}^* \right)-\frac{a\omega}{2\epsilon_{\rm F}}\gamma \delta \Delta^*_{(-)},
\end{align}
\end{widetext}
where we have used Eq.~(\ref{gamma}) and introduced abbreviations
\begin{gather}
\mathcal{A}\equiv \frac{\omega^2-2\Delta_{\rm eq}^2}{2}\overline{\lambda}_0-\frac{v_{\rm F}^2q^2}{4}(\overline{\lambda}_0+\overline{\lambda}_1) , \\
\mathcal{B}\equiv \frac{\omega^2-2\Delta_{\rm eq}^2}{2}\overline{\lambda}_1-\frac{v_{\rm F}^2q^2}{4}(\overline{\lambda}_1+\overline{\lambda}_2) .
\end{gather}
We subtract and add Eq. (\ref{gapeq9}) and Eq. (\ref{gapeq10}), and Eq. (\ref{gapeq11}) and Eq. (\ref{gapeq12}) to obtain the matrix equation,
\begin{eqnarray}
    \label{gapeq13}
    M(\bm q,\omega)\delta{\bm D}(\bm q,\omega)={\bm v}_{\rm ex},
\end{eqnarray}
where the vectors of the order parameter fluctuations and the driving force are given by
\begin{align}
    \delta{\bm D}=\begin{pmatrix}
        \delta \mathcal{D}_+\\
        \delta \mathcal{E}_+\\
        \delta \mathcal{D}_-\\
        \delta \mathcal{E}_-
    \end{pmatrix},~
    {\bm v}_{\rm ex}=\begin{pmatrix}_
       \frac{av_{\rm ex0}}{\epsilon_{\rm F}}\left(\omega\Delta_{\rm eq}\overline{\lambda}_0+ 2\Delta_{\rm eq}(\gamma-X_0)\right)\\
       \frac{av_{\rm ex0}}{\epsilon_{\rm F}}\left(\omega\Delta_{\rm eq}\overline{\lambda}_1- 2\Delta_{\rm eq}X_1\right)\\
        2\omega \Delta_{\rm eq}\overline{\lambda}_0 v_{\rm ex0}\\
        2\omega \Delta_{\rm eq}\overline{\lambda}_1 v_{\rm ex0}
    \end{pmatrix},
\end{align}
respectively, and the matrix $M(\bm q,\omega)$ is
\begin{widetext}
\begin{align}
    M(\bm q,\omega)\equiv
    \left(\begin{smallmatrix}
        \frac{\omega^2-4\Delta_{\rm eq}^2}{2}\overline{\lambda}_0-\frac{v_{\rm F}^2q^2 (\overline{\lambda}_0+\overline{\lambda}_1)}{4}&&
        \frac{\omega^2-4\Delta_{\rm eq}^2}{2}\overline{\lambda}_1-\frac{v_{\rm F}^2q^2(\overline{\lambda}_1+\overline{\lambda}_2)}{4}&&
        \frac{a\omega}{2\epsilon_{\rm F}}\left[\frac{\omega^2-4\Delta_{\rm eq}^2}{2}\overline{\lambda}_0-\frac{v_{\rm F}^2q^2(\overline{\lambda}_0+\overline{\lambda}_1)}{4}+\gamma \right]&&
        \frac{a\omega}{2\epsilon_{\rm F}}\left[\frac{\omega^2-4\Delta_{\rm eq}^2}{2}\overline{\lambda}_1-\frac{v_{\rm F}^2q^2(\overline{\lambda}_1+\overline{\lambda}_2)}{4}\right]\\
        \frac{\omega^2-4\Delta_{\rm eq}^2}{2}\overline{\lambda}_1-\frac{v_{\rm F}^2q^2(\overline{\lambda}_1+\overline{\lambda}_2)}{4}&&
        \frac{\omega^2\overline{\lambda}_0-4\Delta_{\rm eq}^2\overline{\lambda}_2}{2}-\frac{v_{\rm F}^2q^2 (\overline{\lambda}_0+\overline{\lambda}_1)}{4}&&
        \frac{a\omega}{2\epsilon_{\rm F}}\left[\frac{\omega^2-4\Delta_{\rm eq}^2}{2}\overline{\lambda}_1-\frac{v_{\rm F}^2q^2(\overline{\lambda}_1+\overline{\lambda}_2)}{4}\right]&&
        \frac{a\omega}{2\epsilon_{\rm F}}\left[\frac{\omega^2-4\Delta_{\rm eq}^2}{2}\overline{\lambda}_0-\frac{v_{\rm F}^2q^2(\overline{\lambda}_0+\overline{\lambda}_1)}{4}+\gamma \right]\\
        \frac{a\omega}{2\epsilon_{\rm F}}\left[\frac{\omega^2-4\Delta_{\rm eq}^2}{2}\overline{\lambda}_0-\frac{v_{\rm F}^2q^2(\overline{\lambda}_0+\overline{\lambda}_1)}{4}+\gamma \right]&&
        \frac{a\omega}{2\epsilon_{\rm F}}\left[\frac{\omega^2-4\Delta_{\rm eq}^2}{2}\overline{\lambda}_1-\frac{v_{\rm F}^2q^2(\overline{\lambda}_1+\overline{\lambda}_2)}{4}\right]&&
        \frac{\omega^2}{2}\overline{\lambda}_0-\frac{v_{\rm F}^2q^2(\overline{\lambda}_0+\overline{\lambda}_1)}{4}&&
        \frac{\omega^2}{2}\overline{\lambda}_1-\frac{v_{\rm F}^2q^2 (\overline{\lambda}_1+\overline{\lambda}_2)}{4}\\
        \frac{a\omega}{2\epsilon_{\rm F}}\left[\frac{\omega^2-4\Delta_{\rm eq}^2}{2}\overline{\lambda}_1-\frac{v_{\rm F}^2q^2(\overline{\lambda}_1+\overline{\lambda}_2)}{4}\right]&&
        \frac{a\omega}{2\epsilon_{\rm F}}\left[\frac{\omega^2-4\Delta_{\rm eq}^2}{2}\overline{\lambda}_0-\frac{v_{\rm F}^2q^2(\overline{\lambda}_0+\overline{\lambda}_1)}{4}+\gamma\right]&&
        \frac{\omega^2}{2}\overline{\lambda}_1-\frac{v_{\rm F}^2q^2(\overline{\lambda}_1+\overline{\lambda}_2)}{4}&&
        \frac{\omega^2\overline{\lambda}_0-4\Delta_{\rm eq}^2(\overline{\lambda}_0-\overline{\lambda}_2)}{2}-\frac{v_{\rm F}^2q^2 (\overline{\lambda}_0+\overline{\lambda}_1)}{4}
    \end{smallmatrix}\right).
\end{align}
As discussed in the main text, the energy of the phase mode is pushed up to the plasmon energy, which is much larger than any other energy scale in superconductors.
The difference of the energy scale allows us to neglect the phase mode in the matrix equation (\ref{gapeq13}).
We finally obtain Eq.~(\ref{eigeneq}) as
\begin{align}
    \label{eigeneq14}
    &\left(\begin{smallmatrix}
        \frac{\omega^2-4\Delta_{\rm eq}^2}{2}\overline{\lambda}_0-\frac{v_{\rm F}^2q^2 (\overline{\lambda}_0+\overline{\lambda}_1)}{4}&&
        \frac{\omega^2-4\Delta_{\rm eq}^2}{2}\overline{\lambda}_1-\frac{v_{\rm F}^2q^2(\overline{\lambda}_1+\overline{\lambda}_2)}{4}&&
        \frac{a\omega}{2\epsilon_{\rm F}}\left[\frac{\omega^2-4\Delta_{\rm eq}^2}{2}\overline{\lambda}_1-\frac{v_{\rm F}^2q^2(\overline{\lambda}_1+\overline{\lambda}_2)}{4}\right]\\
        \frac{\omega^2-4\Delta_{\rm eq}^2}{2}\overline{\lambda}_1-\frac{v_{\rm F}^2q^2(\overline{\lambda}_1+\overline{\lambda}_2)}{4}&&
        \frac{\omega^2\overline{\lambda}_0-4\Delta_{\rm eq}^2\overline{\lambda}_2}{2}-\frac{v_{\rm F}^2q^2 (\overline{\lambda}_0+\overline{\lambda}_1)}{4}&&
        \frac{a\omega}{2\epsilon_{\rm F}}\left[\frac{\omega^2-4\Delta_{\rm eq}^2}{2}\overline{\lambda}_0-\frac{v_{\rm F}^2q^2(\overline{\lambda}_0+\overline{\lambda}_1)}{4}+\gamma \right]\\
        \frac{a\omega}{2\epsilon_{\rm F}}\left[\frac{\omega^2-4\Delta_{\rm eq}^2}{2}\overline{\lambda}_1-\frac{v_{\rm F}^2q^2(\overline{\lambda}_1+\overline{\lambda}_2)}{4}\right]&&
        \frac{a\omega}{2\epsilon_{\rm F}}\left[\frac{\omega^2-4\Delta_{\rm eq}^2}{2}\overline{\lambda}_0-\frac{v_{\rm F}^2q^2(\overline{\lambda}_0+\overline{\lambda}_1)}{4}+\gamma\right]&&
        \frac{\omega^2\overline{\lambda}_0-4\Delta_{\rm eq}^2(\overline{\lambda}_0-\overline{\lambda}_2)}{2}-\frac{v_{\rm F}^2q^2 (\overline{\lambda}_0+\overline{\lambda}_1)}{4}
    \end{smallmatrix}\right)
    \begin{pmatrix}
        \delta \mathcal{D}_+\\
        \delta \mathcal{E}_+\\
        \delta \mathcal{E}_-
    \end{pmatrix}
    =
    \begin{pmatrix}_
       \frac{av_{\rm ex0}}{\epsilon_{\rm F}}\left(\omega\Delta_{\rm eq}\overline{\lambda}_0+ 2\Delta_{\rm eq}(\gamma-X_0)\right)\\
       \frac{av_{\rm ex0}}{\epsilon_{\rm F}}\left(\omega\Delta_{\rm eq}\overline{\lambda}_1- 2\Delta_{\rm eq}X_1\right)\\
        2\omega \Delta_{\rm eq}\overline{\lambda}_1 v_{\rm ex0}
    \end{pmatrix}.
\end{align}
\end{widetext}

%\end{appendix}
\bibliography{AAGE_ref.bib}

\begin{thebibliography}{103}
\expandafter\ifx\csname natexlab\endcsname\relax\def\natexlab#1{#1}\fi
\expandafter\ifx\csname bibnamefont\endcsname\relax
  \def\bibnamefont#1{#1}\fi
\expandafter\ifx\csname bibfnamefont\endcsname\relax
  \def\bibfnamefont#1{#1}\fi
\expandafter\ifx\csname citenamefont\endcsname\relax
  \def\citenamefont#1{#1}\fi
\expandafter\ifx\csname url\endcsname\relax
  \def\url#1{\texttt{#1}}\fi
\expandafter\ifx\csname urlprefix\endcsname\relax\def\urlprefix{URL }\fi
\providecommand{\bibinfo}[2]{#2}
\providecommand{\eprint}[2][]{\url{#2}}

\bibitem[{\citenamefont{Read and Green}(2000)}]{read2000paired}
\bibinfo{author}{\bibfnamefont{N.}~\bibnamefont{Read}} \bibnamefont{and}
  \bibinfo{author}{\bibfnamefont{D.}~\bibnamefont{Green}},
  \bibinfo{journal}{Phys. Rev. B} \textbf{\bibinfo{volume}{61}},
  \bibinfo{pages}{10267} (\bibinfo{year}{2000}).

\bibitem[{\citenamefont{Goswami and Balicas}(2013)}]{goswami2013topological}
\bibinfo{author}{\bibfnamefont{P.}~\bibnamefont{Goswami}} \bibnamefont{and}
  \bibinfo{author}{\bibfnamefont{L.}~\bibnamefont{Balicas}},
  \bibinfo{journal}{arXiv:1312.3632}  (\bibinfo{year}{2013}).

\bibitem[{\citenamefont{Goswami and
  Nevidomskyy}(2015)}]{goswami2015topological}
\bibinfo{author}{\bibfnamefont{P.}~\bibnamefont{Goswami}} \bibnamefont{and}
  \bibinfo{author}{\bibfnamefont{A.~H.} \bibnamefont{Nevidomskyy}},
  \bibinfo{journal}{Phys. Rev. B} \textbf{\bibinfo{volume}{92}},
  \bibinfo{pages}{214504} (\bibinfo{year}{2015}).

\bibitem[{\citenamefont{Kallin and Berlinsky}(2016)}]{kallin2016chiral}
\bibinfo{author}{\bibfnamefont{C.}~\bibnamefont{Kallin}} \bibnamefont{and}
  \bibinfo{author}{\bibfnamefont{J.}~\bibnamefont{Berlinsky}},
  \bibinfo{journal}{Rep. Prog. Phys.} \textbf{\bibinfo{volume}{79}},
  \bibinfo{pages}{054502} (\bibinfo{year}{2016}).

\bibitem[{\citenamefont{Volovik}(2003)}]{volovik}
\bibinfo{author}{\bibfnamefont{G.~E.} \bibnamefont{Volovik}},
  \emph{\bibinfo{title}{The Universe in a Helium Droplet}}
  (\bibinfo{publisher}{Oxford}, \bibinfo{year}{2003}).

\bibitem[{\citenamefont{Alicea}(2012)}]{alicea2012new}
\bibinfo{author}{\bibfnamefont{J.}~\bibnamefont{Alicea}},
  \bibinfo{journal}{Rep. Prog. Phys.} \textbf{\bibinfo{volume}{75}},
  \bibinfo{pages}{076501} (\bibinfo{year}{2012}).

\bibitem[{\citenamefont{Sato and Fujimoto}(2016)}]{sato2016majorana}
\bibinfo{author}{\bibfnamefont{M.}~\bibnamefont{Sato}} \bibnamefont{and}
  \bibinfo{author}{\bibfnamefont{S.}~\bibnamefont{Fujimoto}},
  \bibinfo{journal}{J. Phys. Soc. Jpn} \textbf{\bibinfo{volume}{85}},
  \bibinfo{pages}{072001} (\bibinfo{year}{2016}).

\bibitem[{\citenamefont{Sato and Ando}(2017)}]{sato2017topological}
\bibinfo{author}{\bibfnamefont{M.}~\bibnamefont{Sato}} \bibnamefont{and}
  \bibinfo{author}{\bibfnamefont{Y.}~\bibnamefont{Ando}},
  \bibinfo{journal}{Rep. Prog. Phys.} \textbf{\bibinfo{volume}{80}},
  \bibinfo{pages}{076501} (\bibinfo{year}{2017}).

\bibitem[{\citenamefont{Fu and Kane}(2008)}]{fu2008superconducting}
\bibinfo{author}{\bibfnamefont{L.}~\bibnamefont{Fu}} \bibnamefont{and}
  \bibinfo{author}{\bibfnamefont{C.~L.} \bibnamefont{Kane}},
  \bibinfo{journal}{Phys. Rev. Lett.} \textbf{\bibinfo{volume}{100}},
  \bibinfo{pages}{096407} (\bibinfo{year}{2008}).

\bibitem[{\citenamefont{Sanno et~al.}(2021)\citenamefont{Sanno, Miyazaki,
  Mizushima, and Fujimoto}}]{sanno2021ab}
\bibinfo{author}{\bibfnamefont{T.}~\bibnamefont{Sanno}},
  \bibinfo{author}{\bibfnamefont{S.}~\bibnamefont{Miyazaki}},
  \bibinfo{author}{\bibfnamefont{T.}~\bibnamefont{Mizushima}},
  \bibnamefont{and} \bibinfo{author}{\bibfnamefont{S.}~\bibnamefont{Fujimoto}},
  \bibinfo{journal}{Phys. Rev. B} \textbf{\bibinfo{volume}{103}},
  \bibinfo{pages}{054504} (\bibinfo{year}{2021}).

\bibitem[{\citenamefont{Kasahara et~al.}(2009)\citenamefont{Kasahara, Shishido,
  Shibauchi, Haga, Matsuda, Onuki, and Matsuda}}]{kasahara2009superconducting}
\bibinfo{author}{\bibfnamefont{Y.}~\bibnamefont{Kasahara}},
  \bibinfo{author}{\bibfnamefont{H.}~\bibnamefont{Shishido}},
  \bibinfo{author}{\bibfnamefont{T.}~\bibnamefont{Shibauchi}},
  \bibinfo{author}{\bibfnamefont{Y.}~\bibnamefont{Haga}},
  \bibinfo{author}{\bibfnamefont{T.}~\bibnamefont{Matsuda}},
  \bibinfo{author}{\bibfnamefont{Y.}~\bibnamefont{Onuki}}, \bibnamefont{and}
  \bibinfo{author}{\bibfnamefont{Y.}~\bibnamefont{Matsuda}},
  \bibinfo{journal}{New J. Phys.} \textbf{\bibinfo{volume}{11}},
  \bibinfo{pages}{055061} (\bibinfo{year}{2009}).

\bibitem[{\citenamefont{Kittaka et~al.}(2016)\citenamefont{Kittaka, Shimizu,
  Sakakibara, Haga, Yamamoto, {\=O}nuki, Tsutsumi, Nomoto, Ikeda, and
  Machida}}]{kittaka2016evidence}
\bibinfo{author}{\bibfnamefont{S.}~\bibnamefont{Kittaka}},
  \bibinfo{author}{\bibfnamefont{Y.}~\bibnamefont{Shimizu}},
  \bibinfo{author}{\bibfnamefont{T.}~\bibnamefont{Sakakibara}},
  \bibinfo{author}{\bibfnamefont{Y.}~\bibnamefont{Haga}},
  \bibinfo{author}{\bibfnamefont{E.}~\bibnamefont{Yamamoto}},
  \bibinfo{author}{\bibfnamefont{Y.}~\bibnamefont{{\=O}nuki}},
  \bibinfo{author}{\bibfnamefont{Y.}~\bibnamefont{Tsutsumi}},
  \bibinfo{author}{\bibfnamefont{T.}~\bibnamefont{Nomoto}},
  \bibinfo{author}{\bibfnamefont{H.}~\bibnamefont{Ikeda}}, \bibnamefont{and}
  \bibinfo{author}{\bibfnamefont{K.}~\bibnamefont{Machida}},
  \bibinfo{journal}{J. Phys. Soc. Jpn.} \textbf{\bibinfo{volume}{85}},
  \bibinfo{pages}{033704} (\bibinfo{year}{2016}).

\bibitem[{\citenamefont{Schemm et~al.}(2015)\citenamefont{Schemm, Baumbach,
  Tobash, Ronning, Bauer, and Kapitulnik}}]{schemm2015evidence}
\bibinfo{author}{\bibfnamefont{E.}~\bibnamefont{Schemm}},
  \bibinfo{author}{\bibfnamefont{R.}~\bibnamefont{Baumbach}},
  \bibinfo{author}{\bibfnamefont{P.}~\bibnamefont{Tobash}},
  \bibinfo{author}{\bibfnamefont{F.}~\bibnamefont{Ronning}},
  \bibinfo{author}{\bibfnamefont{E.}~\bibnamefont{Bauer}}, \bibnamefont{and}
  \bibinfo{author}{\bibfnamefont{A.}~\bibnamefont{Kapitulnik}},
  \bibinfo{journal}{Phys. Rev. B} \textbf{\bibinfo{volume}{91}},
  \bibinfo{pages}{140506} (\bibinfo{year}{2015}).

\bibitem[{\citenamefont{Schemm et~al.}(2014)\citenamefont{Schemm, Gannon,
  Wishne, Halperin, and Kapitulnik}}]{schemm2014observation}
\bibinfo{author}{\bibfnamefont{E.}~\bibnamefont{Schemm}},
  \bibinfo{author}{\bibfnamefont{W.}~\bibnamefont{Gannon}},
  \bibinfo{author}{\bibfnamefont{C.}~\bibnamefont{Wishne}},
  \bibinfo{author}{\bibfnamefont{W.}~\bibnamefont{Halperin}}, \bibnamefont{and}
  \bibinfo{author}{\bibfnamefont{A.}~\bibnamefont{Kapitulnik}},
  \bibinfo{journal}{Science} \textbf{\bibinfo{volume}{345}},
  \bibinfo{pages}{190} (\bibinfo{year}{2014}).

\bibitem[{\citenamefont{Tsutsumi et~al.}(2013)\citenamefont{Tsutsumi, Ishikawa,
  Kawakami, Mizushima, Sato, Ichioka, and Machida}}]{tsutsumi2013upt3}
\bibinfo{author}{\bibfnamefont{Y.}~\bibnamefont{Tsutsumi}},
  \bibinfo{author}{\bibfnamefont{M.}~\bibnamefont{Ishikawa}},
  \bibinfo{author}{\bibfnamefont{T.}~\bibnamefont{Kawakami}},
  \bibinfo{author}{\bibfnamefont{T.}~\bibnamefont{Mizushima}},
  \bibinfo{author}{\bibfnamefont{M.}~\bibnamefont{Sato}},
  \bibinfo{author}{\bibfnamefont{M.}~\bibnamefont{Ichioka}}, \bibnamefont{and}
  \bibinfo{author}{\bibfnamefont{K.}~\bibnamefont{Machida}},
  \bibinfo{journal}{J. Phys. Soc. Jpn.} \textbf{\bibinfo{volume}{82}},
  \bibinfo{pages}{113707} (\bibinfo{year}{2013}).

\bibitem[{\citenamefont{Yanase}(2016)}]{yanase2016nonsymmorphic}
\bibinfo{author}{\bibfnamefont{Y.}~\bibnamefont{Yanase}},
  \bibinfo{journal}{Phys. Rev. B} \textbf{\bibinfo{volume}{94}},
  \bibinfo{pages}{174502} (\bibinfo{year}{2016}).

\bibitem[{\citenamefont{MacLaughlin et~al.}(1984)\citenamefont{MacLaughlin,
  Tien, Clark, Lan, Fisk, Smith, and Ott}}]{maclaughlin1984nuclear}
\bibinfo{author}{\bibfnamefont{D.}~\bibnamefont{MacLaughlin}},
  \bibinfo{author}{\bibfnamefont{C.}~\bibnamefont{Tien}},
  \bibinfo{author}{\bibfnamefont{W.}~\bibnamefont{Clark}},
  \bibinfo{author}{\bibfnamefont{M.}~\bibnamefont{Lan}},
  \bibinfo{author}{\bibfnamefont{Z.}~\bibnamefont{Fisk}},
  \bibinfo{author}{\bibfnamefont{J.}~\bibnamefont{Smith}}, \bibnamefont{and}
  \bibinfo{author}{\bibfnamefont{H.}~\bibnamefont{Ott}},
  \bibinfo{journal}{Phys. Rev. Lett.} \textbf{\bibinfo{volume}{53}},
  \bibinfo{pages}{1833} (\bibinfo{year}{1984}).

\bibitem[{\citenamefont{Heffner et~al.}(1990)\citenamefont{Heffner, Smith,
  Willis, Birrer, Baines, Gygax, Hitti, Lippelt, Ott, Schenck
  et~al.}}]{PhysRevLett.65.2816}
\bibinfo{author}{\bibfnamefont{R.~H.} \bibnamefont{Heffner}},
  \bibinfo{author}{\bibfnamefont{J.~L.} \bibnamefont{Smith}},
  \bibinfo{author}{\bibfnamefont{J.~O.} \bibnamefont{Willis}},
  \bibinfo{author}{\bibfnamefont{P.}~\bibnamefont{Birrer}},
  \bibinfo{author}{\bibfnamefont{C.}~\bibnamefont{Baines}},
  \bibinfo{author}{\bibfnamefont{F.~N.} \bibnamefont{Gygax}},
  \bibinfo{author}{\bibfnamefont{B.}~\bibnamefont{Hitti}},
  \bibinfo{author}{\bibfnamefont{E.}~\bibnamefont{Lippelt}},
  \bibinfo{author}{\bibfnamefont{H.~R.} \bibnamefont{Ott}},
  \bibinfo{author}{\bibfnamefont{A.}~\bibnamefont{Schenck}},
  \bibnamefont{et~al.}, \bibinfo{journal}{Phys. Rev. Lett.}
  \textbf{\bibinfo{volume}{65}}, \bibinfo{pages}{2816} (\bibinfo{year}{1990}).

\bibitem[{\citenamefont{Jin et~al.}(1994)\citenamefont{Jin, Rosenbaum, Kim, and
  Stewart}}]{jin1994low}
\bibinfo{author}{\bibfnamefont{D.}~\bibnamefont{Jin}},
  \bibinfo{author}{\bibfnamefont{T.}~\bibnamefont{Rosenbaum}},
  \bibinfo{author}{\bibfnamefont{J.}~\bibnamefont{Kim}}, \bibnamefont{and}
  \bibinfo{author}{\bibfnamefont{G.}~\bibnamefont{Stewart}},
  \bibinfo{journal}{Phys. Rev. B} \textbf{\bibinfo{volume}{49}},
  \bibinfo{pages}{1540} (\bibinfo{year}{1994}).

\bibitem[{\citenamefont{Golding et~al.}(1985)\citenamefont{Golding, Bishop,
  Batlogg, Haemmerle, Fisk, Smith, and Ott}}]{golding1985observation}
\bibinfo{author}{\bibfnamefont{B.}~\bibnamefont{Golding}},
  \bibinfo{author}{\bibfnamefont{D.}~\bibnamefont{Bishop}},
  \bibinfo{author}{\bibfnamefont{B.}~\bibnamefont{Batlogg}},
  \bibinfo{author}{\bibfnamefont{W.}~\bibnamefont{Haemmerle}},
  \bibinfo{author}{\bibfnamefont{Z.}~\bibnamefont{Fisk}},
  \bibinfo{author}{\bibfnamefont{J.}~\bibnamefont{Smith}}, \bibnamefont{and}
  \bibinfo{author}{\bibfnamefont{H.}~\bibnamefont{Ott}},
  \bibinfo{journal}{Physical Rev. Lett.} \textbf{\bibinfo{volume}{55}},
  \bibinfo{pages}{2479} (\bibinfo{year}{1985}).

\bibitem[{\citenamefont{Batlogg et~al.}(1985)\citenamefont{Batlogg, Bishop,
  Golding, Varma, Fisk, Smith, and Ott}}]{PhysRevLett.55.1319}
\bibinfo{author}{\bibfnamefont{B.}~\bibnamefont{Batlogg}},
  \bibinfo{author}{\bibfnamefont{D.}~\bibnamefont{Bishop}},
  \bibinfo{author}{\bibfnamefont{B.}~\bibnamefont{Golding}},
  \bibinfo{author}{\bibfnamefont{C.~M.} \bibnamefont{Varma}},
  \bibinfo{author}{\bibfnamefont{Z.}~\bibnamefont{Fisk}},
  \bibinfo{author}{\bibfnamefont{J.~L.} \bibnamefont{Smith}}, \bibnamefont{and}
  \bibinfo{author}{\bibfnamefont{H.~R.} \bibnamefont{Ott}},
  \bibinfo{journal}{Phys. Rev. Lett.} \textbf{\bibinfo{volume}{55}},
  \bibinfo{pages}{1319} (\bibinfo{year}{1985}).

\bibitem[{\citenamefont{Machida}(2018{\natexlab{a}})}]{machida2018spin}
\bibinfo{author}{\bibfnamefont{K.}~\bibnamefont{Machida}}, \bibinfo{journal}{J.
  Phys. Soc. Jpn.} \textbf{\bibinfo{volume}{87}}, \bibinfo{pages}{033703}
  (\bibinfo{year}{2018}{\natexlab{a}}).

\bibitem[{\citenamefont{Biswas et~al.}(2013)\citenamefont{Biswas, Luetkens,
  Neupert, St\"urzer, Baines, Pascua, Schnyder, Fischer, Goryo, Lees
  et~al.}}]{PhysRevB.87.180503}
\bibinfo{author}{\bibfnamefont{P.~K.} \bibnamefont{Biswas}},
  \bibinfo{author}{\bibfnamefont{H.}~\bibnamefont{Luetkens}},
  \bibinfo{author}{\bibfnamefont{T.}~\bibnamefont{Neupert}},
  \bibinfo{author}{\bibfnamefont{T.}~\bibnamefont{St\"urzer}},
  \bibinfo{author}{\bibfnamefont{C.}~\bibnamefont{Baines}},
  \bibinfo{author}{\bibfnamefont{G.}~\bibnamefont{Pascua}},
  \bibinfo{author}{\bibfnamefont{A.~P.} \bibnamefont{Schnyder}},
  \bibinfo{author}{\bibfnamefont{M.~H.} \bibnamefont{Fischer}},
  \bibinfo{author}{\bibfnamefont{J.}~\bibnamefont{Goryo}},
  \bibinfo{author}{\bibfnamefont{M.~R.} \bibnamefont{Lees}},
  \bibnamefont{et~al.}, \bibinfo{journal}{Phys. Rev. B}
  \textbf{\bibinfo{volume}{87}}, \bibinfo{pages}{180503}
  (\bibinfo{year}{2013}).

\bibitem[{\citenamefont{Fischer et~al.}(2014)\citenamefont{Fischer, Neupert,
  Platt, Schnyder, Hanke, Goryo, Thomale, and Sigrist}}]{PhysRevB.89.020509}
\bibinfo{author}{\bibfnamefont{M.~H.} \bibnamefont{Fischer}},
  \bibinfo{author}{\bibfnamefont{T.}~\bibnamefont{Neupert}},
  \bibinfo{author}{\bibfnamefont{C.}~\bibnamefont{Platt}},
  \bibinfo{author}{\bibfnamefont{A.~P.} \bibnamefont{Schnyder}},
  \bibinfo{author}{\bibfnamefont{W.}~\bibnamefont{Hanke}},
  \bibinfo{author}{\bibfnamefont{J.}~\bibnamefont{Goryo}},
  \bibinfo{author}{\bibfnamefont{R.}~\bibnamefont{Thomale}}, \bibnamefont{and}
  \bibinfo{author}{\bibfnamefont{M.}~\bibnamefont{Sigrist}},
  \bibinfo{journal}{Phys. Rev. B} \textbf{\bibinfo{volume}{89}},
  \bibinfo{pages}{020509} (\bibinfo{year}{2014}).

\bibitem[{\citenamefont{Sumiyoshi and Fujimoto}(2014)}]{sumiyoshi2014giant}
\bibinfo{author}{\bibfnamefont{H.}~\bibnamefont{Sumiyoshi}} \bibnamefont{and}
  \bibinfo{author}{\bibfnamefont{S.}~\bibnamefont{Fujimoto}},
  \bibinfo{journal}{Phys. Rev. B} \textbf{\bibinfo{volume}{90}},
  \bibinfo{pages}{184518} (\bibinfo{year}{2014}).

\bibitem[{\citenamefont{Yamashita et~al.}(2015)\citenamefont{Yamashita,
  Shimoyama, Haga, Matsuda, Yamamoto, Onuki, Sumiyoshi, Fujimoto, Levchenko,
  Shibauchi et~al.}}]{yamashita2015colossal}
\bibinfo{author}{\bibfnamefont{T.}~\bibnamefont{Yamashita}},
  \bibinfo{author}{\bibfnamefont{Y.}~\bibnamefont{Shimoyama}},
  \bibinfo{author}{\bibfnamefont{Y.}~\bibnamefont{Haga}},
  \bibinfo{author}{\bibfnamefont{T.}~\bibnamefont{Matsuda}},
  \bibinfo{author}{\bibfnamefont{E.}~\bibnamefont{Yamamoto}},
  \bibinfo{author}{\bibfnamefont{Y.}~\bibnamefont{Onuki}},
  \bibinfo{author}{\bibfnamefont{H.}~\bibnamefont{Sumiyoshi}},
  \bibinfo{author}{\bibfnamefont{S.}~\bibnamefont{Fujimoto}},
  \bibinfo{author}{\bibfnamefont{A.}~\bibnamefont{Levchenko}},
  \bibinfo{author}{\bibfnamefont{T.}~\bibnamefont{Shibauchi}},
  \bibnamefont{et~al.}, \bibinfo{journal}{Nat. Phys.}
  \textbf{\bibinfo{volume}{11}}, \bibinfo{pages}{17} (\bibinfo{year}{2015}).

\bibitem[{\citenamefont{Stewart et~al.}(1984)\citenamefont{Stewart, Fisk,
  Willis, and Smith}}]{stewart1984possibility}
\bibinfo{author}{\bibfnamefont{G.~R.} \bibnamefont{Stewart}},
  \bibinfo{author}{\bibfnamefont{Z.}~\bibnamefont{Fisk}},
  \bibinfo{author}{\bibfnamefont{J.~O.} \bibnamefont{Willis}},
  \bibnamefont{and} \bibinfo{author}{\bibfnamefont{J.~L.} \bibnamefont{Smith}},
  \bibinfo{journal}{Phys. Rev. Lett.} \textbf{\bibinfo{volume}{52}},
  \bibinfo{pages}{679} (\bibinfo{year}{1984}).

\bibitem[{\citenamefont{Adenwalla et~al.}(1990)\citenamefont{Adenwalla, Lin,
  Ran, Zhao, Ketterson, Sauls, Taillefer, Hinks, Levy, and
  Sarma}}]{adenwalla1990phase}
\bibinfo{author}{\bibfnamefont{S.}~\bibnamefont{Adenwalla}},
  \bibinfo{author}{\bibfnamefont{S.}~\bibnamefont{Lin}},
  \bibinfo{author}{\bibfnamefont{Q.}~\bibnamefont{Ran}},
  \bibinfo{author}{\bibfnamefont{Z.}~\bibnamefont{Zhao}},
  \bibinfo{author}{\bibfnamefont{J.~B.} \bibnamefont{Ketterson}},
  \bibinfo{author}{\bibfnamefont{J.~A.} \bibnamefont{Sauls}},
  \bibinfo{author}{\bibfnamefont{L.}~\bibnamefont{Taillefer}},
  \bibinfo{author}{\bibfnamefont{D.}~\bibnamefont{Hinks}},
  \bibinfo{author}{\bibfnamefont{M.}~\bibnamefont{Levy}}, \bibnamefont{and}
  \bibinfo{author}{\bibfnamefont{B.~K.} \bibnamefont{Sarma}},
  \bibinfo{journal}{Phys. Rev. Lett.} \textbf{\bibinfo{volume}{65}},
  \bibinfo{pages}{2298} (\bibinfo{year}{1990}).

\bibitem[{\citenamefont{Hasselbach et~al.}(1989)\citenamefont{Hasselbach,
  Taillefer, and Flouquet}}]{hasselbach1989critical}
\bibinfo{author}{\bibfnamefont{K.}~\bibnamefont{Hasselbach}},
  \bibinfo{author}{\bibfnamefont{L.}~\bibnamefont{Taillefer}},
  \bibnamefont{and} \bibinfo{author}{\bibfnamefont{J.}~\bibnamefont{Flouquet}},
  \bibinfo{journal}{Phys. Rev. Lett.} \textbf{\bibinfo{volume}{63}},
  \bibinfo{pages}{93} (\bibinfo{year}{1989}).

\bibitem[{\citenamefont{Ott et~al.}(1983)\citenamefont{Ott, Rudigier, Fisk, and
  Smith}}]{ott1983u}
\bibinfo{author}{\bibfnamefont{H.}~\bibnamefont{Ott}},
  \bibinfo{author}{\bibfnamefont{H.}~\bibnamefont{Rudigier}},
  \bibinfo{author}{\bibfnamefont{Z.}~\bibnamefont{Fisk}}, \bibnamefont{and}
  \bibinfo{author}{\bibfnamefont{J.}~\bibnamefont{Smith}},
  \bibinfo{journal}{Phys. Rev. Lett.} \textbf{\bibinfo{volume}{50}},
  \bibinfo{pages}{1595} (\bibinfo{year}{1983}).

\bibitem[{\citenamefont{Smith et~al.}(1984)\citenamefont{Smith, Fisk, Willis,
  Batlogg, and Ott}}]{smith1984impurities}
\bibinfo{author}{\bibfnamefont{J.}~\bibnamefont{Smith}},
  \bibinfo{author}{\bibfnamefont{Z.}~\bibnamefont{Fisk}},
  \bibinfo{author}{\bibfnamefont{J.}~\bibnamefont{Willis}},
  \bibinfo{author}{\bibfnamefont{B.}~\bibnamefont{Batlogg}}, \bibnamefont{and}
  \bibinfo{author}{\bibfnamefont{H.}~\bibnamefont{Ott}}, \bibinfo{journal}{J.
  Appl. Phys.} \textbf{\bibinfo{volume}{55}}, \bibinfo{pages}{1996}
  (\bibinfo{year}{1984}).

\bibitem[{\citenamefont{Ott et~al.}(1985)\citenamefont{Ott, Rudigier, Fisk, and
  Smith}}]{ott1985phase}
\bibinfo{author}{\bibfnamefont{H.}~\bibnamefont{Ott}},
  \bibinfo{author}{\bibfnamefont{H.}~\bibnamefont{Rudigier}},
  \bibinfo{author}{\bibfnamefont{Z.}~\bibnamefont{Fisk}}, \bibnamefont{and}
  \bibinfo{author}{\bibfnamefont{J.}~\bibnamefont{Smith}},
  \bibinfo{journal}{Phys. Rev. B} \textbf{\bibinfo{volume}{31}},
  \bibinfo{pages}{1651} (\bibinfo{year}{1985}).

\bibitem[{\citenamefont{Kim et~al.}(1991)\citenamefont{Kim, Andraka, and
  Stewart}}]{kim1991investigation}
\bibinfo{author}{\bibfnamefont{J.}~\bibnamefont{Kim}},
  \bibinfo{author}{\bibfnamefont{B.}~\bibnamefont{Andraka}}, \bibnamefont{and}
  \bibinfo{author}{\bibfnamefont{G.}~\bibnamefont{Stewart}},
  \bibinfo{journal}{Phys. Rev. B} \textbf{\bibinfo{volume}{44}},
  \bibinfo{pages}{6921} (\bibinfo{year}{1991}).

\bibitem[{\citenamefont{Rauchschwalbe et~al.}(1987)\citenamefont{Rauchschwalbe,
  Bredi, Steglich, Maki, and Fulde}}]{rauchschwalbe1987phase}
\bibinfo{author}{\bibfnamefont{U.}~\bibnamefont{Rauchschwalbe}},
  \bibinfo{author}{\bibfnamefont{C.}~\bibnamefont{Bredi}},
  \bibinfo{author}{\bibfnamefont{F.}~\bibnamefont{Steglich}},
  \bibinfo{author}{\bibfnamefont{K.}~\bibnamefont{Maki}}, \bibnamefont{and}
  \bibinfo{author}{\bibfnamefont{P.}~\bibnamefont{Fulde}},
  \bibinfo{journal}{Europhys. Lett.} \textbf{\bibinfo{volume}{3}},
  \bibinfo{pages}{757} (\bibinfo{year}{1987}).

\bibitem[{\citenamefont{Sauls}(1994)}]{sauls1994order}
\bibinfo{author}{\bibfnamefont{J.}~\bibnamefont{Sauls}}, \bibinfo{journal}{Adv.
  Phys.} \textbf{\bibinfo{volume}{43}}, \bibinfo{pages}{113}
  (\bibinfo{year}{1994}).

\bibitem[{\citenamefont{Tsutsumi et~al.}(2012)\citenamefont{Tsutsumi, Machida,
  Ohmi, and Ozaki}}]{tsutsumi2012spin}
\bibinfo{author}{\bibfnamefont{Y.}~\bibnamefont{Tsutsumi}},
  \bibinfo{author}{\bibfnamefont{K.}~\bibnamefont{Machida}},
  \bibinfo{author}{\bibfnamefont{T.}~\bibnamefont{Ohmi}}, \bibnamefont{and}
  \bibinfo{author}{\bibfnamefont{M.-a.} \bibnamefont{Ozaki}},
  \bibinfo{journal}{J. Phys. Soc. Jpn.} \textbf{\bibinfo{volume}{81}},
  \bibinfo{pages}{074717} (\bibinfo{year}{2012}).

\bibitem[{\citenamefont{Izawa et~al.}(2014)\citenamefont{Izawa, Machida, Itoh,
  So, Ota, Haga, Yamamoto, Kimura, Onuki, Tsutsumi et~al.}}]{izawa2014pairing}
\bibinfo{author}{\bibfnamefont{K.}~\bibnamefont{Izawa}},
  \bibinfo{author}{\bibfnamefont{Y.}~\bibnamefont{Machida}},
  \bibinfo{author}{\bibfnamefont{A.}~\bibnamefont{Itoh}},
  \bibinfo{author}{\bibfnamefont{Y.}~\bibnamefont{So}},
  \bibinfo{author}{\bibfnamefont{K.}~\bibnamefont{Ota}},
  \bibinfo{author}{\bibfnamefont{Y.}~\bibnamefont{Haga}},
  \bibinfo{author}{\bibfnamefont{E.}~\bibnamefont{Yamamoto}},
  \bibinfo{author}{\bibfnamefont{N.}~\bibnamefont{Kimura}},
  \bibinfo{author}{\bibfnamefont{Y.}~\bibnamefont{Onuki}},
  \bibinfo{author}{\bibfnamefont{Y.}~\bibnamefont{Tsutsumi}},
  \bibnamefont{et~al.}, \bibinfo{journal}{J. Phys. Soc. Jpn.}
  \textbf{\bibinfo{volume}{83}}, \bibinfo{pages}{061013}
  (\bibinfo{year}{2014}).

\bibitem[{\citenamefont{Avers et~al.}(2020)\citenamefont{Avers, Gannon, Kuhn,
  Halperin, Sauls, DeBeer-Schmitt, Dewhurst, Gavilano, Nagy, Gasser
  et~al.}}]{avers}
\bibinfo{author}{\bibfnamefont{K.~E.} \bibnamefont{Avers}},
  \bibinfo{author}{\bibfnamefont{W.~J.} \bibnamefont{Gannon}},
  \bibinfo{author}{\bibfnamefont{S.~J.} \bibnamefont{Kuhn}},
  \bibinfo{author}{\bibfnamefont{W.~P.} \bibnamefont{Halperin}},
  \bibinfo{author}{\bibfnamefont{J.~A.} \bibnamefont{Sauls}},
  \bibinfo{author}{\bibfnamefont{L.}~\bibnamefont{DeBeer-Schmitt}},
  \bibinfo{author}{\bibfnamefont{C.~D.} \bibnamefont{Dewhurst}},
  \bibinfo{author}{\bibfnamefont{J.}~\bibnamefont{Gavilano}},
  \bibinfo{author}{\bibfnamefont{G.}~\bibnamefont{Nagy}},
  \bibinfo{author}{\bibfnamefont{U.}~\bibnamefont{Gasser}},
  \bibnamefont{et~al.}, \bibinfo{journal}{Nat. Phys.}
  \textbf{\bibinfo{volume}{16}}, \bibinfo{pages}{531} (\bibinfo{year}{2020}).

\bibitem[{\citenamefont{Shimizu et~al.}(2017)\citenamefont{Shimizu, Kittaka,
  Nakamura, Sakakibara, Aoki, Homma, Nakamura, and Machida}}]{shimizu17}
\bibinfo{author}{\bibfnamefont{Y.}~\bibnamefont{Shimizu}},
  \bibinfo{author}{\bibfnamefont{S.}~\bibnamefont{Kittaka}},
  \bibinfo{author}{\bibfnamefont{S.}~\bibnamefont{Nakamura}},
  \bibinfo{author}{\bibfnamefont{T.}~\bibnamefont{Sakakibara}},
  \bibinfo{author}{\bibfnamefont{D.}~\bibnamefont{Aoki}},
  \bibinfo{author}{\bibfnamefont{Y.}~\bibnamefont{Homma}},
  \bibinfo{author}{\bibfnamefont{A.}~\bibnamefont{Nakamura}}, \bibnamefont{and}
  \bibinfo{author}{\bibfnamefont{K.}~\bibnamefont{Machida}},
  \bibinfo{journal}{Phys. Rev. B} \textbf{\bibinfo{volume}{96}},
  \bibinfo{pages}{100505} (\bibinfo{year}{2017}).

\bibitem[{\citenamefont{Machida}(2018{\natexlab{b}})}]{machidaJPSJ18}
\bibinfo{author}{\bibfnamefont{K.}~\bibnamefont{Machida}}, \bibinfo{journal}{J.
  Phys. Soc. Jpn.} \textbf{\bibinfo{volume}{87}}, \bibinfo{pages}{033703}
  (\bibinfo{year}{2018}{\natexlab{b}}).

\bibitem[{\citenamefont{Mizushima and Nitta}(2018)}]{mizushima2018topology}
\bibinfo{author}{\bibfnamefont{T.}~\bibnamefont{Mizushima}} \bibnamefont{and}
  \bibinfo{author}{\bibfnamefont{M.}~\bibnamefont{Nitta}},
  \bibinfo{journal}{Phys. Rev. B} \textbf{\bibinfo{volume}{97}},
  \bibinfo{pages}{024506} (\bibinfo{year}{2018}).

\bibitem[{\citenamefont{Aoki et~al.}(2019{\natexlab{a}})\citenamefont{Aoki,
  Ishida, and Flouquet}}]{aoki2019review}
\bibinfo{author}{\bibfnamefont{D.}~\bibnamefont{Aoki}},
  \bibinfo{author}{\bibfnamefont{K.}~\bibnamefont{Ishida}}, \bibnamefont{and}
  \bibinfo{author}{\bibfnamefont{J.}~\bibnamefont{Flouquet}},
  \bibinfo{journal}{J. Phys. Soc. Jpn.} \textbf{\bibinfo{volume}{88}},
  \bibinfo{pages}{022001} (\bibinfo{year}{2019}{\natexlab{a}}).

\bibitem[{\citenamefont{Huy et~al.}(2007)\citenamefont{Huy, Gasparini, de~Nijs,
  Huang, Klaasse, Gortenmulder, de~Visser, Hamann, G\"orlach, and
  L\"ohneysen}}]{PhysRevLett.99.067006}
\bibinfo{author}{\bibfnamefont{N.~T.} \bibnamefont{Huy}},
  \bibinfo{author}{\bibfnamefont{A.}~\bibnamefont{Gasparini}},
  \bibinfo{author}{\bibfnamefont{D.~E.} \bibnamefont{de~Nijs}},
  \bibinfo{author}{\bibfnamefont{Y.}~\bibnamefont{Huang}},
  \bibinfo{author}{\bibfnamefont{J.~C.~P.} \bibnamefont{Klaasse}},
  \bibinfo{author}{\bibfnamefont{T.}~\bibnamefont{Gortenmulder}},
  \bibinfo{author}{\bibfnamefont{A.}~\bibnamefont{de~Visser}},
  \bibinfo{author}{\bibfnamefont{A.}~\bibnamefont{Hamann}},
  \bibinfo{author}{\bibfnamefont{T.}~\bibnamefont{G\"orlach}},
  \bibnamefont{and} \bibinfo{author}{\bibfnamefont{H.~v.}
  \bibnamefont{L\"ohneysen}}, \bibinfo{journal}{Phys. Rev. Lett.}
  \textbf{\bibinfo{volume}{99}}, \bibinfo{pages}{067006}
  (\bibinfo{year}{2007}).

\bibitem[{\citenamefont{Mineev}(2017)}]{mineev2017phase}
\bibinfo{author}{\bibfnamefont{V.}~\bibnamefont{Mineev}},
  \bibinfo{journal}{Phys. Rev. B} \textbf{\bibinfo{volume}{95}},
  \bibinfo{pages}{104501} (\bibinfo{year}{2017}).

\bibitem[{\citenamefont{Aoki et~al.}(2001)\citenamefont{Aoki, Huxley,
  Ressouche, Braithwaite, Flouquet, Brison, Lhotel, and
  Paulsen}}]{aoki2001coexistence}
\bibinfo{author}{\bibfnamefont{D.}~\bibnamefont{Aoki}},
  \bibinfo{author}{\bibfnamefont{A.}~\bibnamefont{Huxley}},
  \bibinfo{author}{\bibfnamefont{E.}~\bibnamefont{Ressouche}},
  \bibinfo{author}{\bibfnamefont{D.}~\bibnamefont{Braithwaite}},
  \bibinfo{author}{\bibfnamefont{J.}~\bibnamefont{Flouquet}},
  \bibinfo{author}{\bibfnamefont{J.-P.} \bibnamefont{Brison}},
  \bibinfo{author}{\bibfnamefont{E.}~\bibnamefont{Lhotel}}, \bibnamefont{and}
  \bibinfo{author}{\bibfnamefont{C.}~\bibnamefont{Paulsen}},
  \bibinfo{journal}{Nature} \textbf{\bibinfo{volume}{413}},
  \bibinfo{pages}{613} (\bibinfo{year}{2001}).

\bibitem[{\citenamefont{Saxena et~al.}(2000)\citenamefont{Saxena, Agarwal,
  Ahilan, Grosche, Haselwimmer, Steiner, Pugh, Walker, Julian, Monthoux
  et~al.}}]{saxena2000superconductivity}
\bibinfo{author}{\bibfnamefont{S.}~\bibnamefont{Saxena}},
  \bibinfo{author}{\bibfnamefont{P.}~\bibnamefont{Agarwal}},
  \bibinfo{author}{\bibfnamefont{K.}~\bibnamefont{Ahilan}},
  \bibinfo{author}{\bibfnamefont{F.}~\bibnamefont{Grosche}},
  \bibinfo{author}{\bibfnamefont{R.}~\bibnamefont{Haselwimmer}},
  \bibinfo{author}{\bibfnamefont{M.}~\bibnamefont{Steiner}},
  \bibinfo{author}{\bibfnamefont{E.}~\bibnamefont{Pugh}},
  \bibinfo{author}{\bibfnamefont{I.}~\bibnamefont{Walker}},
  \bibinfo{author}{\bibfnamefont{S.}~\bibnamefont{Julian}},
  \bibinfo{author}{\bibfnamefont{P.}~\bibnamefont{Monthoux}},
  \bibnamefont{et~al.}, \bibinfo{journal}{Nature}
  \textbf{\bibinfo{volume}{406}}, \bibinfo{pages}{587} (\bibinfo{year}{2000}).

\bibitem[{\citenamefont{Mineev}(2002)}]{PhysRevB.66.134504}
\bibinfo{author}{\bibfnamefont{V.~P.} \bibnamefont{Mineev}},
  \bibinfo{journal}{Phys. Rev. B} \textbf{\bibinfo{volume}{66}},
  \bibinfo{pages}{134504} (\bibinfo{year}{2002}).

\bibitem[{\citenamefont{Hattori et~al.}(2012)\citenamefont{Hattori, Ihara,
  Nakai, Ishida, Tada, Fujimoto, Kawakami, Osaki, Deguchi, Sato
  et~al.}}]{PhysRevLett.108.066403}
\bibinfo{author}{\bibfnamefont{T.}~\bibnamefont{Hattori}},
  \bibinfo{author}{\bibfnamefont{Y.}~\bibnamefont{Ihara}},
  \bibinfo{author}{\bibfnamefont{Y.}~\bibnamefont{Nakai}},
  \bibinfo{author}{\bibfnamefont{K.}~\bibnamefont{Ishida}},
  \bibinfo{author}{\bibfnamefont{Y.}~\bibnamefont{Tada}},
  \bibinfo{author}{\bibfnamefont{S.}~\bibnamefont{Fujimoto}},
  \bibinfo{author}{\bibfnamefont{N.}~\bibnamefont{Kawakami}},
  \bibinfo{author}{\bibfnamefont{E.}~\bibnamefont{Osaki}},
  \bibinfo{author}{\bibfnamefont{K.}~\bibnamefont{Deguchi}},
  \bibinfo{author}{\bibfnamefont{N.~K.} \bibnamefont{Sato}},
  \bibnamefont{et~al.}, \bibinfo{journal}{Phys. Rev. Lett.}
  \textbf{\bibinfo{volume}{108}}, \bibinfo{pages}{066403}
  (\bibinfo{year}{2012}).

\bibitem[{\citenamefont{Tada et~al.}(2013)\citenamefont{Tada, Fujimoto,
  Kawakami, Hattori, Ihara, Ishida, Deguchi, Sato, and Satoh}}]{tada2013spin}
\bibinfo{author}{\bibfnamefont{Y.}~\bibnamefont{Tada}},
  \bibinfo{author}{\bibfnamefont{S.}~\bibnamefont{Fujimoto}},
  \bibinfo{author}{\bibfnamefont{N.}~\bibnamefont{Kawakami}},
  \bibinfo{author}{\bibfnamefont{T.}~\bibnamefont{Hattori}},
  \bibinfo{author}{\bibfnamefont{Y.}~\bibnamefont{Ihara}},
  \bibinfo{author}{\bibfnamefont{K.}~\bibnamefont{Ishida}},
  \bibinfo{author}{\bibfnamefont{K.}~\bibnamefont{Deguchi}},
  \bibinfo{author}{\bibfnamefont{N.}~\bibnamefont{Sato}}, \bibnamefont{and}
  \bibinfo{author}{\bibfnamefont{I.}~\bibnamefont{Satoh}}, \bibinfo{journal}{J.
  Phys.: Conf. Ser.} \textbf{\bibinfo{volume}{449}}, \bibinfo{pages}{012029}
  (\bibinfo{year}{2013}).

\bibitem[{\citenamefont{Aoki et~al.}(2019{\natexlab{b}})\citenamefont{Aoki,
  Nakamura, Honda, Li, Homma, Shimizu, Sato, Knebel, Brison, Pourret
  et~al.}}]{Aoki_UTe2}
\bibinfo{author}{\bibfnamefont{D.}~\bibnamefont{Aoki}},
  \bibinfo{author}{\bibfnamefont{A.}~\bibnamefont{Nakamura}},
  \bibinfo{author}{\bibfnamefont{F.}~\bibnamefont{Honda}},
  \bibinfo{author}{\bibfnamefont{D.}~\bibnamefont{Li}},
  \bibinfo{author}{\bibfnamefont{Y.}~\bibnamefont{Homma}},
  \bibinfo{author}{\bibfnamefont{Y.}~\bibnamefont{Shimizu}},
  \bibinfo{author}{\bibfnamefont{Y.~J.} \bibnamefont{Sato}},
  \bibinfo{author}{\bibfnamefont{G.}~\bibnamefont{Knebel}},
  \bibinfo{author}{\bibfnamefont{J.-P.} \bibnamefont{Brison}},
  \bibinfo{author}{\bibfnamefont{A.}~\bibnamefont{Pourret}},
  \bibnamefont{et~al.}, \bibinfo{journal}{J. Phys. Soc. Jpn.}
  \textbf{\bibinfo{volume}{88}}, \bibinfo{pages}{043702}
  (\bibinfo{year}{2019}{\natexlab{b}}).

\bibitem[{\citenamefont{Ran et~al.}(2019)\citenamefont{Ran, Eckberg, Ding,
  Furukawa, Metz, Saha, Liu, Zic, Kim, Paglione et~al.}}]{ran2019nearly}
\bibinfo{author}{\bibfnamefont{S.}~\bibnamefont{Ran}},
  \bibinfo{author}{\bibfnamefont{C.}~\bibnamefont{Eckberg}},
  \bibinfo{author}{\bibfnamefont{Q.-P.} \bibnamefont{Ding}},
  \bibinfo{author}{\bibfnamefont{Y.}~\bibnamefont{Furukawa}},
  \bibinfo{author}{\bibfnamefont{T.}~\bibnamefont{Metz}},
  \bibinfo{author}{\bibfnamefont{S.~R.} \bibnamefont{Saha}},
  \bibinfo{author}{\bibfnamefont{I.-L.} \bibnamefont{Liu}},
  \bibinfo{author}{\bibfnamefont{M.}~\bibnamefont{Zic}},
  \bibinfo{author}{\bibfnamefont{H.}~\bibnamefont{Kim}},
  \bibinfo{author}{\bibfnamefont{J.}~\bibnamefont{Paglione}},
  \bibnamefont{et~al.}, \bibinfo{journal}{Science}
  \textbf{\bibinfo{volume}{365}}, \bibinfo{pages}{684} (\bibinfo{year}{2019}).

\bibitem[{\citenamefont{Ishihara et~al.}(2021)\citenamefont{Ishihara, Roppongi,
  Kobayashi, Mizukami, Sakai, Haga, Hashimoto, and
  Shibauchi}}]{ishihara2021chiral}
\bibinfo{author}{\bibfnamefont{K.}~\bibnamefont{Ishihara}},
  \bibinfo{author}{\bibfnamefont{M.}~\bibnamefont{Roppongi}},
  \bibinfo{author}{\bibfnamefont{M.}~\bibnamefont{Kobayashi}},
  \bibinfo{author}{\bibfnamefont{Y.}~\bibnamefont{Mizukami}},
  \bibinfo{author}{\bibfnamefont{H.}~\bibnamefont{Sakai}},
  \bibinfo{author}{\bibfnamefont{Y.}~\bibnamefont{Haga}},
  \bibinfo{author}{\bibfnamefont{K.}~\bibnamefont{Hashimoto}},
  \bibnamefont{and}
  \bibinfo{author}{\bibfnamefont{T.}~\bibnamefont{Shibauchi}},
  \bibinfo{journal}{arXiv:2105.13721}  (\bibinfo{year}{2021}).

\bibitem[{\citenamefont{Hayes et~al.}(2020)\citenamefont{Hayes, Wei, Metz,
  Zhang, Eo, Ran, Saha, Collini, Butch, Agterberg et~al.}}]{hayes2020weyl}
\bibinfo{author}{\bibfnamefont{I.~M.} \bibnamefont{Hayes}},
  \bibinfo{author}{\bibfnamefont{D.~S.} \bibnamefont{Wei}},
  \bibinfo{author}{\bibfnamefont{T.}~\bibnamefont{Metz}},
  \bibinfo{author}{\bibfnamefont{J.}~\bibnamefont{Zhang}},
  \bibinfo{author}{\bibfnamefont{Y.~S.} \bibnamefont{Eo}},
  \bibinfo{author}{\bibfnamefont{S.}~\bibnamefont{Ran}},
  \bibinfo{author}{\bibfnamefont{S.~R.} \bibnamefont{Saha}},
  \bibinfo{author}{\bibfnamefont{J.}~\bibnamefont{Collini}},
  \bibinfo{author}{\bibfnamefont{N.~P.} \bibnamefont{Butch}},
  \bibinfo{author}{\bibfnamefont{D.~F.} \bibnamefont{Agterberg}},
  \bibnamefont{et~al.}, \bibinfo{journal}{arXiv:2002.02539}
  (\bibinfo{year}{2020}).

\bibitem[{\citenamefont{Knebel et~al.}(2019)\citenamefont{Knebel, Knafo,
  Pourret, Niu, Vali{\v s}ka, Braithwaite, Lapertot, Nardone, Zitouni, Mishra
  et~al.}}]{knebel_Ute2}
\bibinfo{author}{\bibfnamefont{G.}~\bibnamefont{Knebel}},
  \bibinfo{author}{\bibfnamefont{W.}~\bibnamefont{Knafo}},
  \bibinfo{author}{\bibfnamefont{A.}~\bibnamefont{Pourret}},
  \bibinfo{author}{\bibfnamefont{Q.}~\bibnamefont{Niu}},
  \bibinfo{author}{\bibfnamefont{M.}~\bibnamefont{Vali{\v s}ka}},
  \bibinfo{author}{\bibfnamefont{D.}~\bibnamefont{Braithwaite}},
  \bibinfo{author}{\bibfnamefont{G.}~\bibnamefont{Lapertot}},
  \bibinfo{author}{\bibfnamefont{M.}~\bibnamefont{Nardone}},
  \bibinfo{author}{\bibfnamefont{A.}~\bibnamefont{Zitouni}},
  \bibinfo{author}{\bibfnamefont{S.}~\bibnamefont{Mishra}},
  \bibnamefont{et~al.}, \bibinfo{journal}{J. Phys. Soc. Jpn.}
  \textbf{\bibinfo{volume}{88}}, \bibinfo{pages}{063707}
  (\bibinfo{year}{2019}).

\bibitem[{\citenamefont{Miyake et~al.}(2019)\citenamefont{Miyake, Shimizu,
  Sato, Li, Nakamura, Homma, Honda, Flouquet, Tokunaga, and
  Aoki}}]{miyake2019metamagnetic}
\bibinfo{author}{\bibfnamefont{A.}~\bibnamefont{Miyake}},
  \bibinfo{author}{\bibfnamefont{Y.}~\bibnamefont{Shimizu}},
  \bibinfo{author}{\bibfnamefont{Y.~J.} \bibnamefont{Sato}},
  \bibinfo{author}{\bibfnamefont{D.}~\bibnamefont{Li}},
  \bibinfo{author}{\bibfnamefont{A.}~\bibnamefont{Nakamura}},
  \bibinfo{author}{\bibfnamefont{Y.}~\bibnamefont{Homma}},
  \bibinfo{author}{\bibfnamefont{F.}~\bibnamefont{Honda}},
  \bibinfo{author}{\bibfnamefont{J.}~\bibnamefont{Flouquet}},
  \bibinfo{author}{\bibfnamefont{M.}~\bibnamefont{Tokunaga}}, \bibnamefont{and}
  \bibinfo{author}{\bibfnamefont{D.}~\bibnamefont{Aoki}}, \bibinfo{journal}{J.
  Phys. Soc. Jpn.} \textbf{\bibinfo{volume}{88}}, \bibinfo{pages}{063706}
  (\bibinfo{year}{2019}).

\bibitem[{\citenamefont{Sumiyoshi and Fujimoto}(2013)}]{sumiyoshi_ATHE}
\bibinfo{author}{\bibfnamefont{H.}~\bibnamefont{Sumiyoshi}} \bibnamefont{and}
  \bibinfo{author}{\bibfnamefont{S.}~\bibnamefont{Fujimoto}},
  \bibinfo{journal}{J. Phys. Soc. Jpn.} \textbf{\bibinfo{volume}{82}},
  \bibinfo{pages}{023602} (\bibinfo{year}{2013}).

\bibitem[{\citenamefont{Nomura et~al.}(2012)\citenamefont{Nomura, Ryu,
  Furusaki, and Nagaosa}}]{nomura2012cross}
\bibinfo{author}{\bibfnamefont{K.}~\bibnamefont{Nomura}},
  \bibinfo{author}{\bibfnamefont{S.}~\bibnamefont{Ryu}},
  \bibinfo{author}{\bibfnamefont{A.}~\bibnamefont{Furusaki}}, \bibnamefont{and}
  \bibinfo{author}{\bibfnamefont{N.}~\bibnamefont{Nagaosa}},
  \bibinfo{journal}{Phys. Rev. Lett.} \textbf{\bibinfo{volume}{108}},
  \bibinfo{pages}{026802} (\bibinfo{year}{2012}).

\bibitem[{\citenamefont{Arfi et~al.}(1988)\citenamefont{Arfi, Bahlouli,
  Pethick, and Pines}}]{Arfi1988}
\bibinfo{author}{\bibfnamefont{B.}~\bibnamefont{Arfi}},
  \bibinfo{author}{\bibfnamefont{H.}~\bibnamefont{Bahlouli}},
  \bibinfo{author}{\bibfnamefont{C.~J.} \bibnamefont{Pethick}},
  \bibnamefont{and} \bibinfo{author}{\bibfnamefont{D.}~\bibnamefont{Pines}},
  \bibinfo{journal}{Phys. Rev. Lett.} \textbf{\bibinfo{volume}{60}},
  \bibinfo{pages}{2206} (\bibinfo{year}{1988}).

\bibitem[{\citenamefont{Ngampruetikorn and
  Sauls}(2020)}]{ngampruetikorn2020impurity}
\bibinfo{author}{\bibfnamefont{V.}~\bibnamefont{Ngampruetikorn}}
  \bibnamefont{and} \bibinfo{author}{\bibfnamefont{J.}~\bibnamefont{Sauls}},
  \bibinfo{journal}{Phys. Rev. Lett.} \textbf{\bibinfo{volume}{124}},
  \bibinfo{pages}{157002} (\bibinfo{year}{2020}).

\bibitem[{\citenamefont{Yip}(2016)}]{yip2016low}
\bibinfo{author}{\bibfnamefont{S.}~\bibnamefont{Yip}},
  \bibinfo{journal}{Supercond. Sci. Tech.} \textbf{\bibinfo{volume}{29}},
  \bibinfo{pages}{085006} (\bibinfo{year}{2016}).

\bibitem[{\citenamefont{Y{\i}lmaz and Yip}(2020)}]{yilmaz2020spontaneous}
\bibinfo{author}{\bibfnamefont{F.}~\bibnamefont{Y{\i}lmaz}} \bibnamefont{and}
  \bibinfo{author}{\bibfnamefont{S.}~\bibnamefont{Yip}},
  \bibinfo{journal}{Phys. Rev. Research} \textbf{\bibinfo{volume}{2}},
  \bibinfo{pages}{023223} (\bibinfo{year}{2020}).

\bibitem[{\citenamefont{Ussishkin et~al.}(2002)\citenamefont{Ussishkin, Sondhi,
  and Huse}}]{ussishkin2002gaussian}
\bibinfo{author}{\bibfnamefont{I.}~\bibnamefont{Ussishkin}},
  \bibinfo{author}{\bibfnamefont{S.~L.} \bibnamefont{Sondhi}},
  \bibnamefont{and} \bibinfo{author}{\bibfnamefont{D.~A.} \bibnamefont{Huse}},
  \bibinfo{journal}{Phys. Rev. Lett.} \textbf{\bibinfo{volume}{89}},
  \bibinfo{pages}{287001} (\bibinfo{year}{2002}).

\bibitem[{\citenamefont{Parmenter}(1953)}]{parmenter1953acousto}
\bibinfo{author}{\bibfnamefont{R.}~\bibnamefont{Parmenter}},
  \bibinfo{journal}{Phys. Rev.} \textbf{\bibinfo{volume}{89}},
  \bibinfo{pages}{990} (\bibinfo{year}{1953}).

\bibitem[{\citenamefont{Weinreich and White}(1957)}]{weinreich1957observation}
\bibinfo{author}{\bibfnamefont{G.}~\bibnamefont{Weinreich}} \bibnamefont{and}
  \bibinfo{author}{\bibfnamefont{H.~G.} \bibnamefont{White}},
  \bibinfo{journal}{Phys. Rev.} \textbf{\bibinfo{volume}{106}},
  \bibinfo{pages}{1104} (\bibinfo{year}{1957}).

\bibitem[{\citenamefont{Weinreich et~al.}(1959)\citenamefont{Weinreich,
  Sanders~Jr, and White}}]{weinreich1959acoustoelectric}
\bibinfo{author}{\bibfnamefont{G.}~\bibnamefont{Weinreich}},
  \bibinfo{author}{\bibfnamefont{T.}~\bibnamefont{Sanders~Jr}},
  \bibnamefont{and} \bibinfo{author}{\bibfnamefont{H.~G.} \bibnamefont{White}},
  \bibinfo{journal}{Phys. Rev.} \textbf{\bibinfo{volume}{114}},
  \bibinfo{pages}{33} (\bibinfo{year}{1959}).

\bibitem[{\citenamefont{Kalameitsev et~al.}(2019)\citenamefont{Kalameitsev,
  Kovalev, and Savenko}}]{kalameitsev2019valley}
\bibinfo{author}{\bibfnamefont{A.}~\bibnamefont{Kalameitsev}},
  \bibinfo{author}{\bibfnamefont{V.}~\bibnamefont{Kovalev}}, \bibnamefont{and}
  \bibinfo{author}{\bibfnamefont{I.}~\bibnamefont{Savenko}},
  \bibinfo{journal}{Phys. Rev. Lett.} \textbf{\bibinfo{volume}{122}},
  \bibinfo{pages}{256801} (\bibinfo{year}{2019}).

\bibitem[{\citenamefont{Sukhachov and
  Rostami}(2020)}]{sukhachov2020acoustogalvanic}
\bibinfo{author}{\bibfnamefont{P.~O.} \bibnamefont{Sukhachov}}
  \bibnamefont{and} \bibinfo{author}{\bibfnamefont{H.}~\bibnamefont{Rostami}},
  \bibinfo{journal}{Phys. Rev. Lett.} \textbf{\bibinfo{volume}{124}},
  \bibinfo{pages}{126602} (\bibinfo{year}{2020}).

\bibitem[{\citenamefont{Vollhardt and Woelfle}(1990)}]{vollhardt1990superfluid}
\bibinfo{author}{\bibfnamefont{D.}~\bibnamefont{Vollhardt}} \bibnamefont{and}
  \bibinfo{author}{\bibfnamefont{P.}~\bibnamefont{Woelfle}},
  \emph{\bibinfo{title}{The Superfluid Phases Of Helium 3}}
  (\bibinfo{publisher}{Taylor \& Francis}, \bibinfo{year}{1990}), ISBN
  \bibinfo{isbn}{9780850664126}.

\bibitem[{\citenamefont{Hirschfeld et~al.}(1992)\citenamefont{Hirschfeld,
  Putikka, and W\"olfle}}]{PJH:1992}
\bibinfo{author}{\bibfnamefont{P.~J.} \bibnamefont{Hirschfeld}},
  \bibinfo{author}{\bibfnamefont{W.~O.} \bibnamefont{Putikka}},
  \bibnamefont{and} \bibinfo{author}{\bibfnamefont{P.}~\bibnamefont{W\"olfle}},
  \bibinfo{journal}{Phys. Rev. Lett.} \textbf{\bibinfo{volume}{69}},
  \bibinfo{pages}{1447} (\bibinfo{year}{1992}).

\bibitem[{\citenamefont{Tewordt}(1999)}]{tewordt}
\bibinfo{author}{\bibfnamefont{L.}~\bibnamefont{Tewordt}},
  \bibinfo{journal}{Phys. Rev. Lett.} \textbf{\bibinfo{volume}{83}},
  \bibinfo{pages}{1007} (\bibinfo{year}{1999}).

\bibitem[{\citenamefont{Higashitani and Nagai}(2000)}]{higashitani}
\bibinfo{author}{\bibfnamefont{S.}~\bibnamefont{Higashitani}} \bibnamefont{and}
  \bibinfo{author}{\bibfnamefont{K.}~\bibnamefont{Nagai}},
  \bibinfo{journal}{Phys. Rev. B} \textbf{\bibinfo{volume}{62}},
  \bibinfo{pages}{3042} (\bibinfo{year}{2000}).

\bibitem[{\citenamefont{Balatsky
  et~al.}(2000{\natexlab{a}})\citenamefont{Balatsky, Kumar, and
  Schrieffer}}]{balatsky00}
\bibinfo{author}{\bibfnamefont{A.}~\bibnamefont{Balatsky}},
  \bibinfo{author}{\bibfnamefont{P.}~\bibnamefont{Kumar}}, \bibnamefont{and}
  \bibinfo{author}{\bibfnamefont{J.}~\bibnamefont{Schrieffer}},
  \bibinfo{journal}{Physica C} \textbf{\bibinfo{volume}{341-348}},
  \bibinfo{pages}{807} (\bibinfo{year}{2000}{\natexlab{a}}).

\bibitem[{\citenamefont{Balatsky
  et~al.}(2000{\natexlab{b}})\citenamefont{Balatsky, Kumar, and
  Schrieffer}}]{balatskyPRL00}
\bibinfo{author}{\bibfnamefont{A.~V.} \bibnamefont{Balatsky}},
  \bibinfo{author}{\bibfnamefont{P.}~\bibnamefont{Kumar}}, \bibnamefont{and}
  \bibinfo{author}{\bibfnamefont{J.~R.} \bibnamefont{Schrieffer}},
  \bibinfo{journal}{Phys. Rev. Lett.} \textbf{\bibinfo{volume}{84}},
  \bibinfo{pages}{4445} (\bibinfo{year}{2000}{\natexlab{b}}).

\bibitem[{\citenamefont{Kee et~al.}(2003)\citenamefont{Kee, Maki, and
  Chung}}]{kee}
\bibinfo{author}{\bibfnamefont{H.-Y.} \bibnamefont{Kee}},
  \bibinfo{author}{\bibfnamefont{K.}~\bibnamefont{Maki}}, \bibnamefont{and}
  \bibinfo{author}{\bibfnamefont{C.~H.} \bibnamefont{Chung}},
  \bibinfo{journal}{Phys. Rev. B} \textbf{\bibinfo{volume}{67}},
  \bibinfo{pages}{180504} (\bibinfo{year}{2003}).

\bibitem[{\citenamefont{Miura et~al.}(2007)\citenamefont{Miura, Higashitani,
  and Nagai}}]{miura}
\bibinfo{author}{\bibfnamefont{M.}~\bibnamefont{Miura}},
  \bibinfo{author}{\bibfnamefont{S.}~\bibnamefont{Higashitani}},
  \bibnamefont{and} \bibinfo{author}{\bibfnamefont{K.}~\bibnamefont{Nagai}},
  \bibinfo{journal}{J. Phys. Soc. Jpn.} \textbf{\bibinfo{volume}{76}},
  \bibinfo{pages}{034710} (\bibinfo{year}{2007}).

\bibitem[{\citenamefont{Sauls et~al.}(2015)\citenamefont{Sauls, Wu, and
  Chung}}]{sauls2015anisotropy}
\bibinfo{author}{\bibfnamefont{J.~A.} \bibnamefont{Sauls}},
  \bibinfo{author}{\bibfnamefont{H.}~\bibnamefont{Wu}}, \bibnamefont{and}
  \bibinfo{author}{\bibfnamefont{S.~B.} \bibnamefont{Chung}},
  \bibinfo{journal}{Front. Phys.} \textbf{\bibinfo{volume}{3}},
  \bibinfo{pages}{36} (\bibinfo{year}{2015}).

\bibitem[{\citenamefont{Kee et~al.}(2000)\citenamefont{Kee, Kim, and
  Maki}}]{Kee2000}
\bibinfo{author}{\bibfnamefont{H.-Y.} \bibnamefont{Kee}},
  \bibinfo{author}{\bibfnamefont{Y.~B.} \bibnamefont{Kim}}, \bibnamefont{and}
  \bibinfo{author}{\bibfnamefont{K.}~\bibnamefont{Maki}},
  \bibinfo{journal}{Phys. Rev. B} \textbf{\bibinfo{volume}{62}},
  \bibinfo{pages}{5877} (\bibinfo{year}{2000}).

\bibitem[{\citenamefont{Serene and Rainer}(1983)}]{SERENE1983221}
\bibinfo{author}{\bibfnamefont{J.}~\bibnamefont{Serene}} \bibnamefont{and}
  \bibinfo{author}{\bibfnamefont{D.}~\bibnamefont{Rainer}},
  \bibinfo{journal}{Phys. Rep.} \textbf{\bibinfo{volume}{101}},
  \bibinfo{pages}{221} (\bibinfo{year}{1983}).

\bibitem[{\citenamefont{Xiao et~al.}(2010)\citenamefont{Xiao, Chang, and
  Niu}}]{RevModPhys.82.1959}
\bibinfo{author}{\bibfnamefont{D.}~\bibnamefont{Xiao}},
  \bibinfo{author}{\bibfnamefont{M.-C.} \bibnamefont{Chang}}, \bibnamefont{and}
  \bibinfo{author}{\bibfnamefont{Q.}~\bibnamefont{Niu}}, \bibinfo{journal}{Rev.
  Mod. Phys.} \textbf{\bibinfo{volume}{82}}, \bibinfo{pages}{1959}
  (\bibinfo{year}{2010}).

\bibitem[{\citenamefont{Mizushima et~al.}(2016)\citenamefont{Mizushima,
  Tsutsumi, Kawakami, Sato, Ichioka, and Machida}}]{mizushimaJPSJ16}
\bibinfo{author}{\bibfnamefont{T.}~\bibnamefont{Mizushima}},
  \bibinfo{author}{\bibfnamefont{Y.}~\bibnamefont{Tsutsumi}},
  \bibinfo{author}{\bibfnamefont{T.}~\bibnamefont{Kawakami}},
  \bibinfo{author}{\bibfnamefont{M.}~\bibnamefont{Sato}},
  \bibinfo{author}{\bibfnamefont{M.}~\bibnamefont{Ichioka}}, \bibnamefont{and}
  \bibinfo{author}{\bibfnamefont{K.}~\bibnamefont{Machida}},
  \bibinfo{journal}{J. Phys. Soc. Jpn.} \textbf{\bibinfo{volume}{85}},
  \bibinfo{pages}{022001} (\bibinfo{year}{2016}).

\bibitem[{\citenamefont{Hirashima and
  Namaizawa}(1987{\natexlab{a}})}]{hirashima1987p1}
\bibinfo{author}{\bibfnamefont{D.~S.} \bibnamefont{Hirashima}}
  \bibnamefont{and}
  \bibinfo{author}{\bibfnamefont{H.}~\bibnamefont{Namaizawa}},
  \bibinfo{journal}{Prog. Theor. Phys.} \textbf{\bibinfo{volume}{77}},
  \bibinfo{pages}{563} (\bibinfo{year}{1987}{\natexlab{a}}).

\bibitem[{\citenamefont{Hirashima and
  Namaizawa}(1987{\natexlab{b}})}]{hirashima1987p2}
\bibinfo{author}{\bibfnamefont{D.~S.} \bibnamefont{Hirashima}}
  \bibnamefont{and}
  \bibinfo{author}{\bibfnamefont{H.}~\bibnamefont{Namaizawa}},
  \bibinfo{journal}{Prog. Theor. Phys.} \textbf{\bibinfo{volume}{77}},
  \bibinfo{pages}{585} (\bibinfo{year}{1987}{\natexlab{b}}).

\bibitem[{\citenamefont{Hsiao}(2019)}]{hsiao}
\bibinfo{author}{\bibfnamefont{W.-H.} \bibnamefont{Hsiao}},
  \bibinfo{journal}{Phys. Rev. B} \textbf{\bibinfo{volume}{100}},
  \bibinfo{pages}{094510} (\bibinfo{year}{2019}).

\bibitem[{\citenamefont{Eschrig}(2000)}]{eschrig2000distribution}
\bibinfo{author}{\bibfnamefont{M.}~\bibnamefont{Eschrig}},
  \bibinfo{journal}{Phys. Rev. B} \textbf{\bibinfo{volume}{61}},
  \bibinfo{pages}{9061} (\bibinfo{year}{2000}).

\bibitem[{\citenamefont{Shapourian et~al.}(2015)\citenamefont{Shapourian,
  Hughes, and Ryu}}]{Shapourianstrain}
\bibinfo{author}{\bibfnamefont{H.}~\bibnamefont{Shapourian}},
  \bibinfo{author}{\bibfnamefont{T.~L.} \bibnamefont{Hughes}},
  \bibnamefont{and} \bibinfo{author}{\bibfnamefont{S.}~\bibnamefont{Ryu}},
  \bibinfo{journal}{Phys. Rev. B} \textbf{\bibinfo{volume}{92}},
  \bibinfo{pages}{165131} (\bibinfo{year}{2015}).

\bibitem[{\citenamefont{Yip and Sauls}(1992)}]{yip1992circular}
\bibinfo{author}{\bibfnamefont{S.}~\bibnamefont{Yip}} \bibnamefont{and}
  \bibinfo{author}{\bibfnamefont{J.~A.} \bibnamefont{Sauls}},
  \bibinfo{journal}{J. Low. Temp. Phys.} \textbf{\bibinfo{volume}{86}},
  \bibinfo{pages}{257} (\bibinfo{year}{1992}).

\bibitem[{\citenamefont{Ueki et~al.}(2018)\citenamefont{Ueki, Ohuchi, and
  Kita}}]{ueki2018charging}
\bibinfo{author}{\bibfnamefont{H.}~\bibnamefont{Ueki}},
  \bibinfo{author}{\bibfnamefont{M.}~\bibnamefont{Ohuchi}}, \bibnamefont{and}
  \bibinfo{author}{\bibfnamefont{T.}~\bibnamefont{Kita}}, \bibinfo{journal}{J.
  Phys. Soc. Jpn} \textbf{\bibinfo{volume}{87}}, \bibinfo{pages}{044704}
  (\bibinfo{year}{2018}).

\bibitem[{\citenamefont{Masaki}(2019)}]{masaki2019vortex}
\bibinfo{author}{\bibfnamefont{Y.}~\bibnamefont{Masaki}},
  \bibinfo{journal}{Phys. Rev. B} \textbf{\bibinfo{volume}{99}},
  \bibinfo{pages}{054512} (\bibinfo{year}{2019}).

\bibitem[{\citenamefont{Koch and W\"olfle}(1981)}]{koch}
\bibinfo{author}{\bibfnamefont{V.~E.} \bibnamefont{Koch}} \bibnamefont{and}
  \bibinfo{author}{\bibfnamefont{P.}~\bibnamefont{W\"olfle}},
  \bibinfo{journal}{Phys. Rev. Lett.} \textbf{\bibinfo{volume}{46}},
  \bibinfo{pages}{486} (\bibinfo{year}{1981}).

\bibitem[{sau()}]{sauls}
\bibinfo{note}{J. A. Sauls, Broken Symmetry and Non-Equilibrium Superfluid
  $^3$He, in {\it Topological Defects and Non-Equilibrium Symmetry Breaking
  Phase Transitions, Lecture Notes for the 1999 Les Houches Winter School},
  edited by H. Godfrin and Y. Bunkov (Elsevier, Amsterdam, 2000), pp. 239-265.}

\bibitem[{\citenamefont{Mast et~al.}(1980)\citenamefont{Mast, Sarma,
  Owers-Bradley, Calder, Ketterson, and Halperin}}]{mast}
\bibinfo{author}{\bibfnamefont{D.~B.} \bibnamefont{Mast}},
  \bibinfo{author}{\bibfnamefont{B.~K.} \bibnamefont{Sarma}},
  \bibinfo{author}{\bibfnamefont{J.~R.} \bibnamefont{Owers-Bradley}},
  \bibinfo{author}{\bibfnamefont{I.~D.} \bibnamefont{Calder}},
  \bibinfo{author}{\bibfnamefont{J.~B.} \bibnamefont{Ketterson}},
  \bibnamefont{and} \bibinfo{author}{\bibfnamefont{W.~P.}
  \bibnamefont{Halperin}}, \bibinfo{journal}{Phys. Rev. Lett.}
  \textbf{\bibinfo{volume}{45}}, \bibinfo{pages}{266} (\bibinfo{year}{1980}).

\bibitem[{\citenamefont{Giannetta et~al.}(1980)\citenamefont{Giannetta, Ahonen,
  Polturak, Saunders, Zeise, Richardson, and Lee}}]{giannetta}
\bibinfo{author}{\bibfnamefont{R.~W.} \bibnamefont{Giannetta}},
  \bibinfo{author}{\bibfnamefont{A.}~\bibnamefont{Ahonen}},
  \bibinfo{author}{\bibfnamefont{E.}~\bibnamefont{Polturak}},
  \bibinfo{author}{\bibfnamefont{J.}~\bibnamefont{Saunders}},
  \bibinfo{author}{\bibfnamefont{E.~K.} \bibnamefont{Zeise}},
  \bibinfo{author}{\bibfnamefont{R.~C.} \bibnamefont{Richardson}},
  \bibnamefont{and} \bibinfo{author}{\bibfnamefont{D.~M.} \bibnamefont{Lee}},
  \bibinfo{journal}{Phys. Rev. Lett.} \textbf{\bibinfo{volume}{45}},
  \bibinfo{pages}{262} (\bibinfo{year}{1980}).

\bibitem[{\citenamefont{Avenel et~al.}(1980)\citenamefont{Avenel, Varoquaux,
  and Ebisawa}}]{Avenel}
\bibinfo{author}{\bibfnamefont{O.}~\bibnamefont{Avenel}},
  \bibinfo{author}{\bibfnamefont{E.}~\bibnamefont{Varoquaux}},
  \bibnamefont{and} \bibinfo{author}{\bibfnamefont{H.}~\bibnamefont{Ebisawa}},
  \bibinfo{journal}{Phys. Rev. Lett.} \textbf{\bibinfo{volume}{45}},
  \bibinfo{pages}{1952} (\bibinfo{year}{1980}).

\bibitem[{\citenamefont{Movshovich et~al.}(1988)\citenamefont{Movshovich,
  Varoquaux, Kim, and Lee}}]{movshovich}
\bibinfo{author}{\bibfnamefont{R.}~\bibnamefont{Movshovich}},
  \bibinfo{author}{\bibfnamefont{E.}~\bibnamefont{Varoquaux}},
  \bibinfo{author}{\bibfnamefont{N.}~\bibnamefont{Kim}}, \bibnamefont{and}
  \bibinfo{author}{\bibfnamefont{D.~M.} \bibnamefont{Lee}},
  \bibinfo{journal}{Phys. Rev. Lett.} \textbf{\bibinfo{volume}{61}},
  \bibinfo{pages}{1732} (\bibinfo{year}{1988}).

\bibitem[{com({\natexlab{a}})}]{com2}
\bibinfo{note}{The PHA effect also arises from the velocity of normal
  electrons, modifying Eq.~(15). The PHA in the velocity of normal electrons
  renormalizes the PHA part of the Keldysh response function in Eq.~(15) as,
  $\underline{g}^{\rm K}_{\rm (1)} \to \kappa \underline{g}^{\rm K}_{\rm (1)}$,
  where $\kappa=1+a_v/a$, with $a_v=\frac{\epsilon_{\rm F}}{v_{\rm
  F}}\frac{\partial v_(\epsilon)}{\partial
  \epsilon}\big|_{\epsilon=\epsilon_{\rm F}}$ appearing due to the expansion of
  the velocity in the energy near the Fermi surface. This renormalization of
  the PHA part of the Keldysh Green's function changes the electric current
  quantitatively but not qualitatively. Hence, in the following, we set
  $\kappa=1$ to focus on the PHA of the DOS.}

\bibitem[{\citenamefont{Kobayashi et~al.}(2018)\citenamefont{Kobayashi,
  Matsushita, Mizushima, Tsuruta, and Fujimoto}}]{kobayashi2018negative}
\bibinfo{author}{\bibfnamefont{T.}~\bibnamefont{Kobayashi}},
  \bibinfo{author}{\bibfnamefont{T.}~\bibnamefont{Matsushita}},
  \bibinfo{author}{\bibfnamefont{T.}~\bibnamefont{Mizushima}},
  \bibinfo{author}{\bibfnamefont{A.}~\bibnamefont{Tsuruta}}, \bibnamefont{and}
  \bibinfo{author}{\bibfnamefont{S.}~\bibnamefont{Fujimoto}},
  \bibinfo{journal}{Phys. Rev. Lett.} \textbf{\bibinfo{volume}{121}},
  \bibinfo{pages}{207002} (\bibinfo{year}{2018}).

\bibitem[{\citenamefont{Uematsu et~al.}(2019)\citenamefont{Uematsu, Mizushima,
  Tsuruta, Fujimoto, and Sauls}}]{uematsu2019chiral}
\bibinfo{author}{\bibfnamefont{H.}~\bibnamefont{Uematsu}},
  \bibinfo{author}{\bibfnamefont{T.}~\bibnamefont{Mizushima}},
  \bibinfo{author}{\bibfnamefont{A.}~\bibnamefont{Tsuruta}},
  \bibinfo{author}{\bibfnamefont{S.}~\bibnamefont{Fujimoto}}, \bibnamefont{and}
  \bibinfo{author}{\bibfnamefont{J.}~\bibnamefont{Sauls}},
  \bibinfo{journal}{Phys. Rev. Lett.} \textbf{\bibinfo{volume}{123}},
  \bibinfo{pages}{237001} (\bibinfo{year}{2019}).

\bibitem[{com({\natexlab{b}})}]{com4}
\bibinfo{note}{See the definition of the Tsuneto function in Appendix B.}

\bibitem[{\citenamefont{Matsushita et~al.}(2022)\citenamefont{Matsushita, Ando,
  Masaki, Mizushima, Fujimoto, and Vekhter}}]{PhysRevLett.128.097001}
\bibinfo{author}{\bibfnamefont{T.}~\bibnamefont{Matsushita}},
  \bibinfo{author}{\bibfnamefont{J.}~\bibnamefont{Ando}},
  \bibinfo{author}{\bibfnamefont{Y.}~\bibnamefont{Masaki}},
  \bibinfo{author}{\bibfnamefont{T.}~\bibnamefont{Mizushima}},
  \bibinfo{author}{\bibfnamefont{S.}~\bibnamefont{Fujimoto}}, \bibnamefont{and}
  \bibinfo{author}{\bibfnamefont{I.}~\bibnamefont{Vekhter}},
  \bibinfo{journal}{Phys. Rev. Lett.} \textbf{\bibinfo{volume}{128}},
  \bibinfo{pages}{097001} (\bibinfo{year}{2022}).

\bibitem[{\citenamefont{Batlogg et~al.}(1986)\citenamefont{Batlogg, Bishop,
  Golding, Bucher, Hufnagl, Fisk, Smith, and Ott}}]{PhysRevB.33.5906}
\bibinfo{author}{\bibfnamefont{B.}~\bibnamefont{Batlogg}},
  \bibinfo{author}{\bibfnamefont{D.~J.} \bibnamefont{Bishop}},
  \bibinfo{author}{\bibfnamefont{B.}~\bibnamefont{Golding}},
  \bibinfo{author}{\bibfnamefont{E.}~\bibnamefont{Bucher}},
  \bibinfo{author}{\bibfnamefont{J.}~\bibnamefont{Hufnagl}},
  \bibinfo{author}{\bibfnamefont{Z.}~\bibnamefont{Fisk}},
  \bibinfo{author}{\bibfnamefont{J.~L.} \bibnamefont{Smith}}, \bibnamefont{and}
  \bibinfo{author}{\bibfnamefont{H.~R.} \bibnamefont{Ott}},
  \bibinfo{journal}{Phys. Rev. B} \textbf{\bibinfo{volume}{33}},
  \bibinfo{pages}{5906} (\bibinfo{year}{1986}).

\bibitem[{\citenamefont{Tinkham}(2004)}]{tinkham}
\bibinfo{author}{\bibfnamefont{M.}~\bibnamefont{Tinkham}},
  \emph{\bibinfo{title}{Introduction to Superconductivity}}
  (\bibinfo{publisher}{Dover Publications}, \bibinfo{year}{2004}).

\bibitem[{\citenamefont{Sigrist and Ueda}(1991)}]{RevModPhys.63.239}
\bibinfo{author}{\bibfnamefont{M.}~\bibnamefont{Sigrist}} \bibnamefont{and}
  \bibinfo{author}{\bibfnamefont{K.}~\bibnamefont{Ueda}},
  \bibinfo{journal}{Rev. Mod. Phys.} \textbf{\bibinfo{volume}{63}},
  \bibinfo{pages}{239} (\bibinfo{year}{1991}).

\bibitem[{\citenamefont{Moores and Sauls}(1993)}]{moores1993transverse}
\bibinfo{author}{\bibfnamefont{G.}~\bibnamefont{Moores}} \bibnamefont{and}
  \bibinfo{author}{\bibfnamefont{J.~A.} \bibnamefont{Sauls}},
  \bibinfo{journal}{J. Low Temp. Phys.} \textbf{\bibinfo{volume}{91}},
  \bibinfo{pages}{13} (\bibinfo{year}{1993}).

\end{thebibliography}
\bibliographystyle{apsrev}
\end{document}